\documentclass[11pt,a4paper]{article}

\usepackage[utf8]{inputenc}
\usepackage{amsmath}
\usepackage{amsfonts}
\usepackage{mathrsfs}
\usepackage{amssymb}
\usepackage[a4paper,top=2cm,bottom=2cm,left=2cm,right=2cm,marginparwidth=1.75cm]{geometry}
\usepackage{url}
\usepackage{hyperref}
\hypersetup{colorlinks, citecolor=black, linkcolor=black, urlcolor=blue}
\usepackage{graphicx}
\usepackage{bm}

\usepackage{subcaption}
\usepackage{array}
\usepackage{cancel}
\usepackage[T1]{fontenc}
 \usepackage{pdfpages}
\usepackage{autobreak}
\usepackage{simpler-wick}
\usepackage{floatrow}
\usepackage{booktabs}
\usepackage{multicol}
\usepackage{simpler-wick}
\usepackage{slashed}
\usepackage{graphicx}
\graphicspath{{images/}}
\usepackage{booktabs}
\usepackage{xcolor}
\usepackage{dsfont}
\usepackage{amsthm}

\usepackage{times}
\newcommand{\g}{{\gamma}}
\newcommand{\G}{{\Gamma}}
\newcommand{\eps}{{\epsilon}}

\newcommand{\ph}{{\phi}}
\newcommand{\phd}{{\phi^\dagger}}
\newcommand{\ps}{{\psi}}
\newcommand{\bps}{{\bar{\psi}}}

\renewcommand{\lg}{{\langle}}
\newcommand{\rg}{{\rangle}}
\newcommand{\tl}{{\tilde{\lambda}}}
\newcommand{\hl}{{\hat{\lambda}}}
\newcommand{\tN}{{\tilde{N}}}
\newcommand{\tb}{{\tilde{b}}}
\newcommand{\sign}{{\text{sign}}}

\newcommand{\perm}[2]{\begin{Bmatrix}#1 \\#2\end{Bmatrix}}
\renewcommand{\d}{\partial}
\newcommand{\ds}{{\slashed{\d}}}

\usepackage[sorting=none]{biblatex}
\numberwithin{equation}{section}
\title{Four-point functions and contact terms from higher-spin Ward identities of Chern-Simons-matter theory}
\author{Trivko Kukolj\footnote{trivko.kukolj@weizmann.ac.il} \\\\
\it \normalsize{Department of Particle Physics and Astrophysics, Weizmann Institute of Science,} \\[3pt] \normalsize{\it Rehovot 7610001, Israel.}}
\date{}

\begin{document}

\maketitle

\begin{abstract}
Large $N$ quasifermionic ($QF$) Chern-Simons-matter theories exhibit weakly-broken higher-spin symmetry and contain an infinite-dimensional algebra of almost-conserved higher-spin currents. By analyzing local higher-spin Ward identities, we constrain the higher-spin algebra of this theory, for operators of spin $s<4$. The solution interpolates smoothly between Ward identities of free-fermionic and critical-bosonic $U(N)$ models, in accordance with the bosonization duality. At finite 't Hooft coupling, we decompose four-point current correlators involving $(\d\cdot J_3)^{QF}$ in terms of free theory structures and identify possible epsilon-transform relations for $QF$ four-point functions. Additionally, we find that higher correlators of the theory develop scheme-independent higher-spin contact terms, stemming from the presence of a background Chern-Simons term in the effective action. These terms are related to similar contact terms of the associated quasibosonic Chern-Simons-matter theory via Legendre transform. The results are supplemented by perturbative checks, including collinear limit computations of $\lg T_{-{}-}J_+J_+\rg_{QF}$, $\lg T_{-{}-}J_+J_+O\rg_{QF}$ and $\lg J_{-{}-{}-}J_+J_+J_+\rg_{QF}$.
\end{abstract}
\newpage
\tableofcontents
\newpage

\section{Introduction}\label{Sec1Intro}

Three dimensional conformal field theories of matter minimally coupled to Chern-Simons gauge interactions exhibit remarkable properties in several aspects. Generalizing the well-understood rank/level duality of pure Chern-Simons theory \cite{Hsin:2016blu}, such models commonly appear in recent literature thanks to an abundance of strong-weak dualities relating bosons and fermions. In its simpler form, this 3d bosonization duality relates non-supersymmetric theories containing fundamental matter with unitary or orthogonal gauge groups \cite{Aharony:2015mjs, Aharony:2018pjn}, although several extensions have been investigated \cite{Aharony:2016jvv, Jain:2013gza}, including supersymmetric theories where the correspondence arises from three-dimensional Seiberg dualities \cite{Benini:2011mf, Giveon:2008zn}. 

More than just field theory duals, Chern-Simons-matter theories arise as holographic duals of interacting higher-spin gravity theories on $AdS_4$. Indeed, the initial reasoning for the study of various Chern-Simons-matter CFTs came from generalizing the duality between type A and type B Vasiliev theories and free/critical bosonic/fermionic vector models \cite{Klebanov:2002ja,Sezgin:2003pt}. These dualities are best understood in the large $N$ limit, where both sides are weakly coupled and can be studied perturbatively. On the gravity side, this appears as the classical limit of higher-spin theory, with a bulk coupling constant proportional to $1/\sqrt{N}$. Conversely, at finite $N$, these dualities should provide us with quantum generalizations of gravity, involving a tower of massless higher-spin particles. Recent works have pointed out the existence of UV complete chiral higher-spin theories, equipped with a local action \cite{Skvortsov:2018uru, Skvortsov:2020wtf}, further motivating the study Chern-Simons-matter theories as toy models of $AdS_4/CFT_3$. 

Higher-spin gauge invariance in the bulk manifests as an infinite-dimensional higher-spin symmetry of the dual CFT. This was shown to be rather constraining, as the only theories with exact higher-spin symmetry are essentially free theories \cite{Maldacena:2011jn}. By introducing Chern-Simons interactions, one breaks the symmetry at subleading orders in $N$. The previously conserved higher-spin currents $J_s$ now develop divergences in the form of multi-trace operators:
\begin{flalign}
	(\d\cdot J_s) \propto \frac{1}{N}[J_{s_1}][J_{s_2}]+\frac{1}{N^2}[J_{s_1}][J_{s_2}][J_{s_3}]+\mathcal{O}\bigg(\frac{1}{N^3}\bigg), \quad\text{ for }s>2
\end{flalign}
Nevertheless, for a specific class of large $N$ theories, slightly broken higher-spin symmetry can still be leveraged to constrain correlation functions \cite{Maldacena:2012sf}. Furthermore, in the usual sense, it allows one to build conserved charges and discuss variations of operators, to leading order in large $N$, giving rise to an algebra of higher-spin currents. The main goal of this work is to investigate the application of these techniques to the theory of Dirac fermions gauged by $U(N)_{k}$ Chern-Simons interactions in the large $N$, 't Hooft limit $N\to\infty, k\to\infty$ with $N/k$ fixed. This theory is often referred to as the quasifermionic (QF) Chern-Simons-matter theory \cite{Giombi:2011kc,Gur-Ari:2012lgt}, and is conjectured to be dual\footnote{Outside the large $N$, 't Hooft limit, the duality is slightly more subtle, as argued in \cite{Aharony:2018pjn, Seiberg:2016gmd}.} to the theory of Wilson-Fisher scalars coupled to a $U(k)_{-\sign(k)N}$ Chern-Simons gauge field. We analyze three- and four-point functions by writing local higher-spin Ward identities of the spin-3 current. At separated points, this leads to a decomposition of QF correlators into a small number of well known kinematic structures. 

Of special interest is the analysis of higher-spin contact terms. QF correlators of spinning currents may develop contributions at coincident points, stemming from the presence of background field Chern-Simons terms in the effective action. As argued in \cite{Closset:2012vp}, the fractional parts of these terms correspond to intrinsic observables of the theory. They already arise when discussing the two-point function of the vector current or the stress tensor, which develop parity-breaking contributions consistent with current conservation. Imposing higher-spin symmetry locally leads to a systematic set of constraints, from which one can answer how higher correlators of the theory are affected by such scheme-independent contact terms. This serves to further probe the dualities between the bosonic and fermionic descriptions at large $N$, as well as to better understand the higher-spin gravity dual.\\

The remainder of Section \ref{Sec1Intro} lists the conventions used throughout the text. Section \ref{Sec2} introduces the basics of general higher-spin algebras and associated Ward identities, as well as their computation in free theories. Section \ref{Sec4} details local Ward identities of the critical boson theory, as the simplest example of an interacting CFT with weakly broken higher-spin symmetry. Section \ref{Sec5} presents the constraints on the quasifermionic algebra, including results for four-point functions and higher-spin contact terms. Section \ref{Sec6} contains collinear limit computations of quasifermionic correlators and perturbative checks of spin-3 Ward identities. Section \ref{Sec7} summarizes the main results and discusses possible extensions of this work. Lengthier technical results are collected in appendices. 

\subsection{Normalizations and Conventions}\label{Sec1}
We will use the following abbreviations to denote operators/correlators of different three-dimensional theories: 
FF = free fermionic $U(N)$ model; FB = free bosonic $U(N)$ model; CF = critical fermionic $U(N)$ model; CB = critical bosonic $U(N)$ model; QF = quasifermionic theory; QB = quasibosonic theory. Our conventions for free theory operators are:\\

\underline{\textbf{Free Boson Theory}}
\begin{flalign}\begin{aligned}\label{1FBCurrents}

    O^{FB}&=8\phd\ph\\
    
    J^{FB}_{\mu}&=\phd\bigg(i\overleftarrow{\d_\mu}-i\overrightarrow{\d_\mu} \bigg)\ph\\
    
    T^{FB}_{\mu\nu}&=\phd\bigg(-\frac{1}{4} \overleftarrow{\d_\mu} \overleftarrow{\d_\nu} - 
    						 \frac{1}{4}\overrightarrow{\d_\mu}\overrightarrow{\d_\nu} + 
    						 \frac{3}{2}\overleftarrow{\d_{(\mu}} \overrightarrow{\d_{\nu)}} + 
    						 \frac{g_{\mu\nu}}{12}\overleftarrow{\d^2} + 
    						 \frac{g_{\mu\nu}}{12}\overrightarrow{\d^2} - 
    						 \frac{g_{\mu\nu}}{2}\overleftarrow{\d_{\sigma}}\overrightarrow{\d^\sigma} \bigg)\ph\\
    
    J^{FB}_{\mu\nu\rho}&=\phd\bigg(\frac{1}{6}\overleftarrow{\d_\mu}\overleftarrow{\d_\nu}\overleftarrow{\d_\rho} - 
    							 \frac{1}{6}\overrightarrow{\d_\mu}\overrightarrow{\d_\nu}\overrightarrow{\d_\rho} + 
    							 \frac{5}{2}\overleftarrow{\d_{(\mu}}\overrightarrow{\d_\nu}\overrightarrow{\d_{\rho)}} - 
    							 \frac{5}{2}\overleftarrow{\d_{(\mu}}\overleftarrow{\d_\nu}\overrightarrow{\d_{\rho)}} - 
    							 g_{(\mu\nu}\overleftarrow{\d^\sigma}\overrightarrow{\d_{\rho)}}\overrightarrow{\d_\sigma} +
    							 g_{(\mu\nu}\overleftarrow{\d_{\rho)}}\overleftarrow{\d^\sigma}\overrightarrow{\d_\sigma}\bigg)\ph 							
    							 \\ & \quad     
    							 + \phd\bigg(-\frac{1}{2}g_{(\mu\nu}\overleftarrow{\d_{\rho)}}\overrightarrow{\d^2} + 
    							 \frac{1}{2}g_{(\mu\nu}\overrightarrow{\d_{\rho)}}\overleftarrow{\d^2} -
    							 \frac{1}{10}g_{(\mu\nu}\overleftarrow{\d_{\rho)}}\overleftarrow{\d^2} +
    							 \frac{1}{10}g_{(\mu\nu}\overrightarrow{\d_{\rho)}}\overrightarrow{\d^2}\bigg)\ph
\end{aligned}\end{flalign}

\underline{\textbf{Free Fermion Theory}}

In three dimensions, the generators of the Dirac algebra are given by Pauli matrices $\gamma^\mu=\sigma^\mu$, for $\mu=1,2,3$. The conjugate Dirac spinor field is now given by $\bps=\ps^\dagger$. The algebra satisfies:
\begin{flalign}\label{1:Dirac}
    [\g^\mu,\g^\nu]=2i\eps^{\mu\nu\rho}\g_\rho; \qquad\{\gamma^\mu,\gamma^\nu\}=2g^{\mu\nu}\mathds{1}_{2\times 2} \text{ with } g^{\mu\nu}=diag(1,1,1) =\delta^{\mu\nu} 
\end{flalign}
The current operators for the theory of a single massless Dirac fermion are defined as:
\begin{flalign}\begin{aligned}\label{1FFCurrents}

    O^{FF}&=\bps\ps\\
    
    J^{FF}_{\mu}&=i\bps\g_\mu\ps\\
    
    T^{FF}_{\mu\nu}&=\bps\bigg(-\frac{1}{2}\g_{(\mu}\overrightarrow{\d_{\nu)}} + 
    							\frac{1}{2}\overleftarrow{\d_{(\nu}} \g_{\mu)} +
   							\frac{1}{6}g_{\mu\nu}\overrightarrow{\ds} - 
   							\frac{1}{6}g_{\mu\nu}\overleftarrow{\d^\sigma}\g_\sigma \bigg)\ps \\
    
    J^{FF}_{\mu\nu\rho}&=\bps\bigg(
    							\frac{1}{2}\overleftarrow{\d_{(\mu}}\overleftarrow{\d_\nu}\g_{\rho)} +
    							\frac{1}{2}\g_{(\mu}\overrightarrow{\d_\nu}\overrightarrow{\d_{\rho)}}   - 
    							\frac{5}{3}\overleftarrow{\d_{(\mu}}\g_\nu\overrightarrow{\d_{\rho)}} + 
    							\frac{g_{(\mu\nu}}{3}\overleftarrow{\d^\sigma}\g_{\rho)}\overrightarrow{\d_\sigma} \bigg)\ps 
    							\\ & \quad 
    							-\bps\bigg(\frac{g_{(\mu\nu}}{5}\overrightarrow{\d_{\rho)}}\overrightarrow{\ds} + 
    							\frac{g_{(\mu\nu}}{5}\overleftarrow{\d_{\rho)}}\overleftarrow{\d^\sigma}\g_\sigma + 
    							\frac{g_{(\mu\nu}}{3}\overleftarrow{\d_{\rho)}}\overrightarrow{\ds} + 
    							\frac{g_{(\mu\nu}}{3}\overleftarrow{\d^\sigma}\g_\sigma\overrightarrow{\d_{\rho)}} - 
    							\frac{g_{(\mu\nu}}{10}\overleftarrow{\d^2} \g_{\rho)} + 
    							\frac{g_{(\mu\nu}}{10}\g_{\rho)}\overrightarrow{\d^2} \bigg)\ps
\end{aligned}\end{flalign}
The normalizations are chosen so that two-point functions of the given currents are identical in FF and CB theories. Symmetrization comes with the appropriate $1/n!$ factor. In Sections \ref{Sec5}, \ref{Sec6} and appendices, the subscripts $\lg ...\rg_{FF}$ and $\lg ...\rg_{FB}$ will denote one half of the value of those in the theory of a single Dirac fermion and complex boson, respectively, and carry an extra $1/2$ factor relative to Sections \ref{Sec1Intro}-\ref{Sec4}. We define all momentum space correlators with $(2\pi)^3\delta^{(3)}(\sum p)$ factors stripped.\\

As a shorthand, in the theory labeled by the subscript $X$, we denote the Ward identity where $(\d\cdot J_s)$ is inserted into the correlation function $\lg O_1O_2...O_n\rg_X$ by $\lg [Q_s,O_1O_2...O_n]\rg_X$.
\section{Higher-spin algebras}\label{Sec2}
Three-dimensional CFTs possessing at least one conserved higher-spin current have been shown to exactly satisfy higher-spin symmetry, with their correlation functions matching those of free field theories \cite{Maldacena:2011jn}. Relaxing these requirements, one can next look at large $N$ theories where higher-spin symmetry is broken by $1/N$ corrections. Two particularly well understood classes of such CFTs were named quasibosonic and quasifermionic theories. 

Such theories contain a spectrum of single-trace operators congruent with Vasiliev theories. They have a large $\tN$ parameter, which plays a role analogous to $N$ in free theories. Depending on the details of the theory\footnote{There are examples where the spectrum is truncated, e.g. $O(N)$ theories with only even-spin currents $J_{2s}$. In Chern-Simons-matter theories, one can also consider extended operators \cite{Gabai:2022vri,Gabai:2022mya,Gabai:2023lax}.}, we generically have an infinite tower of conformal primaries $J_s$ with integer spin $s>0$ and scaling dimension $\Delta_s=s+1+\mathcal{O}(1/\tN)$. In addition, QB theories contain a scalar $J_0$ of dimension $\Delta_0=1+\mathcal{O}(1/\tN)$, while QF theories contain a scalar $\tilde{J_0}$ of dimension $\Delta_{\tilde{0}}=2+\mathcal{O}(1/\tN)$. To leading order the currents are conserved, therefore for each spin-$s$ current $J_{\mu_1\mu_2..\mu_s}$ we can build a (pseudo)charge by integrating its zeroth component over a closed, equal-time surface $\Sigma$:
\begin{flalign}
Q_{\mu_1\mu_2...\mu_{s-1}}= \int_\Sigma * J = \int_{\Sigma} J^0_{\ \ \mu_1\mu_2...\mu_{s-1}}
\end{flalign}
We will mostly consider the full operator $Q_{\mu_1\mu_2...\mu_{s-1}}$, however its specific components can be accessed by first contracting the free indices with a specified conformal Killing tensor $\zeta^{\mu_1\mu_2...\mu_{s-1}}$ and then integrating. An example of this in the lightcone frame is given in \cite{Maldacena:2011jn}. Each charge $Q_s$ carries spin $S=s-1$ and scaling dimension $\Delta=s-1+\mathcal{O}(1/\tN)$. Even though QB and QF theories do not have an exact higher-spin symmetry, to leading order, one can write down Ward identities, as derived in \cite{Maldacena:2012sf}. We emphasize the difference between \textit{local} and \textit{integrated} Ward identities written therein. The latter comes from integrating the former over the insertion of $(\d\cdot J_{s\geq 3})$, however this difference becomes important when discussing theories with weakly-broken higher-spin symmetry.

As in \cite{Maldacena:2012sf}, we define the action of higher-spin charges on local operators $[Q_s,O_i]$, which will appear in integrated higher-spin Ward identities. The action of the charge is defined as the finite part of a closed spatial surface integral over the selected operator\footnote{Note that here we assume the absence of any other nearby operators.} $O_i$, with the insertion of the corresponding higher-spin current $J_s$. Furthermore, $[Q_s,O_i]$ is not sensitive to the choice of this closed surface, so we can imagine a tiny sphere around $O_i$ whose radius is then taken to zero.
\begin{flalign}\label{2:[Q,J]def}
	[Q_{\mu_1\mu_2...\mu_{s-1}},O_i]=\int_{|x|=r} \hat{n}\cdot J_s \times O_i \vert_{\text{finite as }r\to 0}
\end{flalign}
This variation can be found directly in the case of free theories, as we will show in Sections \ref{Sec2.2}, \ref{Sec2.3}, however one can already draw constraints from using the CFT data, as we will do in Section \ref{Sec2.1}.

\subsection{\texorpdfstring{$[Q_s,J_{s'}]$}{} from conformal quantum numbers}\label{Sec2.1}
Assuming a general QF or QB theory, we outline an algorithm for constraining the action of the (pseudo)charge of a slightly conserved higher-spin current $Q_{l+1}=Q_{\mu_1\mu_2..\mu_l}$ on a general current primary $J_{s}=J_{\alpha_1\alpha_2...\alpha_s}$, up to coefficients, which may be constants or general functions of parameters of the theory. We proceed under the assumption that both the QF scalar current $\tilde{J_0}$ and the QB scalar current $J_0$ are present in the spectrum. The end result for the QF theory is then simply obtained by setting $J_0=0$ and for the QB theory, by setting $\tilde{J_0}=0$. 
 
The anomalous dimensions of current primaries receive corrections only at subleading order in large $\tN$, which will not play a role here. The (pseudo)charge $Q_{l+1}$ has dimension $\Delta=l$ and spin $S=l$, while a spin $s$ current $J_s$ has dimension $\Delta=s+1$ and spin $S=s$. The QF scalar $\tilde{J_0}$ has dimension $\Delta=2$, but spin $S=0$, so we will treat the case $[Q_s,\tilde{J_0}]$ separately of the other generic cases.  A general $[Q_{l+1},J_s]$ has total scaling dimension $\Delta=l+s+1$ and total spin $S=l+s$. Schematically, we can write 
\begin{flalign}\begin{aligned}
	[Q_{\mu_1..\mu_l},J_{\alpha_1..\alpha_s}]=&
	
	A^1_{\nu_1\nu_2...\nu_{s+l}\rho_1...\rho_{s+l}}\sum_{k=\max\{s-l,0\}}^{s+l}\d^{\rho_1}\d^{\rho_2}...\d^{\rho_k}(J_{s+l-k})^{\rho_{k+1}...\rho_{s+l}}\\&
	
	+B^1_{\nu_1\nu_2...\nu_{s+l}\rho_1...\rho_{s+l-1}}\d^{\rho_1}\d^{\rho_2}...\d^{\rho_{s+l-1}}\tilde{J_0}\\&
	
	+\frac{1}{\tN}A^2_{\nu_1\nu_2...\nu_{s+l}\rho_1...\rho_{s+l-1}}\mathcal{O}_2^{\rho_1...\rho_{s+l-1}}([J_{s_1}][J_{s_2}])\\&
	
	+\frac{1}{\tN}B^2_{\nu_1\nu_2...\nu_{s+l}\rho_1...\rho_{s+l-2}}\mathcal{O}_2^{\rho_1...\rho_{s+l-2}}([J_{s_1}][\tilde{J_{0}}])\\&	
	
	+\frac{1}{N}C^2_{\nu_1\nu_2...\nu_{s+l}\rho_1...\rho_{s+l-3}}\mathcal{O}_2^{\rho_1...\rho_{s+l-3}}([\tilde{J_{0}}][\tilde{J_{0}}])
	
	+\frac{1}{\tN^2}(\text{Triple-trace operators } \mathcal{O}_3) +...

\end{aligned}\end{flalign}
Contributing single-trace operators are explicitly written in the first two lines on the right-hand side. The lower bound on the right-hand side sum comes from arguments presented in \cite{Maldacena:2011jn}, while the upper bound comes from exhausting all single trace primaries. By $\mathcal{O}_k([A_1][A_2]...[A_k])$ we denote $k$-trace operators formed by multiplying single trace primaries $A_i$, with derivatives $\d^{\rho_j}$ sprinkled in. For fixed $s$ and $l$, there is a finite number of such terms. For each $k$ we form all possible $\mathcal{O}_k$ by matching spin and scaling to the total spin and scaling dimension of the left-hand side. After listing all double-trace contributions $\mathcal{O}_2$, we list all triple-trace contributions $\mathcal{O}_3$, and so forth, until we exhaust all possible $\mathcal{O}_n$. 

By construction, all $\rho$'s are dummy indices, which we seek to replace by $\{\nu_1,\nu_2,...,\nu_{s+l}\}=\{(\mu_1,\mu_2...,\mu_l),$ $(\alpha_1,\alpha_2,...\alpha_s)\}$\footnote{The $()$ brackets here denote symmetrizing over $\alpha$'s and $\rho$'s separately.}. $A^1$, $B^1$, $A^2$, $B^2$, $C^2$, and so forth represent general Lorentz invariant structures that we can use to make these replacements, each giving us a specific term on the right-hand side. These are built out of the metric and Levi-Civita symbols. Every such structure can be represented by a set of numbers $(a,b,c,d,e,f,g)$ with each letter corresponding to the number of appearances of a term from Table \ref{2.1:Table}.
\begin{table}[h]
	\caption{Building blocks of Lorentz structures $A^1$, $B^1$, etc. accompanying  multitrace operators that appear in $[Q_{l+1},J_s]$}
	\label{2.1:Table}
    \centering
    \begin{tabular}{|c c c c c c c|}
        \hline
        Number of terms & Term & $\rho$'s lowered & $\nu$'s used & Symmetry & Antisymmetry & $\neq 0$ for $\nu_i=-$ \\
        \hline
        $a$ & $g_{\nu_i\nu_j}$ & 0 & 2 & $\nu_i\leftrightarrow\nu_j$ &  &  \\
        
        $b$ & $\eps_{\nu_i\nu_j\nu_k}$ & 0 & 3 &  & $\nu_i\leftrightarrow\nu_j\leftrightarrow\nu_k$ &  \\
        
        $c$ & $g_{\nu_i\rho_j}$ & 1 & 1 & $\nu_i\leftrightarrow\rho_j$ &  & $\times$ \\
        
        $d$ & $\eps_{\nu_i\nu_j\rho_k}$ & 1 & 2 &  & $\nu_i\leftrightarrow\nu_j\leftrightarrow\rho_k$ &  \\
        
        $e$ & $g_{\rho_i\rho_j}$ & 2 & 0 & $\rho_i\leftrightarrow\rho_j$ &  & $\times$ \\
        
        $f$ & $\eps_{\nu_i\rho_j\rho_k}$ & 2 & 1 &  & $\nu_i\leftrightarrow\rho_j\leftrightarrow\rho_k$ & $\times$ \\
        
        $g$ & $\eps_{\rho_i\rho_j\rho_k}$ & 3 & 0 &  & $\rho_i\leftrightarrow\rho_j\leftrightarrow\rho_k$ & $\times$ \\
        \hline
    \end{tabular}
\end{table}
For each $A^1$, $B^1$, $A^2$, and so on, finding all possible terms in $[Q_{l+1},J_s]$ now reduces to finding all possible sets of integers $a-g$ satisfying certain constraints. Let's focus on $A^1$, with the other terms following analogously. $A^1$ has $(s+l)$ $\nu$ indices and $(s+l)$ $\rho$ indices. These have to be split between $a-g$ terms, therefore we have two equations: 
\begin{flalign}
c+d+2e+2f+3g=s+l; \qquad 2a+3b+c+2d+f=s+l
\end{flalign}
Additionally, we have constraints on each of the integers: 
\begin{itemize}
  \item $a\geq 0$, $c\geq 0$, $e\geq 0$ with no further constraints on these integers.
  \item Out of three $\nu$ indices, two have to be either $\alpha$'s or $\mu$'s. However since $[Q_{l+1},J_s]$ is symmetric with respect to these separately, any term antisymmetric in three $\nu$ indices will not appear. Hence we find $b=0$. Note that this also holds generally for all the Lorentz structures.
  \item We need at most one epsilon symbol, since the product of two epsilon symbols can always be written as a sum of products of $g_{\nu_i\nu_j}$, $g_{\nu_i\rho_j}$ and $g_{\rho_i\rho_j}$. Therefore $d\geq 0$, $f\geq 0$, $g\geq 0$ and $d+f+g\leq 1$. These constraints also hold generally.
  \item Further constraints can be imposed, but these are term-specific. For example, $A_1$ multiplies \newline $\d^{\rho_1}\d^{\rho_2}...\d^{\rho_k}(J_{s+l-k})^{\rho_{k+1}...\rho_{s+l}}$. Therefore multiplying it by any term that is antisymmetric in three $\rho$ indices will yield zero. In this case we can say that $g=0$.
\end{itemize}

We define the twist of each operator to be in the lightcone-minus direction as $\tau=\Delta-s$, where $\Delta$ is the scaling dimension and $s$ is the number of minus indices. When all $\nu$ indices are $\nu_i=-$, the pseudo-charge has twist $\tau=0$, and only operators with $\tau=1$ can contribute on the right-hand side. This makes all Lorentz structures multiplying multi-trace terms vanish, as well as $B^1$, while, as discussed in \cite{Maldacena:2011jn}, the structure $A^1$ can be either $g_{-\rho_1}g_{-\rho_2}...g_{-\rho_{s+l}}$ or zero. Referring to table \ref{2.1:Table}, this constraint forbids all terms $A^1$ that satisfy $c+2e+2f+3g=s+l$ and $c+f=s+l$, except $(a,b,c,d,e,f,g)=(0,0,s+l,0,0,0,0)$. However one can show that indeed this system of equation has no other solutions. For $B^1$, $A^2$, etc. the constraint forbids all solutions that satisfy $c+2e+2f+3g\leq 0$ and $c+f=s+l$. However, this system also does not have any would-be-forbidden solutions. Overall, we see that twist conservation imposes no further constraints. \\

$[Q_{l+1},\tilde{J_0}]$ can be constrained similarly, by treating $\tilde{J_0}$ as a spin-1 current, but with one less Lorentz index. The difference being that the Lorentz structures we described now contain one less $\nu$ index, e.g. $A^1_{\nu_1\nu_2...\nu_{s+l-1}\rho_1...\rho_{s+l}}$ and so forth. The rest of the discussion follows analogously.

To each term we assign a separate coefficient. The last step is to use the fact that the currents are traceless, bosonic operators. This implies:
\begin{flalign}
	g^{\mu_i\mu_j}[Q_{\mu_1..\mu_i..\mu_j..\mu_l},J_{\alpha_1..\alpha_s}]=0 \text{,} \quad\quad  
	g^{\alpha_i\alpha_j}[Q_{\mu_1..\mu_l},J_{\alpha_1..\alpha_i..\alpha_j..\alpha_s}]=0 \text{,}\quad\quad 
	 \eps_{\sigma_1\sigma_2\sigma_3}J_s^{\sigma_1} J_s^{\sigma_2} =0
\end{flalign}
Which can make certain terms vanish, or relate coefficients of different terms. In some cases, the analysis may be helped by symmetries of the theory. For example, any terms that are not invariant under $\vec{x}\to -\vec{x}$ would be forbidden if the theory is parity invariant. The computation now reduces to finding all solutions to the system of constraints. This is easily done using computation packages such as Mathematica, even for larger values of $l$ and $s$.

The action of the spin-3 (pseudo)charge on scalar and vector currents will be of primary interest. To simplify the discussion, we assume the CFT has discrete $\mathbb{Z}_2$ symmetry\footnote{In the FF theory, this is just part of the larger charge conjugation symmetry, as can be checked using \eqref{1:Dirac}, \eqref{1FFCurrents} and the charge conjugation matrix $C=\gamma^2$. This symmetry is still present in the full QF Chern-Simons-matter theory, discussed in Section \ref{Sec5}.} under which $J_s\to(-1)^s J_s$. We write out these commutators for QF and QB theories:
\begin{flalign}\begin{aligned}\label{2.1:[Q,J]qf}
[Q_{\mu_1\mu_2},\tilde{J_0}]_{QF}=& c_0^{QF} \eps_{\rho_1\rho_2(\mu_1}\d_{\mu_2)}\d^{\rho_1}J^{\rho_2}\\

[Q_{\mu_1\mu_2},J_\alpha]_{QF}=& c_1^{QF}\eps_{\alpha\rho(\mu_1}\d_{\mu_2)}\d^\rho \tilde{J_0}+c_2^{QF}\d_{(\mu_1}J_{2\mu_2)\alpha}+c_3^{QF}\d_{\alpha}J_{2\mu_1\mu_2}+\frac{c_4^{QF}}{\tN}\eps_{\alpha\rho(\mu_1}J_{1\mu_2)}J_1^\rho \qquad\quad
\end{aligned}\end{flalign}
\begin{flalign}\begin{aligned}\label{2.1:[Q,J]qb}
[Q_{\mu_1\mu_2},J_0]_{QB}=& c_0^{QB} \d_{(\mu_1}J_{\mu_2)}\\

[Q_{\mu_1\mu_2},J_\alpha]_{QB}=& 
c_1^{QB}g_{\mu_1\mu_2}\d_\alpha\d^2 J_0 + c_2^{QB}g_{\alpha(\mu_1}\d_{\mu_2)}\d^2 J_0 - (3c_1^{QB}+c_2^{QB})\d_{\mu_1}\d_{\mu_2}\d_\alpha J_0 + c_3^{QB}\d_{(\mu_1}J_{2\mu_2)\alpha} \\& 

+ c_4^{QB}\d_{\alpha}J_{2\mu_1\mu_2} + \frac{c_5^{QB}}{\tN}\eps_{\rho\alpha(\mu_1}J_0 \d_{\mu_2)}\d^\rho J_0 + \frac{c_6^{QB}}{\tN}\eps_{\rho\alpha(\mu_1}\d^\rho J_0 \d_{\mu_2)} J_0 + \frac{c_{7}^{QB}}{\tN}\eps_{\alpha\rho(\mu_1}J_{1\mu_2)}J_1^\rho
\end{aligned}\end{flalign}
The FB and FF algebras contain the same contributions as those appearing in QB and QF theories, respectively. In the case of FB theories, we denote the algebra coefficients in place of $c_i^{QB}$ by $b_i$, while for FF theories, we replace the $c_i^{QF}$ coefficients by $a_i$. We fix down the constants $a_i$ and $b_i$ by direct calculation in Sections \ref{Sec2.2} and \ref{Sec2.3}. One can similarly constrain the algebra for higher spins. For example, the action of the spin-4 (pseudo)charge on the scalar operator of QF theories is given by:
\begin{flalign}\begin{aligned}
[Q_{\mu_1\mu_2\mu_3},\tilde{J_0}]_{QF} =& 
d_1^{QF}\eps_{\rho_1\rho_2(\mu_2}\d_{\mu_2}\d^{\rho_1}J_{2\mu_3)}^{\rho_2}-2d_2^{QF}\d_{\mu_1}\d_{\mu_2}\d_{\mu_3}\tilde{J_0}+d_2^{QF}g_{(\mu_1\mu_2}\d_{\mu_3)}\d^2\tilde{J_0}\\&

-\frac{d_3^{QF}}{5\tN}g_{(\mu_1\mu_2}(J_1\cdot\d)J_{1\mu_3)} -\frac{d_3^{QF}}{5\tN}g_{(\mu_1\mu_2}J_1^\rho\d_{\mu_3)}J_{1\rho}+\frac{d_3^{QF}}{\tN}J_{1(\mu_1}\d_{\mu_2}J_{1\mu_3)}
\end{aligned}\end{flalign}
And so forth, with the number of possible terms increasing quickly, as we increase $l$ and $s$. The analysis also applies to charges built from spin-1 and spin-2 currents. In free theories, with operators given by \eqref{1FBCurrents}, \eqref{1FFCurrents}, it's easy to see that all the current operators remain uncharged under the $U(1)$ symmetry, while the translation operators built from stress-tensors act as $P_\nu^{FB,FF} = \d_\nu$. To simplify notation, we will denote the spin-0,1,2 currents by $O, J_\mu, T_{\mu\nu}$ respectively, in further text.

\subsection{Free Fermion algebra from OPE}\label{Sec2.2}
Free theories are important building blocks for the study of higher-spin symmetry in any number of dimensions. In this context, we will encounter them as limits of Chern-Simons-matter theories. Hence we first study the higher-spin algebra of a three-dimensional theory of a single massless Dirac fermion $\ps$. In this case we know the form of all higher-spin currents explicitly \cite{Giombi:2016zwa}, and the algebra can be computed directly using \eqref{2:[Q,J]def} and Stokes theorem:
\begin{flalign}\label{2.2:FFOPE}
	[Q_{\mu_1\mu_2...\mu_{s-1}},O_i(0)]=\int_{|x|=r} \hat{n}\cdot J_s \times O_i \vert_{\text{finite as }r\to 0} = \int_{V} d^3x \ \ \d_x^\mu \big(J_{\mu\mu_1\mu_2...\mu_{s}}(x)\times O_i(0) \big)_{\text{finite as }r\to 0}
\end{flalign}
Which reduces to computing free fermion OPEs. The operator $O_i$ is placed at the origin, while $V$ is the volume of the small sphere encompassing the product, whose radius $r$ is taken to zero. Since higher-spin symmetry is exact, $\d\cdot J_s = 0$, we only expect the OPE to produce terms diverging as delta functions. Then upon volume integration we only end up with finite terms in the $r\to 0$ limit.
We make use of the basic properties of the free fermion propagator:
\begin{flalign}\begin{aligned}\label{2.2:ffprop}
S(x)&=\lg\bps(x)\ps(0)\rg = \lg\ps(x)\bps(0)\rg =-\frac{1}{4\pi}\ds\bigg(\frac{1}{|x|}\bigg); \qquad\qquad
\ds S(x)= \d^\sigma S(x)\g_\sigma = \delta(x)
\end{aligned}\end{flalign}
Along with the free fermion equation of motion $\ds\ps(x)=\d^\sigma\bps(x)\g_\sigma = 0$, as well as properties of the Dirac algebra defined in \eqref{1:Dirac}. We confine ourselves to operators defined in Section \ref{Sec1}, however the procedure applies quite generally to currents and charges of any spin. We first analyze the action of the spin-3 charge on the fundamental fields $\ps$ and $\bps$. The former reads:
\begin{flalign}\begin{aligned}
     [Q_{\nu\rho},\ps(0)] 
     &= \int_V \d^\mu \bigg(J_{\mu\nu\rho} (x) \times \ps(0)\bigg)\\ 
     &= \int_V \d^\mu \bigg[
		  -\frac{1}{2}S(x)\g_{(\mu}\d_\nu\d_{\rho)}\ps(x) 
            - \frac{1}{2}\d_{\mu}\d_\nu S(x)\g_{\rho)}\ps(x) 
		  +\frac{5}{3}\d_{\mu}S(x)\g_\nu\d_{\rho}\ps(x)\\ & \quad \quad \quad \quad 
    
            - \frac{1}{3}g_{(\mu\nu} \d^\sigma S(x)\g_{\rho)}\d_\sigma \ps(x) 
		  + \frac{g_{(\mu\nu}}{5}S(x)\d_{\rho)}\ds\ps(x) 
            + \frac{g_{(\mu\nu}}{5}\d_{\rho)}\d^\sigma S(x)\g_\sigma\ps(x) \\ & \quad \quad \quad \quad 
		  - \frac{g_{(\mu\nu}}{3}\d_{\rho)}S(x)\ds\ps(x) 
            - \frac{g_{(\mu\nu}}{3}\d^\sigma S(x)\g_\sigma\d_{\rho)}\ps(x) \\ & \quad \quad \quad \quad
			
            +\frac{g_{(\mu\nu}}{10}S(x)\g_{\rho)}\d^2\ps(x) + \frac{g_{(\mu\nu}}{10}\d^2 S(x)\g_{\rho)}\ps(x)\bigg] + \cancelto{\to 0 \text{ as }r\to 0}{\text{regular}}
\end{aligned}\end{flalign}
Terms proportional to the equation of motion cancel, and the expression simplifies further after taking the derivative $\d^\mu$. We find:
\begin{flalign}\begin{aligned}
     [Q_{\nu\rho},\ps(0)] &= -\frac{1}{3}\int_V \delta(x) \bigg(
     \frac{8}{3}\d_\nu\d_\rho\ps(x) + \frac{8}{3}\g_\alpha\g_{(\nu}\d_{\rho)}\d^\alpha\ps(x)\bigg)
\end{aligned}\end{flalign}
The second term contains a part proportional to the equation of motion for $\ps(x)$. By using the properties of \eqref{1:Dirac}, we rewrite it as $\g_\alpha\g_{(\nu}\d_{\rho)}\d^\alpha \ps(x) = \{\g_\alpha,\g_{(\nu}\}\d_{\rho)}\d^\alpha\ps(x) - \g_{(\nu}\d_{\rho)}\ds \ps(x)$. Using $\frac{1}{2}\{\g_\alpha,\g_{(\nu}\}=g_{\alpha(\nu}$, we find:
\begin{flalign}\label{2.2:Qps}
[Q_{\nu\rho},\ps]_{FF}&=-\frac{8}{3}\d_\nu\d_\rho\ps
\end{flalign}
Following an analogous derivation, we find the action of the spin-3 charge on the Dirac conjugate field $\bps$:
\begin{flalign}\begin{aligned}\label{2.2:Qbps}
     [Q_{\nu\rho},\bps(0)] &= 
     \int_V \d^\mu \bigg(J_{\mu\nu\rho} (x) \times \bps(0)\bigg) = 
     \frac{8}{3}\d_\nu\d_\rho\bps(0)
\end{aligned}\end{flalign}

We find that the coefficients of \eqref{2.2:Qps} and \eqref{2.2:Qbps} differ by a minus sign. Once we've fixed down the action of the higher-spin charges on the fundamental fields, we can use this to find the action of the charge on composite operators. Using the definitions of operators \eqref{1FFCurrents}, we find commutator of the spin-3 charge with the scalar current $O^{FF}=\bps\ps$:
\begin{flalign}\begin{aligned}\label{2.2:QO}
     [Q_{\nu\rho},O]_{FF} &= [Q_{\nu\rho},\bps]\ps + \bps[Q_{\nu\rho},\ps] 
     = \frac{8}{3} \eps_{\alpha\sigma(\nu}\d_{\rho)}\d^\alpha J^\sigma
\end{aligned}\end{flalign}
This is consistent with the expression we find in \eqref{2.1:[Q,J]qf} and furthermore, this is the only possible term allowed by conformal invariance. We now analyze the commutator of the spin-3 charge with the vector current:
\begin{flalign}\begin{aligned}
     [Q_{\nu\rho},J_\alpha]_{FF} &= i[Q_{\nu\rho},\bps]\g_\alpha\ps + i\bps\g_\alpha[Q_{\nu\rho},\ps]  
     = \frac{8i}{3}\d_{(\nu}\bigg(\d_{\rho)}\bps\g_\alpha\ps - \bps\g_\alpha\d_{\rho)}\ps\bigg)
\end{aligned}\end{flalign}
The resulting expression can be written in terms of single-trace primaries, as in \eqref{2.1:[Q,J]qf}. Using the definitions of currents \eqref{1FFCurrents} and disregarding any terms proportional to the equations of motion, we find:
\begin{flalign}\begin{aligned}\label{2.2:QJ}
     [Q_{\nu\rho},J_\alpha]_{FF} &= 
     -\frac{4}{3}\eps_{\sigma\alpha(\nu}\d_{\rho)}\d^\sigma O 
     + \frac{16i}{3}\d_{(\nu}T_{\rho)\alpha}
\end{aligned}\end{flalign}
Comparing to \eqref{2.1:[Q,J]qf}, it's interesting to note that the double-trace spin-1 contribution as well as the $T_{\nu\rho}$ term do not contribute in the free fermion theory. However these terms may appear in the QF Chern-Simons-matter theory and we will analyze them in Section \ref{Sec5}. \\

Equivalently, instead of finding the action of higher-spin charges on currents by using commutation relations of fundamental fields, one may compute the OPE \eqref{2.2:FFOPE} directly. It is easy to verify that the \eqref{2.2:QO} and \eqref{2.2:QJ} can be obtained by computing:
\begin{flalign}
     [Q_{\nu\rho},O(0)] = \int_V \d^\mu \bigg(J_{\mu\nu\rho} (x) \times :\bps\ps:(0)\bigg); \qquad 
     [Q_{\nu\rho},J_\alpha(0)] = \int_V \d^\mu \bigg(J_{\mu\nu\rho} (x) \times :i\bps\g_\alpha\ps:(0)\bigg) 
\end{flalign}
By the same procedure, we can obtain all other commutators of the free fermion higher-spin algebra. 

For comparisons to the remainder of this text, we write the results of this Section as:
\begin{flalign}\begin{aligned}[t]
    &[Q_{\nu\rho},\ps]_{FF} = a \d_\nu\d_\rho \ps;\\&
    
     [Q_{\nu\rho},\bps]_{FF} = -a \d_\nu\d_\rho \bps;\\&
     
     [Q_{\nu\rho},O]_{FF} = a_0 \eps_{\alpha\sigma(\nu}\d_{\rho)}\d^\alpha J^\sigma;\\&

    [Q_{\nu\rho},J_\alpha]_{FF} = 
     a_1\eps_{\sigma\alpha(\nu}\d_{\rho)}\d^\sigma O 
     + a_2\d_{(\nu}T_{\rho)\alpha}; \qquad\qquad
\end{aligned}
\begin{aligned}[t]
     &a=-\frac{8}{3}\\[10pt]&
     a_0=-a\\&
    a_1=\frac{a}{2},\; a_2=-2i a
\end{aligned}\label{2.2:FFOPEfinal}
\end{flalign}

\subsection{Free Boson algebra from OPE}\label{Sec2.3}
We now discuss the CFT of a single free complex boson $\ph$ in three dimensions. The computation is set up analogously to the free fermion case, thus we now evaluate \eqref{2.2:FFOPE} using FB currents \eqref{1FBCurrents}. We use properties of the free boson propagator:
\begin{flalign}\label{2.3:fbprop}
	F(x) = \lg\phd(x)\ph(0)\rg = \lg\ph(x)\phd(0)\rg = \frac{1}{4\pi}\bigg(\frac{1}{|x|}\bigg); \qquad
	- \d^2 F(x) = - \frac{1}{4\pi}\d^2 \bigg(\frac{1}{|x|}\bigg) = \delta(x)
\end{flalign}
We first analyze the action of the spin-3 charge on the fields $\ph$ and $\phd$:
\begin{flalign}\begin{aligned}
       [Q_{\nu\rho},\ph(0)] &= \int_V \d^\mu \bigg(J_{\mu\nu\rho} (x) \times \ph(0)\bigg)\\
       & = \int_V \d^\mu \bigg[
       		\frac{1}{6}\d_\mu\d_\nu\d_\rho F(x) \ph(x) -
       		\frac{1}{6}F(x)\d_\mu\d_\nu\d_\rho \ph(x) + 
       		\frac{5}{2}\d_{(\mu}F(x)\d_\nu\d_{\rho)}\ph(x)  \\ & \quad \quad \quad \quad 
         
       		-\frac{5}{2}\d_{(\mu}\d_\nu F(x) \d_{\rho)}\ph(x)      		   	
      		-g_{(\mu\nu} \d^\sigma F(x) \d_{\rho)}\d_\sigma \ph(x) +
       		g_{(\mu\nu}\d_{\rho)}\d^\sigma F(x) \d_\sigma \ph(x)  \\ & \quad \quad \quad \quad 
                
                -\frac{g_{(\mu\nu}}{2}\d_{\rho)}F(x)\d^2\ph(x) +
       		\frac{g_{(\mu\nu}}{2}\d^2 F(x) \d_{\rho)}\ph(x)
       		-\frac{g_{(\mu\nu}}{10}\d_{\rho)}\d^2 F(x)\ph(x) \\ & \quad \quad \quad \quad  
                +\frac{g_{(\mu\nu}}{10} F(x)\d_{\rho)}\d^2\ph(x)
		\bigg] + \cancelto{\to 0 \text{ as }r\to 0}{\text{regular}}
\end{aligned}\end{flalign}
After simplifying and using \eqref{2.3:fbprop}, we only end up with terms diverging as delta functions. We find:
\begin{flalign}\label{2.3:Qph}
       [Q_{\nu\rho},\ph(0)] &= - \int_V \bigg(
			\frac{1}{6}\d_\nu\d_\rho\delta(x)\ph(x) - 
			\frac{10}{6}\d_{(\nu}\ph(x)\d_{\rho)}\delta(x) +
			\frac{5}{6}\delta(x)\d_\nu\d_\rho\ph(x) \bigg) = -\frac{8}{3}\d_\nu\d_\rho\ph(0)
\end{flalign}
We denote the coefficient multiplying the right-hand side of \eqref{2.3:Qph} by $b=-\frac{8}{3}$, which we will use to express other coefficients of the FB higher-spin algebra. Analogously, we find the action of the spin-3 charge on the conjugate field $\phd$:
\begin{flalign}\label{2.3:Qphd}
       [Q_{\nu\rho},\phd(0)] &= \int_V \d^\mu \bigg(J_{\mu\nu\rho} (x) \times \phd(0)\bigg) = \frac{8}{3}\d_\nu\d_\rho\phd(0)
\end{flalign}
It is curious to note that, after normalizing the spin-3 current so that its two point functions are the same in FF and FB theories, we find that the action of the spin-3 charge on the fundamental fields \eqref{2.2:Qps} and \eqref{2.2:Qbps} has the same coefficients as \eqref{2.3:Qph} and \eqref{2.3:Qphd}. 
Using these relations we can now find the action of the spin-3 charge on conserved currents of the FB theory. For the scalar current, we have:
\begin{flalign}\begin{aligned}\label{2.3:QO}
[Q_{\nu\rho},O]_{FB}&= 
	8[Q_{\nu\rho},\phd]\ph + 8\phd[Q_{\nu\rho},\ph] =
	-\frac{64i}{3}\d_{(\nu}\bigg(i\d_{\rho)}\phd\ph-i\phd\d_{\rho)}\ph\bigg) =
	-\frac{64i}{3}\d_{(\nu}J_{\rho)}
\end{aligned}\end{flalign}
This is consistent with the expression found in \eqref{2.1:[Q,J]qb}. We now analyze the action of the spin-3 charge on the vector current. The charge commutes with the generator of translations $[Q_{\nu\rho},\d]=0$, therefore:
\begin{flalign}\begin{aligned}
	[Q_{\nu\rho},J_\alpha]_{FB} &= i\d_\alpha[Q_{\nu\rho},\phd]\ph + i\d_\alpha\phd[Q_{\nu\rho},\ph]- i[Q_{\nu\rho},\phd]\d_\alpha\ph -i\phd\d_\alpha[Q_{\nu\rho},\ph] \\&
	=-\frac{8i}{3}\bigg(\d_\alpha\phd\d_\nu\d_\rho\ph + \d_\nu\d_\rho\phd\d_\alpha \ph\bigg) + \frac{8i}{3}\bigg(\d_\alpha\d_\nu\d_\rho\phd\ph + \phd\d_\alpha\d_\nu\d_\rho\ph \bigg)
\end{aligned}\end{flalign}
The result can be rewritten in terms of single-trace primaries, defined in \eqref{1FBCurrents}. Putting it together, the commutator $[Q_{\nu\rho},J_\alpha]_{FB}$ can be written as:
%
%
\begin{flalign}\label{2.3:QJ}
	[Q_{\nu\rho},J_\alpha]_{FB} &= 
	\frac{i}{6}\d_\nu\d_\rho\d_\alpha O - \frac{i}{6}g_{\alpha(\nu}\d_{\rho)}\d^2 O
	-\frac{16i}{3}\d_{(\nu}T_{\rho)\alpha}
\end{flalign}
The expression \eqref{2.3:QJ} is consistent with the constraints obtained in \eqref{2.1:[Q,J]qb}, and we find that only single-trace terms appear. \\
For comparisons to the remainder of this text, we write the results of this Section as:
\begin{flalign}\begin{aligned}[t]
    & [Q_{\nu\rho},\ph]_{FB} = b\d_\nu\d_\rho\ph;\\&
    
     [Q_{\nu\rho},\phd]_{FB} = -b\d_\nu\d_\rho\phd;\\&
     
     [Q_{\nu\rho},O]_{FB} = b_0\d_{(\nu}J_{\rho)};\\&

    [Q_{\nu\rho},J_\alpha]_{FB} = 
	b_1\d_\nu\d_\rho\d_\alpha O + b_2 g_{\alpha(\nu}\d_{\rho)}\d^2 O
	+ b_3\d_{(\nu}T_{\rho)\alpha}; \qquad\qquad
\end{aligned}
\begin{aligned}[t]
     &b=-\frac{8}{3}\\[10pt]&
     b_0=8ib\\&
    b_1=-b_2=-\frac{ib}{16},\; b_3=2i b
\end{aligned}\label{2.3:FBOPEfinal}
\end{flalign}
\subsection{Higher-spin Ward identities of free and interacting theories}\label{Sec3}
We make some broad remarks about the structure of higher-spin Ward identities, both in theories with exact and slightly broken higher-spin symmetry. Suppose we denote all the dynamical fields of the theory by $\phi$ and that the spectrum of single-trace operators contains a tower of (slightly-)conserved higher-spin currents $J_s$, which are sourced by $\mathcal{A}_s$. The path integral of the theory reads: 
\begin{flalign}\label{3:PathInt}
Z[\phi,\mathcal{A}_i]=\int_\phi \mathcal{D}\phi e^{-S[\phi,\mathcal{A}_i]-\mathcal{A}_0 J_0-\mathcal{A}_{1\mu} J_1^\mu-\mathcal{A}_{2\mu\nu} J_2^{\mu\nu}-...}
\end{flalign}
Correlators involving $J_s(x)$ can now be built by taking derivatives $\delta/\delta \mathcal{A}_s(x)$ and then setting $\mathcal{A}_i$ to their background values.
In theories with slightly broken higher-spin symmetry, the exponent in \eqref{3:PathInt} should be thought of as up to $\mathcal{O}(1/N)$ corrections. Even so, one can build conserved charges to leading order. Furthermore, we may assume the spin-s current develops a divergence in the form of some local operator $\d\cdot J_s \propto\frac{1}{N}X$, which we can formally replace by $\d\cdot\mathcal{J}_s =(\d\cdot J_s - \frac{1}{N}X)$ and discuss Ward identities of $\mathcal{J}_s$ in standard fashion, interpreting the resulting terms as variations of local operators.

Notably, the action $S[\phi,\mathcal{A}_i]$ may also contain non-linear terms in the sources. As a simple example, consider a theory with a global U(1) symmetry. The action in \eqref{3:PathInt} may contain a term\footnote{This term is a global symmetry analogue of the seagull vertex in scalar QED.} like $-\mathcal{A}_{1\mu}\mathcal{A}_1^\mu J_0$. Such a term would then contribute contact terms in all U(1) Ward identities containing at least two insertions of $J_1$ and one insertion of $J_0$ (see Appendix \ref{sec:AppC3}).

The (almost-)conservation of a higher spin current $J_s^{\mu_1..\mu_s}$ results from the invariance of the generating functional, to leading order in large $N$, under the corresponding global transformation, parametrized by a uniform parameter $\varepsilon$: $\delta_\varepsilon Z = \mathcal{O}(1/N)$. Let us now promote $\varepsilon$ to a spacetime-dependent parameter $\varepsilon(x)$. Under such a transformation, the operators of the theory transform as\footnote{We suppress Lorentz indices on the variation as well as the infinitesimal parameter $\varepsilon$ itself, which should be taken schematically.} $O_i(x_i)\to O_i(x_i)+ \delta_\varepsilon O_i(x_i)$, so that:
\begin{flalign}\label{3:localvar}
\delta_\varepsilon O_i(x_i) = \varepsilon(x) \delta^{(0)} O_i(x_i) + \d^{\alpha_1}\varepsilon(x) \delta^{(1)}_{\alpha_1} O_i(x_i) + \d^{\alpha_1}\d^{\alpha_2}\varepsilon(x) \delta^{(2)}_{\alpha_1\alpha_2} O_i(x_i) + ...
\end{flalign}
Here $\delta^{(n)} O_i(x_i)$ stand in for local operators that define the variation. The higher-spin Ward identity is obtained by varying $\int_\phi \mathcal{D}\phi e^{-S}O_1(x_1)O_2(x_2)... O_n(x_n)$. Expanding to leading order in small $\epsilon(x)$ and large $N$, by using manipulations under the integral sign  we find: 
\begin{flalign}\begin{aligned}
0 = \int_\phi \mathcal{D}\phi e^{-S}\int_x \varepsilon(x)\bigg[&
    \d\cdot \mathcal{J}_s(x)\prod_{i=1}^n O_i(x_i)\\&
    -\sum_{i=1}^n \bigg(\delta(x-x_i)\delta^{(0)}O_i(x_i)-\delta^{\alpha_1}\delta(x-x_i)\delta^{(1)}_{\alpha_1}O_i(x_i) + ...\bigg)\prod_{i\neq j}O_j(x_j)\bigg]
\end{aligned}\end{flalign}
Where $\mathcal{J}_s$ is the effective spin-s current defined previously, and is exactly conserved at leading order in large $N$. We can now still require that the variation vanishes at leading order, only for $\varepsilon=const$. However, note that the contact terms on the right-hand side contain contributions where derivatives act on delta functions. Since this holds for any uniform $\varepsilon$, we arrive at the \textit{local} higher-spin Ward identity:
\begin{flalign}\label{3:LocalWI}
	\lg (\d\cdot J_s(x))\prod_{i=1}^n O_i(x_i)\rg = \sum_{i=1}^n \lg \delta O_i(x_i)\prod_{i\neq j}O_j(x_j)\rg + \frac{1}{N}\lg :X:\prod_{i=1}^n O_j(x_j)\rg
\end{flalign}
The \textit{local variation} of each operator is given by $\delta O_i(x_i) = \delta(x-x_i)\delta^{(0)}O_i(x_i) - \d^{\alpha_1}\delta(x-x_i)\delta^{(1)}_{\alpha_1} O_i(x_i)+...$ where the operators $\delta^{(n)}O_i$ is defined as in \eqref{3:localvar}. Inside correlation functions, this represents how a local operator $O_i(x_i)$ is affected by the insertion of $(\d\cdot J_s(x))$. The contact terms involving derivatives acting on delta functions highlight the spacetime nature of higher-spin symmetry. 

As we integrate the Ward identity \eqref{3:LocalWI}, the variation should match the action of the spin-s charge $\int_x \delta O_i = [Q_s,O_i]$. This implies that the expressions we find in Section \ref{Sec2} fix the local variations of operators up to a linear sum of contact terms. For example, comparing to \eqref{2.3:QO}, the local variation of the scalar current $O^{FB}(x_1)$ due to the insertion of $((\d\cdot J)_{\mu\nu}(x))^{FB}$ is:
\begin{flalign}
	\delta O^{FB}(x_1) &= -\frac{64i}{3}\bigg(k\d_{(\mu}\delta(x_1-x)J_{\nu)}(x) + (1-k)\delta(x_1-x)\d_{(\mu}J_{\nu)}(x) - (\text{$\mu=\nu$ Trace terms}) \bigg)
\end{flalign}
For some $k\in\mathbb{C}$, and analogously for other operators. As we will see in Section \ref{Sec4}, in cases where higher-spin symmetry is broken, it will be important to understand the higher-spin algebra and associated Ward identities in the unintegrated form \eqref{3:LocalWI}. In the remainder of this section, we analyze Ward identities of free theories. In this case, $X=0$ and one can integrate the equation over insertions of $(\d\cdot J_s(x))$ without creating any divergences \cite{Maldacena:2012sf}. Hence, in subsections \ref{Sec3.1} and \ref{Sec3.2} we ignore the subtleties and analyze the integrated equations directly.

\subsubsection{Free-fermionic Ward Identities}\label{Sec3.1}
Integrated higher-spin Ward identities take a simple form in the free fermion theory \cite{Maldacena:2011jn,Maldacena:2012sf}:
\begin{flalign}\label{3.1:FFWI}
 0 = \sum_{i=1}^n \lg O_1(x_1)O_2(x_2)...[Q_{\mu\nu},O_i(x_i)]...O_n(x_n)\rg_{FF} 
\end{flalign}
In Appendix \ref{C:FFWI}, by using free field contractions, we verify that coefficients obtained in \eqref{2.2:QO} and \eqref{2.2:QJ} lead to consistent relations between correlators constructed of scalar and vector currents, for general $n$-point functions. For simplicity, we focus on matching only non-contact terms, as all contact terms can be recovered by using the precise form of current operators \eqref{1FFCurrents} and carefully tracking the derivations. As an illustration, we analyze a simple vertex Ward identity in the main text.\\

$\underline{\bm{\lg [Q_{\mu\nu},\ps O\bps]\rg_{FF}} \textbf{ Ward identity}}$
\vspace{3pt}

Although one cannot write a gauge-invariant vertex in the full QF Chern-Simons-matter theory, it is interesting to analyze the vertex Ward identities in the free theory limit. For simplicity, we start from\footnote{The ratio of coefficients in $[Q_{\mu\nu},\psi]$ and $[Q_{\mu\nu},\bps]$ is easily found from analyzing the action of $Q_{\mu\nu}$ on $S_{ij}$} \eqref{2.2:Qps}, \eqref{2.2:Qbps} and relate this to the coefficient appearing in \eqref{2.2:QO}. The fermion propagator from $x_i$ to $x_j$ is denoted by $S_{ij}$ and defined similarly to \eqref{2.2:ffprop}. Note that translation invariance implies $\d_i S_{ij}=-\d_j S_{ij}$. We have:
\begin{flalign}\label{3.1:0vert1}
     a_0\eps_{\rho_1\rho_2(\mu}\d_{2\nu)}\d_{2}^{\rho_1}
     \lg \ps(x_1)J^{\rho_2}(x_2)\bps(x_3)\rg_{FF} - 
     \big(a\d_{3\mu}\d_{3\nu} - a\d_{1\mu}\d_{1\nu}\big)
     \lg \ps(x_1)O(x_2)\bps(x_3)\rg_{FF} = 0    
\end{flalign}
Using the form of the operators \eqref{1FFCurrents}:
\begin{flalign}\begin{aligned}
     \lg \ps(x_1)O(x_2)\bps(x_3)\rg_{FF} =& -\lg\ps(x_1)\bps\ps(x_2)\bps(x_3)\rg_{FF}=-S_{12}S_{23}\\
	 \lg \ps(x_1)J_\mu(x_2)\bps(x_3)\rg_{FF} =& -\lg\ps(x_1) i\bps\g_\mu\ps(x_2)\bps(x_3)\rg_{FF}=-i S_{12}\g_\mu S_{23}
\end{aligned}\end{flalign}
The term involving the spin-1 vertex can be rewritten using the properties of \eqref{1:Dirac}:
\begin{flalign}\begin{aligned}\label{3.1:0vert2}
	-ia_0 \d_{2(\nu} \d_{2}^{\rho_1} \big( S_{12}\eps_{\mu)\rho_1\rho_2}\g^{\rho_2} S_{23}\big) &= 
         
         -\frac{a_0}{2} \d_{2(\nu}\big( \d_{2}^{\rho_1}S_{12}[\g_\mu,\g_{\rho_1}]S_{23} + S_{12}[\g_\mu,\g_{\rho_1}]\d_{2}^{\rho_1}S_{23}\big) \\&\to 
         
         -\frac{a_0}{2} \d_{2(\nu}\big( \d_1^{\rho_1} S_{12}\{\g_\mu,\g_{\rho_1}\}S_{23} - S_{12}\{\g_{\mu},\g_{\rho_1}\}\d_2^{\rho_1}S_{23}\big) \\
        
         &= -a_0 \d_{2\mu} \d_{2\nu} S_{12}S_{23}+ a_0 S_{12}\d_{2\mu}\d_{2\nu}S_{23}  
\end{aligned}\end{flalign}
The first and second line are equivalent up to EoM terms. The spin-0 vertex contributions can be rewritten as:
\begin{flalign}\label{3.1:0vert3}
         -\big(a\d_{3\mu}\d_{3\nu}-a\d_{1\mu}\d_{1\nu}\big) \big(-S_{12}S_{23}\big) &= a S_{12}\d_{2\mu}\d_{2\nu}S_{23}- a\d_{2\mu}\d_{2\nu}S_{12}S_{23}
\end{flalign}
Here we've used translation invariance to write all derivatives as $\d_2$. Inserting \eqref{3.1:0vert2} and \eqref{3.1:0vert3} into \eqref{3.1:0vert1}, we find the equation is satisfied for $a_0=-a$, consistent with the free-fermionic OPE calculation \eqref{2.2:FFOPEfinal}. Note that free theory Ward identities are homogeneous, hence they constrain the coefficients up to an overall constant, which needs to fixed from the OPE.

\subsubsection{Free-bosonic Ward Identities}\label{Sec3.2}
Ward Identities of the spin-3 current in the free boson theory take an analogous form \cite{Maldacena:2011jn,Maldacena:2012sf}:
\begin{flalign}\label{3.2:FBWI}
 0 = \sum_{i=1}^n \lg O_1(x_1)O_2(x_2)...[Q_{\mu\nu},O_i(x_i)]...O_n(x_n)\rg_{FB} 
\end{flalign}
The free-bosonic algebra will become important when discussing the critical limit ($\tl\to\infty$) of QF Chern-Simons-matter theory, in the following section. For our purposes, we confine ourselves to verifying the spin-3 Ward identities of two- and three-point functions involving the scalar current $O^{FB}$ and vector current $J_\mu^{FB}$ defined in \eqref{1FBCurrents}.  Again, we will focus on matching only non-contact terms, as contact terms can be included straightforwardly by following the derivation. We write out the two-point function Ward identity, and refer to Appendix \ref{C:FBWI} for FB Ward identities of three-point functions.\\

$\underline{\bm{\lg [Q_{\mu\nu},J_\alpha O]\rg_{FB}} \textbf{ Ward identity}}$
\vspace{3pt}

The only non-trivial Ward identity at the level of two-point functions comes from the action of the spin-3 charge on $\lg J_\alpha O\rg_{FB}$. For simplicity, we assume only terms that appear in \eqref{2.3:QO} and \eqref{2.3:QJ} with general coefficients, but it is easy to verify that even starting from the most general ansatz \eqref{2.1:[Q,J]qf} leads to the same constraints. Analogously to \eqref{2.3:fbprop}, we denote the free boson propagator from $x_i$ to $x_j$ by $F_{ij}$. Since $\lg J_s J_{s'}\rg \propto \delta_{ss'}$, we have:
\begin{flalign}\begin{aligned}\label{3.2:2ptWI}
	b_1 \d_{1\mu}\d_{1\nu}\d_{1\alpha}\lg O(x_1)O(x_2)\rg_{FB} + 
	 b_2 g_{\alpha(\mu}\d_{1\nu)}\d_1^2 \lg O(x_1)O(x_2)\rg_{FB}
	+ b_0 \d_{2(\mu}\lg J_\alpha(x_1)J_{\nu)}(x_2)\rg_{FB} = 0
\end{aligned}\end{flalign}
We use $\lg O(x_1)O(x_2)\rg_{FB} = 64 F_{12}^2$ and rewrite the $\lg OO\rg_{FB}$ terms as:
\begin{flalign}\begin{aligned}\label{3.2:2ptWIe1}
	&\d_{1\mu}\d_{1\nu}\d_{1\alpha}\lg O(x_1)O(x_2)\rg_{FB} = \\[5pt]&
	128 (\d_{1\mu}\d_{1\nu}F_{12}\d_{1\alpha}F_{12} + \d_{1\mu}\d_{1\alpha}F_{12}\d_{1\nu}F_{12} 
		+ \d_{1\nu}\d_{1\alpha}F_{12}\d_{1\mu}F_{12} + F_{12}\d_{1\mu}\d_{1\nu}\d_{1\alpha}F_{12})\\[5pt]
	
	&g_{\alpha(\mu}\d_{1\nu)}\d_1^2\lg O(x_1)O(x_2)\rg_{FB} = 256 g_{\alpha(\mu}\d_1^\sigma F_{12}\d_{1\nu)} \d_{1\sigma} F_{12}
\end{aligned}\end{flalign}
The spin-1 two point function is can be written as $\lg J_\alpha(x_1) J_\nu(x_2)\rg_{FB} = 2\d_{1\alpha} F_{12}\d_{1\nu}F_{12} - 2F_{12}\d_{1\alpha}\d_{1\nu} F_{12}$. We find:
\begin{flalign}
	&\d_{2(\mu} \lg J_\alpha(x_1) J_{\nu)}(x_2)\rg = - \d_{1(\mu} \lg J_\alpha(x_1) J_{\nu)}(x_2)\rg = -2\d_{1\alpha}\d_{1(\mu}F_{12}\d_{1\nu)}F_{12} + 2 F_{12}\d_{1\mu}\d_{1\nu}\d_{1\alpha}F_{12}
\end{flalign}
Up to contact terms $\lg T_{\mu\nu}(x_1)O(x_2)\rg_{FB}=0$, which leads to an additional relation between terms:
\begin{flalign}\label{3.2:2ptWIe2}
	3\d_{1\mu}F_{12}\d_{1\nu}F_{12}- g_{\mu\nu}\d_{1\sigma}F_{12}\d_1^\sigma F_{12} - F_{12}\d_{1\mu}\d_{1\nu}F_{12} = 0
\end{flalign}
Combining \eqref{3.2:2ptWIe1} - \eqref{3.2:2ptWIe2} with the Ward identity \eqref{3.2:2ptWI}, we find:
\begin{flalign}\begin{aligned}
	&\bigg(128b_1 - 2b_0 + 384b_2\bigg)\d_{1\alpha}F_{12}\d_{1\mu}\d_{1\nu}F_{12} +
	\bigg(256b_1  + 256b_2 \bigg)\d_{1(\mu}F_{12}\d_{1\nu)}\d_{1\alpha}F_{12}  \\
	&+\bigg(128b_1 + 2b_0 -128b_2 \bigg)F_{12}\d_{1\mu}\d_{1\nu}\d_{1\alpha}F_{12} = 0
\end{aligned}\end{flalign}
Since the terms are linearly independent, this is only satisfied if $b_2=-b_1=\frac{b_0}{128}$, which matches the free-boson OPE results \eqref{2.3:FBOPEfinal}.\\
\section{Critical theory Ward idenities}\label{Sec4}

The FB theory has an exact higher-spin symmetry with a tower of conserved higher-spin currents. In this section, we consider a free scalar $U(N)$ model, deformed by a quartic interaction $\frac{\lambda_4}{2N}(\phd\ph)^2$. In $d=3$, this is a relevant deformation, hence by tuning the physical mass to zero and flowing to the IR we reach an interacting conformal fixed point, referred to as the critical boson (CB) theory \cite{Giombi:2016ejx}. This theory plays a further role in our analysis as the $\tilde{\lambda}\to\infty$ limit of QF Chern-Simons-matter theory.

A key difference when considering higher-spin Ward identities of the CB theory relative to free theories is the fact that higher-spin symmetry is broken. In particular, the spin-3 current now develops a divergence at subleading orders in large $N$ in the form of double- and triple-trace operators. The divergence $(\d\cdot J)_{\mu\nu}$ is itself a conformal primary, and is therefore constrained to the form found in \cite{Giombi:2016zwa}:
\begin{flalign}\label{4:dJ3}
	(\d\cdot J)^{CB}_{\mu\nu} = \frac{A}{N}\bigg(\frac{2}{5} O\d_{(\mu}J_{\nu)} - \frac{3}{5} J_{(\mu}\d_{\nu)}O + \frac{g_{\mu\nu}}{5}J^\sigma\d_\sigma O \bigg)
\end{flalign}
The factor $A$ is related to the choice of normalization of $J_3, J_1, O$. The same expression was obtained in \cite{Giombi:2011kc}, by analyzing the classical equations of motion to leading order in the 't Hooft coupling, and then taking the critical theory limit $\tl\to\infty$.

At the FB fixed point the scalar has conformal dimension $\Delta_0=1$, while at the critical fixed point its dimension is given by $\Delta_{\tilde{0}} = 2+\mathcal{O}(\frac{1}{N})$. Correlators of the critical theory are related to FB correlators by a Legendre transformation with respect to the scalar current. As shown in \cite{Aharony:2012nh}, at large $N$, momentum space relations between two- and three-point functions are remarkably simple. With the FB currents normalized as in \eqref{1FBCurrents}, we have:
\setlength{\jot}{5pt}
\begin{flalign}\begin{aligned}\label{4:FBtoCB}
	&\lg O(p)O(-p) \rg_{CB} = -\lg O(p)O(-p) \rg^{-1}_{FB} \\
	
	&\lg J_s(p)J_s(-p) \rg_{CB} = \lg J_s(p)J_s(-p) \rg_{FB} \\
	
	&\lg O(p_1)O(p_2)O(p_3) \rg_{CB} = \lg O(p_1)O(p_2)O(p_3) \rg_{FB} \lg O(p_1)O(-p_1) \rg^{-1}_{FB} \lg O(p_2)O(-p_2) \rg^{-1}_{FB} \lg O(p_3)O(-p_3) \rg^{-1}_{FB} \\
	
	&\lg J_{s_1}(p_1)O(p_2)O(p_3) \rg_{CB} = \lg J_{s_1}(p_1)O(p_2)O(p_3) \rg_{FB} \lg O(p_2)O(-p_2) \rg^{-1}_{FB} \lg O(p_3)O(-p_3) \rg^{-1}_{FB} \\
	
	&\lg J_{s_1}(p_1)J_{s_2}(p_2)O(p_3) \rg_{CB} = \lg J_{s_1}(p_1)J_{s_2}(p_2)O(p_3) \rg_{FB} \lg O(p_3)O(-p_3) \rg^{-1}_{FB} \\
	
	&\lg J_{s_1}(p_1)J_{s_2}(p_2)J_{s_3}(p_3) \rg_{CB} = \lg J_{s_1}(p_1)J_{s_2}(p_2)J_{s_3}(p_3) \rg_{FB} 
\end{aligned}\end{flalign}
\setlength{\jot}{0pt}

Unlike in previous sections, it will be necessary to analyze higher-spin Ward identities of CB theory locally. To emphasize this, we briefly analyze the three-point function $\lg(\d\cdot J)_{\mu\nu}(x)J_\alpha(x_1) O(x_2)\rg_{CB}$. This correlator contains non-local terms, determined by \eqref{4:dJ3} and therefore if we integrate it over $x$, we may expect to pick up divergent contributions as $x\to x_{1,2}$.

On the other hand, we can first analyze the $\lg(\d\cdot J)_{\mu\nu}(x)J_\alpha(x_1) O(x_2)\rg_{FB}$ Ward identity, and then perform the Legendre transform. In doing this, we pick up non-local terms reflecting the fact that higher-spin symmetry is broken, in addition to contact terms, which we now interpret as variations of operators in the CB theory. By ensuring that the two ways of doing the calculation lead to the same result, we will obtain constraints on the unintegrated variations \eqref{3:localvar} of $\delta O$ and $\delta J_\alpha$ in the FB and CB theories.\\

By viewing the CB theory as a special QF theory, we can already form an expectation for its higher-spin algebra following the arguments of Section \ref{Sec2.1}. The theory is invariant under parity, which further constrains the algebra to:
\begin{flalign}\begin{aligned}\label{4:cbalg}
	&[Q_{\mu\nu},O]_{CB}= \varnothing \\
	&[Q_{\mu\nu},J_\alpha]_{CB} = \tb_2 \d_{(\mu}T_{\nu)\alpha} + \tb_3 \d_{\alpha}T_{\mu\nu} + \tb_4 \eps_{\alpha\rho(\mu}J_{1\nu)}J_1^\rho 
\end{aligned}\end{flalign}
With our task being to determine the overall constants $\tb_{2,3,4}$ as well as constraints on contact terms that integrate to \eqref{4:cbalg}. We find that this leads to consistent CB Ward identities for two- and three-point functions.

\subsection{\texorpdfstring{$\lg (\d\cdot J)_{\mu\nu} J_\alpha O \rg_{CB}$}{} calculation. CB algebra.}\label{Sec4.1}

\begin{figure}[htbp]
\centering
\includegraphics[width=0.5\textwidth]{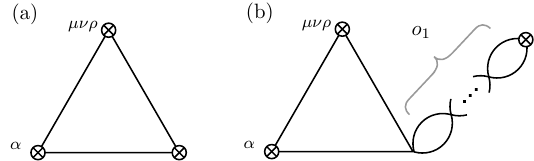}
\caption{(a) $\lg J_{\mu\nu\rho}(p)J_\alpha(p_1)O(p_2)\rg_{FB}$ is obtained by summing the triangle diagram with its reflection, to account for permutations of vertices. (b) The CB correlator is obtained by summing displayed diagrams for $o_1=0, 1, 2, 3,...$, also with vertex permutations.}
\label{fig:310cb}
\end{figure}

To capture higher-spin contact terms in $\lg (ip\cdot J)_{\mu\nu}(p)J_\alpha(p_1)O(p_2)\rg_{FB}$ we perform a 1-loop calculation using the operators defined in \eqref{1FBCurrents}, this time including the exact form of the operators, even up to the equations of motion. To then obtain the CB result, we sum over scalar bubbles, as displayed in Figure \ref{fig:310cb}(b). 

In momentum space, we denote the insertion of a spin-$s$ current of momentum $p$ with $V_s(p,k)$, where $k$ is the outgoing loop momentum. $\lg J_3 J_1 O\rg_{CB}$ can be computed by using vertex expressions for \eqref{1FBCurrents} and usual Feynman integrals:
\setlength{\jot}{5pt}
\begin{flalign}\begin{aligned}\label{4.1:310FB}
		&\lg J_{\mu\nu\rho}(p)J_\alpha(p_1)O(p_2)\rg_{FB} =\\& N \int \frac{d^3 k}{(2\pi)^3} 
		\bigg(\frac{V_{\mu\nu\rho}(p,k-p_1)V_\alpha(p_1,k)V_0(p_2,k+p_2)}{(k-p_1)^2k^2(k+p_2)^2} + \frac{V_{\mu\nu\rho}(p,k-p_2)V_0(p_2,k)V_\alpha(p_1,k+p_1)}{(k-p_2)^2k^2(k+p_1)^2}\bigg)
\end{aligned}\end{flalign}
\setlength{\jot}{0pt}

The resulting expression is tedious, containing a large number of terms. However, we are not necessarily interested in the full result for $\lg J_{\mu\nu\rho}(p)J_\alpha(p_1)O(p_2)\rg_{FB}$, rather, we want to understand how the scalar and vector currents are affected by the insertion of $(ip\cdot J)_{\mu\nu}(p)$. After taking the divergence of $J_{\mu\nu\rho}$, the expression \eqref{4.1:310FB} reduces to a tractable number of terms. As the FB theory has exact higher-spin symmetry, we expect all terms to be contact terms, however, since the spin-3 current is not conserved in the critical theory, we expect the Legendre transform \eqref{4:FBtoCB} to introduce non-contact terms. The computation gives:
\setlength{\jot}{5pt}
\begin{flalign}\begin{aligned}
	&\lg (ip \cdot J)_{\mu\nu}(p) J_\alpha(p_1) O (p_2)\rg_{FB} = \\&
	
	\frac{64i}{3}\bigg(\frac{2}{5}ip_{1(\mu} 
	- \frac{3}{5}ip_{2(\mu}\bigg)\lg J_\alpha(p_1)J_{\nu)}(-p_1)\rg_{FB} \\& 
 
	+ \bigg(-\frac{1}{10}p_{1(\mu}g_{\nu)\alpha}p_2^2 
	- \frac{1}{20}p_{1(\mu}p_{2\nu)}p_{2\alpha} 
	+ \frac{3}{40}p_{1\alpha}p_{2\mu}p_{2\nu} 		
	+ \frac{1}{40}p_{1\mu}p_{1\nu}p_{2\alpha} 	
	- \frac{3}{20}p_{1\alpha}p_{1(\mu}p_{2\nu)} \\& \qquad
	+ \frac{1}{40}p_{1\alpha}p_{1\mu}p_{1\nu} 	
	+ \frac{1}{15}p_{2(\mu}g_{\nu)\alpha}p_2^2 
	+ \frac{1}{120}p_{2\mu}p_{2\nu}p_{2\alpha}\bigg)\lg O(-p_2)O(p_2)\rg_{FB} 	
 	- \text{Trace terms} 
\end{aligned}\end{flalign}
\setlength{\jot}{0pt}
Where we've used the two-point functions in a theory of a single complex boson:
\begin{flalign}\label{4.1:2ptJJOO}
	\lg O(-p_2) O(p_2)\rg_{FB} = \frac{8}{p_2}; \qquad\quad
	\lg J_\alpha(p_1) J_\nu (-p_1)\rg_{FB} = \frac{p_{1\alpha}p_{1\nu}-g_{\alpha\nu}p_1^2}{16p_1}
\end{flalign}
Since all terms are polynomials of either $p_1$ or $p_2$, all terms are contact terms. Trace terms are included so that the expression is exactly traceless over $\mu=\nu$, consistent with the tracelessness of $(ip\cdot J)_{\mu\nu}$. To emphasize the presence of different kinds of contact terms, we write the local Ward identity \eqref{3:LocalWI} in position space:
\begin{flalign}\label{4.1:310fbwi}
	\lg (\d \cdot J)_{\mu\nu}(x) J_\alpha(x_1) O(x_2) \rg_{FB} =  \lg J_\alpha (x_1) \delta O^{FB}(x_2) \rg_{FB} + \lg \delta J^{FB}_\alpha (x_1) O(x_2) \rg_{FB}
\end{flalign}

With the \textit{local variations} of the spin-0 current $\delta O^{FB}(x_2)$ and the spin-1 current $\delta J^{FB}_\alpha(x_1)$ in the FB theory due to the insertion of $(\d\cdot J)_{\mu\nu}(x)$ are given by:
\setlength{\jot}{5pt}
\begin{flalign}
	\delta O^{FB}(x_2) &= 
	-\frac{64i}{3}\bigg( \frac{3}{5}\d_{2(\mu}\delta(x_2-x)J_{\nu)}(x) 
	+ \frac{2}{5}\delta(x_2-x)\d_{(\mu}J_{\nu)}(x) - \text{Trace} \bigg) \label{4.1:Ofbvar}\\
	
	\delta J^{FB}_\alpha(x_1) &= 
	-\frac{i}{10}\d_{1(\mu}\delta(x_1-x)g_{\nu)\alpha}\d^2 O(x) 
	- \frac{i}{20}\d_{1(\mu}\delta(x_1-x)\d_{\nu)}\d_{\alpha}O(x) 	\nonumber\\& \quad					
        + \frac{3i}{40}\d_{1\alpha}\delta(x_1-x)\d_{\mu}\d_{\nu}O(x)

	- \frac{i}{40}\d_{1\mu}\d_{1\nu}\delta(x_1-x)\d_{\alpha}O(x)  \nonumber \\& \quad
	+ \frac{3i}{20}\d_{1\alpha}\d_{1(\mu}\delta(x_1-x)\d_{\nu)}O(x) 
	+ \frac{i}{40}\d_{1\alpha}\d_{1\mu}\d_{1\nu}\delta(x_1-x)O(x) \label{4.1:Jfbvar}\\& \quad 
	
	- \frac{i}{15}\delta(x_1-x)g_{\alpha(\nu}\d_{\mu)}\d^2 O(x) 
	- \frac{i}{120}\delta(x_1-x)\d_{\mu}\d_{\nu}\d_{\alpha}O(x) \nonumber\\& \quad 
	+ \delta_{T}J^{FB}_\alpha(x_1) + \cancel{\frac{1}{N}(OO(x)\text{ terms})} - \text{Trace}  \nonumber 
		
\end{flalign}
\setlength{\jot}{0pt}

We will denote the single trace terms in \eqref{4.1:Ofbvar} proportional to the scalar current by $\delta_O J_\alpha^{FB}$. From the OPE we expect $\delta J_\alpha^{FB}$ to also contain terms proportional to the stress tensor, which we denote by $\delta_{\text{T}}J^{FB}_\alpha$. However such terms do not show up in this calculation, since they would be proportional to the two-point function $\lg TO\rg_{FB} = 0$. On the other hand, $\delta J_\alpha^{FB}$ may also double- and triple-trace contact terms, compatible with \eqref{2.1:[Q,J]qb} under the condition that such terms integrate to zero. Combining this with charge conjugation, we find that only double-trace terms going like $:OO:$ could in principle appear in $\delta J_\alpha^{FB}$, though as we will see, further analysis will render them to zero.

Even though the OPE calculation is not sensitive to the different types of contact terms, it can easily be checked that integrating the variations leads to results consistent with \eqref{2.3:QO} and \eqref{2.3:QJ} and hence, up to contact terms, we obtain the integrated FB Ward identity \eqref{3.2:2ptWI}. From consistency with the OPE, we can guess the stress-tensor terms that appear in $\delta J^{FB}_\alpha$ up to an unfixed constant $w_{1}^{FB}\in\mathbb{C}$:
\begin{flalign}\begin{aligned}\label{4.1:dTJ}
	\delta_{T} J^{FB}_\alpha(x_1) =&  
	-\frac{16i}{3}\delta(x_1-x)\d_{(\mu}T_{\nu)\alpha}(x)
 
        + w_1^{FB}\bigg(\delta(x_1-x)\d_{\alpha}T_{\mu\nu}(x) 
        + \d_{\alpha}\delta(x_1-x)T_{\mu\nu}(x)\bigg)
\end{aligned}\end{flalign}
Note that terms involving $T_{\mu\nu}$ may still appear within the local variation $\delta J_\alpha^{FB}$. Equation \eqref{2.3:QJ} only restricts such terms from appearing after integration, however we can write a linear sum of contact terms that vanishes after integrating Ward identities.

Including the correct contact terms in free theories is essential for the higher-spin Ward identities to be satisfied locally. Generally, the Legendre transformation can generate non-contact terms. These non-contact terms should match the double-trace contributions developed by $(\d\cdot J)_{\mu\nu}^{CB}$ as we break higher-spin symmetry. The local CB Ward identity $\lg[Q_{\mu\nu},J_\alpha O]\rg_{CB}$ is obtained simply by Legendre transforming the FB result:
\setlength{\jot}{5pt}
\begin{flalign}\begin{aligned}\label{4.1:310cbwi}
	\lg (ip\cdot J)_{\mu\nu}(p)J_\alpha(p_1)O(p_2)\rg_{CB} 
			&= \lg (ip\cdot J)_{\mu\nu}(p)J_\alpha(p_1)O(p_2)\rg_{FB}\lg O(p_2)O(-p_2)\rg^{-1}_{FB}\\	
			&= \lg \delta J^{FB}_\alpha(p_1)O(p_2)\rg_{FB}\lg O(p_2)O(-p_2)\rg^{-1}_{FB} \\&\quad 
                + \lg J_\alpha(p_1) \delta O^{FB}(p_2)\rg_{FB}\lg O(p_2)O(-p_2)\rg^{-1}_{FB}
\end{aligned}\end{flalign}
\setlength{\jot}{0pt}

The two terms on the right-hand side can be evaluated using \eqref{4.1:2ptJJOO} and the variations \eqref{4.1:Ofbvar}, \eqref{4.1:Jfbvar}. We then find that the second term contributes only as a sum of contact terms in $p_1$ and $p_2$. By choosing a specific regularization scheme, these terms can all be removed from $\lg (ip\cdot J)_{\mu\nu}(p)J_\alpha(p_1)O(p_2)\rg_{CB}$. As discussed in Section \ref{Sec3}, the appearance of contact terms in Ward identities can be traced back to non-linear source terms in the generating functional of the theory. Thus, we set $ \lg J_\alpha(p_1) \delta O^{FB}(p_2)\rg_{FB} \lg O(p_2)O(-p_2)\rg^{-1}_{FB} = 0$, which assumes adding appropriate counterterms to the CB theory. 

Let's examine how the equation behaves when integrated in position space over the insertion of $(\d\cdot J)_{\mu\nu}(x)$, which is equivalent to setting $p=0$ in \eqref{4.1:310cbwi}. The left-hand side can  now be separated into two contributions, the first coming from the divergence \eqref{4:dJ3} of $J_3$ and the second, coming from the CB algebra. 
As $(ip\cdot J)_{\mu\nu}(p)$ is no longer a zero-operator, we generically expect to find divergences as $p_{1,2}$ approach $p$. This is precisely compensated by the first term on the right-hand side of \eqref{4.1:310cbwi}:
\begin{flalign}
	\lg (ip \cdot J)_{\mu\nu}(p)J_\alpha(p_1)O(p_2)\rg_{CB} \vline_{p = 0 \atop \text{divergent as } p_2 = - p_1 \to p} = \lg \delta J^{FB}_\alpha(p_1)O(p_2)\rg_{FB}\lg O(p_2)O(-p_2)\rg^{-1}_{FB} 
\end{flalign}

From here we find the normalization constant $A=\frac{64i}{3}$ in \eqref{4:dJ3}.

The remaining terms are finite as $p_{1,2}\to p$ and by \eqref{2:[Q,J]def} correspond precisely to the CB higher-spin algebra:
\setlength{\jot}{5pt}
\begin{flalign}\begin{aligned}
	&\lg [Q_{\mu\nu},J_\alpha(p_1)]O(p_2)\rg_{CB} + \lg J_\alpha(p_1)[Q_{\mu\nu},O(p_2)] \rg_{CB}  = 
	\lg (ip\cdot J)_{\mu\nu}(p)J_\alpha(p_1)O(p_2)\rg_{CB} \vline_{p = 0 \atop \text{finite as } p_2 = 
 - p_1 \to p} \\& 
        = \lg J_\alpha(p_1) \delta O^{FB}(p_2)\rg_{FB}\lg O(p_2)O(-p_2)\rg^{-1}_{FB} = 0
\end{aligned}\end{flalign}
\setlength{\jot}{0pt}

Which is consistent with our expectation \eqref{4:cbalg}. We can write the \textit{local variations} of the spin-0 current $\delta O^{CB}(x_2)$ and the spin-1 current $\delta J^{CB}_\alpha(x_1)$ in the CB theory, under the insertion of $(\d\cdot J)_{\mu\nu}(x)$, as:
\begin{flalign}\begin{aligned}\label{4.1:cbcon1}
	\delta O^{CB}(x_2) &= \varnothing \\	
	\delta J^{CB}_\alpha(x_1) &= \delta_T J^{CB}_\alpha(x_1) + \tb_4 \delta(x_1-x) \eps_{\alpha\rho(\mu}:J_{\nu)}J^\rho:(x) 
\end{aligned}\end{flalign}
$\delta J^{CB}_\alpha$ can have terms proportional to the stress-tensor which we write as $\delta_T J^{CB}_\alpha$. Furthermore, as the scalar current now has dimension $\Delta_{\tilde{0}}=2+\mathcal{O}(1/N)$, the only other contribution allowed by conformal invariance is the double-trace term $:J_1J_1:$. Since the corresponding term in \eqref{4:cbalg} does not have any derivatives, within the unintegrated variation \eqref{4.1:cbcon1} it shows up as a unique contact term. To constrain the algebra further, we consider Ward identities of three-point functions.

\subsection{CB resummation. Bowtie diagrams.}\label{Sec4.2}

There are four possible three-point function Ward identities that we can analyze, involving the spin-3 (pseudo) charge $Q_{\mu\nu}$ acting on a combination of scalar and vector currents. By interpreting these as diagrammatic summation statements, we obtain further constraints on the CB algebra. Two of the four Ward identities simplify considerably, as some of the appearing correlators vanish.\\

$\underline{\bm{\lg [Q_{\mu\nu},OOO]\rg_{CB}} \textbf{ Ward identity}}$
\vspace{3pt}

Since the scalar current remains unaffected by by the spin-3 current, we may only check that the Ward identity is satisfied, however it does not depend on the higher-spin algebra. Using \eqref{3:LocalWI}, we write out the equation in position space:
\setlength{\jot}{5pt}
\begin{flalign}\begin{aligned}\label{4.2:allOcbwi}
	&\lg (\d \cdot J)_{\nu\rho}(x)O(x_1)O(x_2)O(x_3)\rg_{CB} =\\&
	\lg \delta O^{CB}(x_1)O(x_2)O(x_3)\rg_{CB}  + 
	
	\frac{64i}{3}\bigg(\frac{2}{5}\lg O(x)O(x_1)\rg_{CB}\d_{(\mu}\lg J_{\nu)}(x)O(x_2)O(x_3)\rg_{CB}\\&
	
	-\frac{3}{5}\d_{(\mu}\lg O(x)O(x_1)\rg_{CB}\lg J_{\nu)}(x)O(x_2)O(x_3)\rg_{CB}
	-\text{ Trace}\bigg)
	+ \begin{Bmatrix}1\leftrightarrow 2\end{Bmatrix} 
	+ \begin{Bmatrix}1\leftrightarrow 3\end{Bmatrix}
\end{aligned}\end{flalign}
\setlength{\jot}{0pt}

The terms proportional to $\delta O^{CB}$ are zero. The terms inside the bracket are the leading order in large $N$ contributions from the local divergence of $(\d\cdot J_3)^{CB}$. Since $\lg J_1 OO\rg_{CB}$ and $\lg J_1 OOO\rg_{CB}$ vanish by charge conjugation, we conclude that the Ward identity trivially consistent.\\

$\underline{\bm{\lg [Q_{\mu\nu},J_\alpha OO]\rg_{CB}} \textbf{ Ward identity}}$
\vspace{3pt}

We next write the local Ward identity coming from the insertion of $(\d\cdot J)_{\mu\nu}^{CB}$ into $\lg J_\alpha OO\rg_{CB}$. Combining \eqref{3:LocalWI} with \eqref{4:dJ3} and \eqref{4.1:cbcon1}, we find:
\setlength{\jot}{5pt}
\begin{flalign}\begin{aligned}\label{4.2:3100cbwi}
	&\lg (ip\cdot J)_{\mu\nu}(p) J_\alpha(p_1) O(p_2) O(p_3) \rg_{CB} = \lg \delta_T J^{CB}_\alpha(p_1) O(p_2) O(p_3) \rg_{CB} + \\&
	
	\frac{64i}{3}\bigg[\bigg(ip_{1(\mu}+\frac{3}{5}ip_{(\mu} - \text{Trace} \bigg)\lg J_{\nu)}(-p_1)J_\alpha(p_1)\rg_{CB}\lg O(p+p_1)O(p_2)O(p_3) \rg_{CB}  \\&\qquad
	
	+ \bigg(ip_{2(\mu}+\frac{3}{5}ip_{(\mu} - \text{Trace} \bigg)\lg O(p_2)O(-p_2)\rg_{CB}\lg J_\alpha(p_1)J_{\nu)}(p+p_2)O(p_3) \rg_{CB} + \begin{Bmatrix}2\leftrightarrow 3\end{Bmatrix}\bigg]
\end{aligned}\end{flalign}
\setlength{\jot}{0pt}

\begin{figure}[htbp]
\centering
\includegraphics[width=0.9\textwidth]{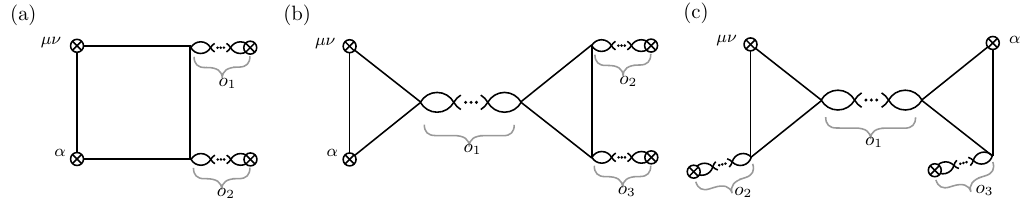}
\caption{Diagrams contributing to $\lg (ip\cdot J)_{\mu\nu}(p)J_\alpha(p_1)O(p_2)O(p_3)\rg$, up to vertex permutations: The FB correlator is the sum of diagrams (a) with $o_{1,2}=0$. The CB correlator is obtained by summing all diagrams for all nonnegative combinations of $o_i$.}
\label{fig:3100cb}
\end{figure}

The first term on the right-hand side comes from the higher-spin algebra of the CB theory \eqref{4.1:cbcon1}, while the remaining terms come from the local divergence \eqref{4:dJ3} developed by the spin-3 current. Our goal is to compare this expression to the Legendre-transformed free theory answer. This is a sum of corrected free theory diagrams and bowtie diagrams\footnote{The minus sign in the second line of \eqref{4.2:3100cb} appears due to resummation of a geometric series.} (Figure \ref{fig:3100cb}):
\setlength{\jot}{5pt}
\begin{flalign}\begin{aligned}\label{4.2:3100cb}
	&\lg (ip\cdot J)_{\mu\nu}(p) J_\alpha(p_1) O(p_2) O(p_3) \rg_{CB} = \\&
			\lg O(p_2)O(-p_2)\rg_{FB}^{-1}\lg O(p_3)O(-p_3)\rg_{FB}^{-1}\lg (ip\cdot J)_{\mu\nu}(p) J_\alpha(p_1) O(p_2) O(p_3) \rg_{FB} \\& 
			
			- \bigg[\frac{\lg (ip\cdot J)_{\mu\nu}(p) J_\alpha(p_1) O(-p-p_1)\rg_{FB}\lg O(p+p_1)O(p_2)O(p_3)\rg_{FB}}{\lg O(-p-p_1)O(p+p_1)
			\rg_{FB}\lg O(p_2)O(-p_2)\rg_{FB}\lg O(p_3)O(-p_3)\rg_{FB}} \\&\qquad
			
			 +\bigg( \frac{\lg (ip\cdot J)_{\mu\nu}(p) O(p_2) O(-p-p_2) \rg_{FB} \lg J_\alpha(p_1) O(p+p_2) O(p_3) \rg_{FB}}{\lg O(-p-p_2)O(p+p_2)
			\rg_{FB}\lg O(p_3)O(-p_3)\rg_{FB}} + \begin{Bmatrix}2\leftrightarrow 3\end{Bmatrix}\bigg)\bigg]
\end{aligned}\end{flalign}
\setlength{\jot}{0pt}

To derive constraints on the algebra, it will prove sufficient to compare terms in \eqref{4.2:3100cbwi} and \eqref{4.2:3100cb} and that are not contact terms\footnote{After integrating over $x$ in position space, i.e. setting $p=0$ in momentum space, such terms are those that remain non-contact terms in the integrated Ward identity.} in at least two of the three momenta $p_{1,2,3}$, while the difference in any other contact terms between the two equations can be included by adding appropriate counterterms to the generating functional of the theory. We will postpone a more careful discussion of higher-spin contact terms to Section \ref{Sec5}, of which the CB theory contact terms are a special case.

In \eqref{4.2:3100cb}, let us first consider the terms appearing in the square brackets. The first term is proportional to the scalar three-point function $\lg OOO\rg_{CB}$. As observed in \cite{Aharony:2012nh}, this correlator is at most a constant in momentum space. Hence the entire term is only a function of $p$ and $p_1$ which we neglect in this analysis. The remaining bowtie diagram terms are proportional to $\lg J_1 OO\rg_{CB}$ which vanishes by charge conjugation. The only non-contact terms on the right-hand side of \eqref{4.2:3100cb} come from expanding $\lg (ip\cdot J)_{\mu\nu}(p) J_\alpha(p_1) O(p_2) O(p_3) \rg_{FB}$ using the FB Ward identity and \eqref{4:FBtoCB}:
\setlength{\jot}{5pt}
\begin{flalign}\begin{aligned}\label{4.2:3100cbe1}
	&\lg (ip\cdot J)_{\mu\nu}(p) J_\alpha(p_1) O(p_2) O(p_3) \rg_{CB} = 	
	\lg O(p_1)O(-p_1)\rg_{CB}\lg \delta_O J^{FB}_\alpha(p_1) O(p_2) O(p_3) \rg_{CB}  \\&
 
	+ \lg \delta_T J^{FB}_\alpha(p_1) O(p_2) O(p_3) \rg_{CB}	
	+\bigg(\lg O(p_2)O(-p_2)\rg_{CB}\lg J_\alpha(p_1) \delta O^{FB}(p_2)O(p_3)\rg_{CB} + \begin{Bmatrix}2\leftrightarrow 3\end{Bmatrix}\bigg) 
\end{aligned}\end{flalign}
\setlength{\jot}{0pt}

Notice that we've left out the terms coming from the double-trace contributions of the FB algebra $\delta J^{FB}\propto \frac{1}{N}:OO:$ on the right-hand side. Schematically, these terms would contribute as non-contact terms of the form $\lg OO\rg_{CB}\lg OO\rg_{CB}$ at leading order in $N$. However, it's easy to verify that one cannot kinematically match such terms to any contributions appearing on the right-hand side of the Ward identity \eqref{4.2:3100cbwi}. Hence we find that the FB algebra \eqref{4.1:Jfbvar} does not contain any double-trace terms.

The term on the right-hand side containing $\delta_O J^{FB}_\alpha$ is proportional to the three-point function $\lg OOO\rg_{CB}$. As discussed, this correlator is a constant in momentum space and the corresponding term is only a function of $p_1$, which we can neglect. By using the FB algebra \eqref{4.1:Ofbvar}, the term in brackets can actually be rewritten as coming from the divergence of $(\d\cdot J_3)^{CB}$, given by \eqref{4:dJ3}:
\begin{flalign}\begin{aligned}\label{4.2:3100cbe2}
&\lg O(p_2)O(-p_2)\rg_{CB}\lg J_\alpha(p_1) \delta O^{FB}(p_2)O(p_3)\rg_{CB} =\\&

\frac{64i}{3}\bigg(\frac{3}{5}ip_{(\mu} + ip_{2(\mu} - \text{Trace} \bigg)\lg O(p_2)O(-p_2)\rg_{CB}\lg J_\alpha(p_1)J_{\nu)}(p+p_2)O(p_3) \rg_{CB}	
\end{aligned}\end{flalign}
We now combine \eqref{4.2:3100cbe1} and \eqref{4.2:3100cbe2} and compare to \eqref{4.2:3100cbwi}. In the Ward identity \eqref{4.2:3100cbwi}, the term on the right hand side proportional to $\lg OOO\rg_{CB}$ may be ignored in this analysis, as it is a contact term in $p_{2,3}$. Matching the non-contact terms in $p_{1,2,3}$ between the two approaches, we end up with:
\begin{flalign}
	\lg \delta_T J^{CB}_\alpha(p_1) O(p_2) O(p_3) \rg_{CB} = \lg \delta_T J^{FB}_\alpha(p_1) O(p_2) O(p_3) \rg_{CB}
\end{flalign}
Which is only consistent if $\delta_T J_\alpha^{CB}=\delta_T J_\alpha^{FB}$. We thus find that terms in the local variation of the spin-1 current that are not related to the scalar current are identical between the free and critical theories. Presumably, this is not surprising, as the transformation between the two theories only affects the scalar $O$, at least at the level of single-trace terms.\\

$\underline{\bm{\lg [Q_{\mu\nu},J_\alpha J_\beta O]\rg_{CB}} \textbf{ Ward identity}}$
\vspace{3pt}

Writing out \eqref{3:LocalWI}, we find contributions both from the CB algebra \eqref{4.1:cbcon1}, and the divergence of the spin-3 current:
\setlength{\jot}{5pt}
\begin{flalign}\begin{aligned}\label{4.2:JJOcbwi}

	&\lg(\d\cdot J)_{\mu\nu}(x)J_\alpha(x_1)J_\alpha(x_2)O(x_3)\rg_{CB} = \lg J_\alpha(x_1)J_\beta(x_2)\delta O(x_3)\rg_{CB} \\&
	
	 + \bigg(\lg \delta_T J_\alpha^{CB}(x_1) J_\beta(x_2) O(x_3)\rg_{CB} + 
	\frac{\tb_4}{N} \delta(x_1-x)\eps_{\alpha\sigma(\mu}\lg J_{\nu)}(x)J^\sigma(x)J_\beta(x_2)O(x_3)\rg_{CB}	
	+ \perm{1\leftrightarrow 2}{\alpha\leftrightarrow \beta}\bigg) \\& 
	
	+\frac{64i}{3}\bigg[\bigg(\frac{2}{5}\lg O(x)O(x_1)\rg_{CB}\d_{(\mu}\lg J_{\nu)}(x)O(x_2)O(x_3)\rg_{CB}    - \frac{3}{5}\d_{(\mu}\lg O(x)O(x_1)\rg_{CB}\lg J_{\nu)}(x)O(x_2)O(x_3)\rg_{CB} \\& \qquad\qquad
 
    + \perm{1\leftrightarrow 2}{\alpha\leftrightarrow\beta}\bigg) 
    + \bigg(\frac{2}{5}\lg O(x)O(x_1)\rg_{CB}\d_{(\mu}\lg J_{\nu)}(x)O(x_2)O(x_3)\rg_{CB} \\&\qquad\qquad
     - \frac{3}{5}\d_{(\mu}\lg O(x)O(x_1)\rg_{CB}\lg J_{\nu)}(x)O(x_2)O(x_3)\rg_{CB} \bigg)
     -\text{Trace terms}\bigg]
\end{aligned}\end{flalign}
\setlength{\jot}{0pt}

The three-point functions appearing on the right-hand side go as $\lg JOO\rg_{CB}$, $\lg TJO\rg_{CB}$, $\lg JJJ\rg_{CB}$. These correlators all vanish, as was argued in \cite{Giombi:2011rz}. The remaining term comes with a $\tb_4/N$ factor, and therefore at leading order only contributes as $\tb_4\lg JJ\rg_{CB}\lg JO\rg_{CB}$. However as $\lg J_sJ_{s'}\rg_{CB}\propto \delta_{ss'}$, this contribution also vanishes, and we find that the right-hand side of the Ward identity is zero. This is consistent, as the left-hand side vanishes by charge conjugation, and the Ward identity is trivial.\\

$\underline{\bm{\lg [Q_{\mu\nu},J_\alpha J_\beta J_\gamma]\rg_{CB}} \textbf{ Ward identity}}$
\vspace{3pt}

The final CB Ward identity constrains the double-trace terms appearing in \eqref{4.1:cbcon1}. We can write out the Ward identity \eqref{3:LocalWI} using the algebra and the divergence \eqref{4:dJ3}. We also use $A=\frac{64i}{3}$, obtained from the $\lg [Q_{\mu\nu},J_\alpha O]\rg_{CB}$ Ward identity, to find:
\setlength{\jot}{5pt}
\begin{flalign}\begin{aligned}\label{4.2:3111cbwi}
	&\lg (ip\cdot J)_{\mu\nu}(p) J_\alpha(p_1) J_\beta(p_2) J_\gamma(p_3) \rg_{CB} = \\&
	\frac{64i}{3} \bigg(ip_{1(\mu}+\frac{3}{5}ip_{(\mu} - \text{Trace}\bigg)   
	\lg J_\alpha(p_1)J_{\nu)}(-p_1)\rg_{CB}\lg O(p+p_1)J_\beta(p_2)J_\gamma(p_3) \rg_{CB} \\&
 
      + \lg \delta_T J^{CB}_\alpha(p_1) J_\beta(p_2) J_\gamma(p_3) \rg_{CB}	
	 + \tb_4 \eps_{\alpha\sigma(\mu} \lg J_{\nu)}(p+p_1)J_\beta(p_2)\rg_{CB}\lg J^\sigma(-p-p_1)J_\gamma(p_3)\rg_{CB}\\&
	+ \perm{1\leftrightarrow 2}{\alpha \leftrightarrow \beta} + \perm{1\leftrightarrow 3}{\alpha \leftrightarrow \gamma}
\end{aligned}\end{flalign}
\setlength{\jot}{0pt}

On the right-hand side, the first term results from the non-zero divergence of the spin-3 current in the CB theory. The remaining terms come from the CB algebra. In particular, note that the double-trace terms of \eqref{4.1:cbcon1} here appear as a disconnected contribution $\lg JJ\rg_{CB}\lg JJ\rg_{CB}$ at leading order in large $N$.

\begin{figure}[htbp]
\centering
\includegraphics[width=0.6\textwidth]{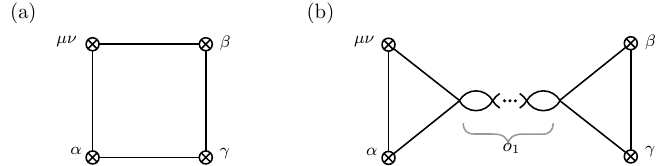}
\caption{$\lg (ip\cdot J)_{\mu\nu}(p)J_\alpha(p_1)J_\beta(p_2)J_\gamma(p_3)\rg$ diagrams, up to vertex permutations: The FB theory diagrams are given by (a), and after Legendre transforming, we add bowtie diagrams (b) for all $o_1\geq 0$. There are no vertex corrections.}
\label{fig:3111cb}
\end{figure}

We can now try to obtain the same correlator Legendre transforming the FB result. Since $\lg (\d\cdot J_3)J_1J_1J_1\rg_{CB}$ does not contain any scalars, there will be no need to renormalize the current vertices. However, we still have contributions coming from bowtie diagrams displayed in Figure \ref{fig:3111cb}(b). We have:
\setlength{\jot}{5pt}
\begin{flalign}\begin{aligned}\label{4.2:3111cb}
	&\lg (ip\cdot J)_{\mu\nu}(p) J_\alpha(p_1) J_\beta(p_2) J_\gamma(p_3) \rg_{CB} =\\&
	
	\lg (ip\cdot J)_{\mu\nu}(p) J_\alpha(p_1) J_\beta(p_2) J_\gamma(p_3) \rg_{FB}\\&
	 
	- \bigg[ \frac{\lg (ip\cdot J)_{\mu\nu}(p) J_\alpha(p_1) O(-p-p_1)\rg_{FB}\lg O(p+p_1)J_\beta(p_2)J_\gamma(p_3)\rg_{FB}}{\lg O(-p-p_1)O(p+p_1)\rg_{FB}}
			
			+ \perm{1\leftrightarrow 2}{\alpha \leftrightarrow \beta}+\perm{1\leftrightarrow 3}{\alpha \leftrightarrow \gamma}\bigg]
\end{aligned}\end{flalign}
\setlength{\jot}{0pt}

Our goal is to fix down the CB algebra by relating it to the FB theory. The first term on the right-hand side of \eqref{4.2:3111cb} involves the zero-operator $(\d\cdot J_3)^{FB}$ so it can only contribute contact terms. Therefore, any non-contact contributions must come from the bowtie diagrams. We can use the $\lg [Q_{\mu\nu},J_\alpha O]\rg_{CB}$ Ward identity and \eqref{4:FBtoCB} to write these in a different form:
\setlength{\jot}{5pt}
\begin{flalign}\begin{aligned}\label{4.2:3111eq1}
	&\bigg[ \frac{\lg (ip\cdot J)_{\mu\nu}(p) J_\alpha(p_1) O(-p-p_1)\rg_{FB}\lg O(p+p_1)J_\beta(p_2)J_\gamma(p_3)\rg_{FB}}{\lg O(-p-p_1)O(p+p_1)\rg_{FB}} 
	+ \perm{1\leftrightarrow 2}{\alpha \leftrightarrow \beta}+\perm{1\leftrightarrow 3}{\alpha \leftrightarrow \gamma}\bigg] = \\& \qquad\qquad 
	
	\bigg(\lg \delta_O J^{FB}_\alpha(p_1) O(-p-p_1)\rg_{FB}\lg O(p+p_1)J_\beta(p_2)J_\gamma(p_3) \rg_{CB} 
	+ \perm{1\leftrightarrow 2}{\alpha \leftrightarrow \beta}+\perm{1\leftrightarrow 3}{\alpha \leftrightarrow \gamma}\bigg) \\ &\qquad\qquad 
	
	+ \bigg(\lg J_\alpha(p_1) \delta_J O^{FB}(-p-p_1)\rg_{FB}\lg O(p+p_1)J_\beta(p_2)J_\gamma(p_3) \rg_{CB} 
	+ \perm{1\leftrightarrow 2}{\alpha \leftrightarrow \beta}+\perm{1\leftrightarrow 3}{\alpha \leftrightarrow \gamma}\bigg)
\end{aligned}\end{flalign}
\setlength{\jot}{0pt}

In the first bracket, we have decomposed the variation of the vector current as $\delta J_\alpha^{FB} = \delta_O J_\alpha^{FB} + \delta_T J_\alpha^{FB}$ with the latter vanishing inside the two-point function. We can now use the expression \eqref{4.1:Jfbvar}, to rewrite the terms in the first bracket as coming from the FB Ward identity:
\setlength{\jot}{5pt}
\begin{flalign}
	&\lg \delta_O J^{FB}_\alpha(p_1) O(-p-p_1)\rg_{FB}\lg O(p+p_1)J_\beta(p_2)J_\gamma(p_3) \rg_{CB}  = 
	
	\lg \delta_O J^{FB}_\alpha(p_1)J_\beta(p_2)J_\gamma(p_3) \rg_{FB} 
\end{flalign}
\setlength{\jot}{0pt}

The terms in the second bracket can be written to match the terms coming from local non-conservation of the spin-3 current in \eqref{4.2:3111cbwi}. To do so, we employ the FB algebra \eqref{4.1:Ofbvar} and the fact that the two point function of the spin-1 current remains the same in the free and critical theory, as in \eqref{4:FBtoCB}:
\setlength{\jot}{5pt}
\begin{flalign}\begin{aligned}\label{4.2:3111eq2}
	&\lg J_\alpha(p_1) \delta_J O^{FB}(-p-p_1)\rg_{FB}\lg O(p+p_1)J_\beta(p_2)J_\gamma(p_3) \rg_{CB} =\\& 
	
	-\frac{64i}{3} \bigg(ip_{1(\mu}+\frac{3}{5}ip_{(\mu} - \text{Trace} \bigg)   
	\lg J_\alpha(p_1)J_{\nu)}(-p_1)\rg_{CB}\lg O(p+p_1)J_\beta(p_2)J_\gamma(p_3) \rg_{CB}
\end{aligned}\end{flalign}
\setlength{\jot}{0pt}

Schematically speaking, this is the same mechanism observed for the two-point function Ward identity of Section \ref{Sec4.1}. Going from the free to the critical theory, the variation of the scalar current changes, but this is compensated by the divergence developed by the spin-3 current. We can insert equations \eqref{4.2:3111eq1} - \eqref{4.2:3111eq2} back into \eqref{4.2:3111cb} and use the FB algebra to write out $\lg(\d\cdot J_3)J_1J_1J_1\rg_{FB}$:
\setlength{\jot}{5pt}
\begin{flalign}\begin{aligned}\label{4.2:3111cb2}
	&\lg (ip\cdot J)_{\mu\nu}(p) J_\alpha(p_1) J_\beta(p_2) J_\gamma(p_3) \rg_{CB} =\\&
	
	\lg (\cancel{\delta_O J_\alpha^{FB}} + \delta_T J_\alpha^{FB}) J_\alpha(p_1) J_\beta(p_2) J_\gamma(p_3) \rg_{FB} -\cancel{\lg \delta_O J_\alpha^{FB} J_\alpha(p_1) J_\beta(p_2) J_\gamma(p_3) \rg_{FB}}\\& 
	
	+ \frac{64i}{3} \bigg(ip_{1(\mu}+\frac{3}{5}ip_{(\mu}- \text{Trace}\bigg)\lg J_\alpha(p_1)J_{\nu)}(-p_1)\rg_{CB} \lg O(p+p_1)J_\beta(p_2)J_\gamma(p_3) \rg_{CB} \\& 
	+ \perm{1\leftrightarrow 2}{\alpha \leftrightarrow \beta}+\perm{1\leftrightarrow 3}{\alpha \leftrightarrow \gamma}
\end{aligned}\end{flalign}
\setlength{\jot}{0pt}

We can now directly compare this expression to the Ward identity \eqref{4.2:3111cbwi}. The only possibility to match kinematic dependences of the two equations leads to $\delta_T J_\alpha^{CB} = \delta_T J_\alpha^{FB}$ and $\tb_4=0$. We find unique algebra by which all higher-spin Ward identities are consistent. Our final results for this section are summed up as: 
\setlength{\jot}{5pt}
\begin{flalign}\begin{aligned}\label{4.2:CBalgebra}
	&\delta O^{CB} = \varnothing \qquad\implies [Q_{\mu\nu},O]_{CB} = \varnothing\\
	&\delta J_\alpha^{CB}= \delta_T J_\alpha^{FB} \implies 
	[Q_{\mu\nu},J_\alpha]_{CB} = -\frac{16i}{3}\d_{(\mu}T_{\mu)\alpha} \\
	
	&(\d\cdot J)^{CB}_{\mu\nu} = \frac{1}{N}\frac{64i}{3}\bigg(\frac{2}{5} O\d_{(\mu}J_{\nu)} - \frac{3}{5} J_{(\mu}\d_{\nu)}O + \frac{g_{\mu\nu}}{5}J^\sigma\d_\sigma O \bigg)
\end{aligned}\end{flalign}
\setlength{\jot}{0pt}

Where $\delta_T J_\alpha^{FB}$ is given by \eqref{4.1:dTJ}. Of course, one should keep in mind the normalization of operators \eqref{1FBCurrents}. Furthermore, the Ward identities also provide constraints on contact terms of the theory. If we wish to preserve higher-spin symmetry, contact terms that we add to three-point functions must satisfy \eqref{4.2:allOcbwi}, \eqref{4.2:3100cbwi},  \eqref{4.2:JJOcbwi} and  \eqref{4.2:3111cbwi}. We expand on this in Section \ref{Sec5}.\\
\section{Local quasifermionic Ward identities}\label{Sec5}

Having understood the Ward identities in the free/critical limits, we analyze the interpolating case, that being theories of matter coupled to a Chern-Simons gauge field in three dimensions. As identified in \cite{Maldacena:2012sf}, such theories have approximate higher-spin symmetry at large $N$, even at strong coupling. There is a number of dualities relating conformal fixed points of various Chern-Simons-matter theories \cite{Aharony:2015mjs, Hsin:2016blu, Aharony:2016jvv}, which can be thought of as generalizations of the level/rank duality that include matter fields. Furthermore, these field theories are interesting as holographic duals of higher-spin gravity theories in $AdS_4$. 

We will particularly be interested in Ward identities of Quasifermionic Chern-Simons-matter theory. This theory admits multiple equivalent descriptions. The primary description is that of a Dirac fermion $\psi$ in the fundamental representation of $U(N)$ coupled to gluons $A_\mu$ with Chern-Simons interactions at level $k$. The action is given by:
\begin{flalign}\label{FF+CS}
 S_{FF+CS} = -\frac{ik}{8\pi}\int d^3 x \eps^{\mu\nu\rho} \bigg(A_\mu^a \d_\nu A_\rho^a + \frac{1}{3}f^{abc} A^a_\mu A^b_\nu A^c_\rho\bigg) + \int d^3 x \bps\g^\mu D_\mu \ps
\end{flalign}
At large $N$, this theory is suggested to be dual to the $U(k)$ model of Legendre-transformed bosons with Chern-Simons interactions at level $-\sign(k)N$. The duality was checked robustly by comparing three-point correlators  \cite{Aharony:2012nh,Gur-Ari:2012lgt} and thermal partition functions \cite{Giombi:2011kc}. The action of the CB+CS theory at level $k_b$ can be written as:
\begin{flalign}\label{CB+CS}
 S_{CB+CS} = -\frac{ik_b}{8\pi}\int d^3 x \eps^{\mu\nu\rho} \bigg(A_\mu^a \d_\nu A_\rho^a + \frac{1}{3}f^{abc} A^a_\mu A^b_\nu A^c_\rho\bigg) + \int d^3 x \bigg(|D_\mu \ph|^2 + \sigma (\phd\ph)\bigg)
\end{flalign}
Here, $\sigma$ is an auxiliary field introduced by performing the Legendre transform with respect to $O^{FB}=\phd\ph$.

Lastly, from the holographic point of view, QF Chern-Simons-matter theory is mapped to a quantum theory of interacting higher-spin gauge fields \cite{Giombi:2011kc}. The equations of motion for a class of such theories were written down by Vasiliev \cite{Vasiliev:1990en}. In \cite{Skvortsov:2018uru}, the cubic vertices of the bulk theory were obtained by bootstrapping from higher-spin symmetry, which further suggested that in certain chiral limits one obtains a quantum-complete, local theory of higher-spin gravity. The (anti-)chiral higher-spin theories\footnote{These theories were analyzed in \cite{Skvortsov:2018jea,Skvortsov:2020gpn,Sharapov:2022wpz,Sharapov:2022nps,Skvortsov:2022syz,Didenko:2022qga,Sharapov:2022faa} and others, including flat space constructions studied in \cite{Metsaev:1991mt,Metsaev:1991nb,Ponomarev:2016lrm}.} were argued to correspond to closed subsectors of QF Chern-Simons-matter theory, although at present there is no clear CFT description. In \cite{Aharony:2024nqs,Jain:2024bza}, it was argued that by analytically continuing the parameters of Chern-Simons-matter theory, one obtains a consistent match between CFT three-point functions and cubic bulk vertices. \\

We will analyze these theories in the 't Hooft limit $N,k\to\infty$ at finite 't Hooft coupling $\lambda=N/k$. Even without understanding the underlying microscopic theory, the presence of weakly broken higher-spin symmetry is sufficient to constrain correlators to a two-variable family\footnote{The analysis also assumes a well-behaved CFT, with a specific spectrum of single-trace operators. For details, see \cite{Maldacena:2012sf}}. In the case of QF Chern-Simons-matter theory, these variables were established via parameters of the theory \cite{Gur-Ari:2012lgt}. In the \eqref{FF+CS} description, we have:
\begin{flalign}\label{5:tNtldef}
\tN = 2N \frac{\sin(\pi\lambda)}{\pi\lambda}, \qquad
\tl =  \tan \bigg(\frac{\pi \lambda}{2}\bigg)
\end{flalign}
Qualitatively, $\tN$ can be thought of as the effective number of degrees of freedom of the theory, while $\tl$ controls the level of symmetry breaking, so that at $\tl=0$ the symmetry is unbroken, and then by turning on the gauge field we increase $\tl$, reaching the critical model in the $\tl\to\infty$ limit.

Correlators of the QF theory are often written in terms of FF and CB correlators. Note that in this section and Section \ref{Sec6}, we will be using a different convention than in Sections \ref{Sec1}-\ref{Sec4}. All QF correlators scale like $\tN$, which we set to one by setting $N=\frac{1}{2}, \lambda\to 0$. Hence we take the subscripts $\lg\rg_{FF}$ (similarly $\lg\rg_{CB}$) in the following sections to mean one half of the correlator in the theory of a single Dirac fermion (or complex boson), and in particular carry an extra $\frac{1}{2}$ factor relative to the ones defined in Section \ref{Sec1}. 

We will first summarize some known results for two- and three-point functions. The scalar two-point function remains the same as in the free theory, up to an overall factor, while two-point functions of spinning operators acquire a parity-odd contact term:
\begin{flalign}\begin{aligned}\label{5:2pt}
	\lg O(p) O(-p)\rg_{QF} =& \tN \lg O(p) O(-p)\rg_{FF} \\
	\lg J_s(p) J_s(-p)\rg_{QF} =& \tN \lg J_s(p) J_s(-p)\rg_{FF} + \tN g_s(\tl)\lg J_s(p) J_s(-p)\rg_{odd} \text{ for $s>0$}
\end{aligned}\end{flalign}
The odd piece can be written as $\lg J_s(p) J_s(-p)\rg_{odd} = \lg(\epsilon\cdot J_s)(p) J_s(-p)\rg_{FF}$, where $\epsilon$ denotes the epsilon-transform \cite{Jain:2022ajd,Jain:2021gwa}:
\begin{flalign}\label{5:epsdef}
	(\epsilon \cdot J)_{\mu_1\mu_2...\mu_s}(p) = \eps_{\alpha\beta(\mu_1}\frac{p^\alpha}{p} J^\beta_{\mu_2...\mu_s)}(p)
\end{flalign}

The functions $g_s(\tl)$ can apriori be any general functions of the coupling. We will only encounter contact terms of the spin-1 two-point function which were computed in \cite{Aharony:2012nh,Gur-Ari:2012lgt}. Curiously, perturbative computation in the FF+CS theory implies $g_1(\tl)=\tl$, while the same computation in the CB+CS theory yields $g_1(\tl)=-\tl^{-1}$. This suggests that the descriptions \eqref{FF+CS} and \eqref{CB+CS} as well as their gravity duals \cite{Skvortsov:2018uru} are not equivalent at the level of contact terms. We explore this question further in Section \ref{Sec5.3} 
 
Three-point functions take the form obtained in \cite{Maldacena:2012sf}, up to contact terms. For $s_i>0$, in the normalization \eqref{5:2pt}, we have:
\begin{flalign}\begin{aligned}\label{5:3pt}
	\lg J_{s_1} O O\rg_{QF} =& \tN \lg J_{s_1} O O\rg_{FF} = \tN \lg J_{s_1} O O\rg_{CB}\\
	\lg J_{s_1} J_{s_2} O\rg_{QF} =& \frac{\tN}{\sqrt{1+\tl^2}}\lg J_{s_1} J_{s_2} O\rg_{FF} + \frac{\tN\tl}{\sqrt{1+\tl^2}}\lg J_{s_1} J_{s_2} O\rg_{CB}\\
	\lg J_{s_1} J_{s_2} J_{s_3}\rg_{QF} =& \frac{\tN}{1+\tl^2}\lg J_{s_1} J_{s_2} J_{s_3}\rg_{FF} + \frac{\tN\tl}{1+\tl^2}\lg J_{s_1} J_{s_2} J_{s_3}\rg_{odd} + \frac{\tN\tl^2}{1+\tl^2}\lg J_{s_1} J_{s_2} J_{s_3}\rg_{CB}
\end{aligned}\end{flalign}
Three-point functions not-involving the scalar current have an additional parity-odd piece, that does not appear in the free/critical limits. As was discussed in \cite{Jain:2021gwa,Jain:2021whr}, the odd piece can be related to the FF and CB pieces via the epsilon-transform \eqref{5:epsdef} on one of the currents. We will highlight this in specific examples.

The QF theory breaks higher-spin symmetry at subleading order in $\tN$. This is evident in the CB limit when $\tl$ is taken to infinity, however it also holds at finite $\tl$. Current operators remain conformal primaries to leading order in large $N$, therefore most generally, we have:
\begin{flalign}\label{5:dJ3}
	(\d\cdot J)^{QF}_{\mu\nu} = \tilde{F}(\tl) (\d\cdot J)^{CB}_{\mu\nu} = \frac{1}{\tN}\frac{64i}{3}F(\tl)\bigg(\frac{2}{5} O\d_{(\mu}J_{\nu)} - \frac{3}{5} J_{(\mu}\d_{\nu)}O + \frac{g_{\mu\nu}}{5}J^\sigma\d_\sigma O \bigg)
\end{flalign}
For some function $F(\tl)$ satisfying $F(0)=0$ and $F(\infty)=1$. We know that $F(\tl)\neq 0$ for finite $\tl$, since the only theories with exact higher-spin symmetry are free theories \cite{Maldacena:2011jn}. Indeed, by considering classical equations of motion of the theory \cite{Giombi:2011kc}, it was derived that $F(\tl)=\tl\sqrt{1+\tl^2}^{-1}$ up to possible subleading corrections. As we will see, the Ward identity analysis consistently constrains this to be the full result to all orders in $\lambda$.\\

One of our tasks for this section is to constrain the higher-spin algebra of QF Chern-Simons-matter theory, confining ourselves to operators with spin $s<4$. We assume that the variations of operators interpolate smoothly between free and critical limits. In the FF theory, we have fixed the action of the spin-3 pseudocharge on scalar and vector currents as \eqref{2.2:QO} and \eqref{2.2:QJ}. This constrains the local variations up to a linear combinations of local contact terms. Under the insertion of $(\d\cdot J)^{FF}_{\mu\nu}(x)$, the variation of the scalar current $O(x_1)$ and the vector current $J_\alpha(x_2)$ are given by:
\begin{flalign}\begin{aligned}\label{5:ffvar}
	&\delta O^{FF}(x_1) = \frac{8}{3}\eps_{\rho_1\rho_2(\mu}\bigg(w_1^{FF}\delta(x-x_1)\d_{\nu)}\d^{\rho_1} - w_2^{FF}\d_{\nu)}\delta(x-x_1)\d^{\rho_1} - w_2^{FF}\d^{\rho_1}\delta(x-x_1)\d_{\nu)}
	\\&\qquad\qquad\qquad\qquad\qquad 
	+(1-w_1^{FF}-2w_2^{FF})\d_{\nu)}\d^{\rho_1}\delta(x-x_1)\bigg)J^{\rho_2}(x)\\[15pt]

 &\delta J_\alpha^{FF}(x_2) = \delta_O J_\alpha^{FF}(x_2) + \delta_T J_\alpha^{FF}(x_2)\\[5pt]
	
	&\delta_O J^{FF}_{\alpha}(x_2) = -\frac{4}{3}\eps_{\alpha\rho(\mu}\bigg(w_3^{FF}\delta(x-x_2)\d_{\nu)}\d^\rho - w_4^{FF}\d^\rho\delta(x-x_2)\d_{\nu)} - w_4^{FF}\d_{\nu)}\delta(x-x_2)\d^\rho
	\\&\qquad\qquad\qquad\qquad\qquad
	+ (1-w_3^{FF}-2w_4^{FF})\d_{\nu)}\d^\rho\delta(x-x_2)\bigg) O(x) \\
	
	&\delta_T J^{FF}_{\alpha}(x_2) = 
        \frac{16i}{3}\delta(x_2-x)\d_{(\mu} T_{\nu)\alpha}(x)  
        + w_5^{FF}\bigg(\delta(x_2-x)\d_{\alpha}+\d_{\alpha}\delta(x_2-x)\bigg)T_{\mu\nu}(x)
\end{aligned}\end{flalign}
The parameters $w_{1-5}^{FF}$ could in principle be fixed by calculating various FF correlators involving $(\d\cdot J_3)^{FF}$ and insertions of spin-0,1,2 currents, which we leave as a topic of future work. In the critical limit, these expressions should go over smoothly to the variations of the critical theory: $\delta O^{FF}\to \delta O^{CB}$ and $\delta J_\alpha^{FF}\to \delta J_\alpha^{CB}$, which were determined in \eqref{4.1:dTJ} and \eqref{4.2:CBalgebra} up to an unfixed constant $w_{1}^{FB}$. We can now write a general ansatz for the algebra of the QF theory as:
\begin{flalign}\begin{aligned}\label{5:QFalgAnsatz}
	&\delta O^{QF}(x_1) = f_0(\tl)\delta O^{FF}(x_1) + h_0(\tl)\delta O^{int}(x_1)\\[10pt]
	
	&\delta J_\alpha^{QF}(x_1) = f_1(\tl)\delta_O J_\alpha^{FF}(x_1) + f_2(\tl)\delta_T J_\alpha^{FF}(x_1) + f_3(\tl)\delta_T J_\alpha^{CB}(x_1) \\[5pt]& \qquad\quad
 
        + \bigg(h_1(\tl)\delta_O J_\alpha^{int}(x_1) + h_2(\tl)\delta_T J_\alpha^{int}(x_1) 
        + \frac{h_3(\tl)}{\tN}\delta(x_1-x)\eps_{\alpha\rho(\mu}J_{\nu)}J^\rho(x)\bigg)  
\end{aligned}\end{flalign}
Here we have also included the possible intermediate terms that can appear at finite $\tl$ but vanish in the free/critical limits. $\delta O^{int}$ is such a contribution in the variation of the QF scalar and can have the form similar to $\delta O^{FF}$ in \eqref{5:ffvar}, but possibly with different coefficients. Similarly, $\delta J_\alpha^{QF}$ can have intermediate terms $\delta_O J_\alpha^{int}$ proportional to the scalar current and terms $\delta_T J_\alpha^{int}$ proportional to the stress-tensor, which are similar to their FF counterparts in \eqref{5:ffvar}, with possibly different coefficients. Furthermore, $\delta J_\alpha^{QF}$ may now contain a double-trace contribution of two spin-1 currents, as explicitly written in \eqref{5:QFalgAnsatz}. Note that in principle each monomial in the intermediate variations could have appeared with a different function of $\tl$, however as we will see, these functions are strongly constrained by the higher-spin Ward identities. Anticipating this, we've written down the ansatz with certain terms grouped, and general functions $h_i(\tl)$ that vanish as $\tl\to 0,\infty$. We expect the functions $f_i(\tl)$ to satisfy:
\begin{flalign}\label{5:QFalgLim}
\lim_{\tl \to 0} f_0(\tl)=1 \qquad \lim_{\tl \to 0} f_1(\tl)=1 \qquad
\lim_{\tl \to 0} f_2(\tl)=1 \qquad \lim_{\tl \to \infty} f_3(\tl) = 1
\end{flalign}
With all other $\tl\to 0,\infty$ limits vanishing. Based on these considerations, the spin-3 pseudocharge should act on $O$ and $J_\alpha$ as:
\begin{flalign}\begin{aligned}\label{5:QFalgAnsatzInt}
	&[Q_{\mu\nu},O]_{QF} = \bigg(f_0(\tl)+h_0(\tl)\bigg)\eps_{\alpha\beta(\mu}\d_{\nu)}\d^\alpha J^\beta \\[10pt]
 
	&[Q_{\mu\nu},J_\alpha]_{QF} = \bigg(f_1(\tl)+h_1(\tl)\bigg)\eps_{\sigma\alpha(\mu}\d_{\nu)} \d^\sigma O + \bigg(f_2(\tl)+f_3(\tl)+h_2(\tl)\bigg)\d_{(\mu}T_{\nu)\alpha} \\[5pt]&
 
    \qquad\qquad\qquad
    + h_2(\tl)\d_\alpha T_{\mu\nu} + \frac{h_3(\tl)}{\tN}\eps_{\alpha\rho(\mu}J_{\nu)}J^\rho 
\end{aligned}\end{flalign}
In Section \ref{Sec5.1}, we fix the unknown functions by analyzing local Ward identities of the spin-3 current in the QF theory. We will use these results to constrain four-point functions in Section \ref{Sec5.2} and discuss higher-spin contact terms in Section \ref{Sec5.3}.  

\subsection{Fixing the QF algebra}\label{Sec5.1}

The simplest correlator to consider is the three-point function of involving the scalar, vector and spin-3 currents. As higher-spin symmetry is broken, we find contributions from both the divergence of the spin-3 current \eqref{5:dJ3} and the local QF algebra \eqref{5:QFalgAnsatz}. In position space, the higher-spin Ward identity reads:
\setlength{\jot}{5pt}
\begin{flalign}\begin{aligned}\label{5.1:310qfwi}
	&\lg (\d\cdot J)_{\mu\nu}(x) J_\alpha(x_1)O(x_2)\rg_{QF} =
	
	\lg \delta J^{QF}_\alpha(x_1) O(x_2) \rg_{QF} + \lg J_\alpha(x_1) \delta O^{QF}(x_2) \rg_{QF} \\&
	
	+ \frac{1}{\tN}\frac{64i}{3}F(\tl) \bigg(-\frac{2}{5}\d_{1(\mu} + \frac{3}{5}\d_{2(\mu} - \text{Trace}\bigg)
	\lg J_{\alpha}(x_1)J_{\nu)}(x)\rg_{QF}\lg O(x)O(x_2)\rg_{QF}
\end{aligned}\end{flalign}
\setlength{\jot}{0pt}

We compare this with the form of the three-point function $\lg J_{\mu\nu\rho}J_\alpha O\rg_{QF}$ implied by higher-spin symmetry. As in \eqref{5:3pt}, up to possible contact terms, we can decompose the correlator into FF and CB pieces. Then, by taking the divergence of the spin-3 current, we have:
\begin{flalign}\begin{aligned}
	&\lg (\d\cdot J)_{\mu\nu}(x) J_\alpha(x_1)O(x_2)\rg_{QF} =\\&
	
	\frac{\tN}{\sqrt{1+\tl^2}}\lg (\d\cdot J)_{\mu\nu}(x) J_\alpha(x_1)O(x_2)\rg_{FF} 
	+ \frac{\tN\tl}{\sqrt{1+\tl^2}}\lg (\d\cdot J)_{\mu\nu}(x) J_\alpha(x_1)O(x_2)\rg_{CB}
\end{aligned}\end{flalign}
We find that the QF Ward identity reduces to a combination of the FF and CB Ward identities, which were analyzed in Sections \ref{Sec3.1} and \ref{Sec4.1}. We can rewrite this as:
\begin{flalign}\begin{aligned}\label{5.1:310mz}
	&\lg (\d\cdot J)_{\mu\nu}(x) J_\alpha(x_1)O(x_2)\rg_{QF} = \\[5pt]&

	\bigg(\frac{\tN}{\sqrt{1+\tl^2}}\lg \delta J^{FF}_\alpha(x_1)O(x_2)\rg_{FF} + \frac{\tN}{\sqrt{1+\tl^2}}\lg J_\alpha(x_1)\delta O^{FF}(x_2)\rg_{FF} \bigg)\\& + 
	\bigg(\cancel{\frac{\tN\tl}{\sqrt{1+\tl^2}}\lg \delta J^{CB}_\alpha(x_1)O(x_2)\rg_{CB}} + \cancel{\frac{\tN\tl}{\sqrt{1+\tl^2}}\lg J_\alpha(x_1)\delta O^{CB}(x_2)\rg_{CB}} \\& 
	
	+ \frac{\tN \tl}{\sqrt{1+\tl^2}}\frac{64i}{3}\bigg(-\frac{2}{5}\d_{1(\mu} + \frac{3}{5}\d_{2(\mu} - \text{Trace}\bigg)
	\lg J_{\alpha}(x_1)J_{\nu)}(x)\rg_{CB}\lg O(x)O(x_2)\rg_{CB}\bigg)
\end{aligned}\end{flalign}
As discussed in Section \ref{Sec4}, the contributions from the CB algebra vanish in this case, since $\delta O^{CB}=0$ and terms coming from $\delta J^{CB}$ would appear in the equation as $\lg TO\rg_{CB}=0$. The expression \eqref{5.1:310mz} should match the Ward identity \eqref{5.1:310qfwi} both in contact terms and non-contact terms. We postpone the discussion of contact terms to Section \ref{Sec5.3} and here focus only on matching pieces that remain non-zero for $x\neq x_1\neq x_2$. In \eqref{5.1:310qfwi}, the terms appearing on the right-hand side involve only two-point functions. Using \eqref{5:2pt}, we can rewrite the algebra terms as as FF correlators, while the disconnected term coming from the local divergence developed by the spin-3 current can be written as a product of of CB correlators. It is straightforward then to check that the only way to match kinematic dependences between \eqref{5.1:310qfwi} and \eqref{5.1:310mz} is to have the higher-spin algebra pieces and the disconnected pieces match among themselves. Comparing to \eqref{5:QFalgAnsatz}, factors of $\tN$ drop out and we find expressions for the unknown functions (we suppress the arguments for brevity):
\begin{flalign}\begin{aligned}\label{5.1:310res}
	&f_0(\tl)=\frac{1}{\sqrt{1+\tl^2}},\; h_0(\tl) = 0 \implies \delta O^{QF} = \frac{1}{\sqrt{1+\tl^2}}\delta O^{FF}\\
	&f_1(\tl)=\frac{1}{\sqrt{1+\tl^2}},\; h_1(\tl) = 0 \implies \delta J^{QF}_\alpha = \frac{1}{\sqrt{1+\tl^2}}\delta_O J^{FF}_\alpha + \delta_T J^{QF} + \frac{h_3(\tl)}{\tN}\delta(x_1-x)\eps_{\alpha\rho(\mu}J_{\nu)}J^\rho \\	
	&F(\tl) = \frac{\tl}{\sqrt{1+\tl^2}}
\end{aligned}\end{flalign}
Here we have collectively denoted all the stress-tensor terms that appear in the second line of \eqref{5:QFalgAnsatz} by $\delta_T J_\alpha^{QF}$. It is interesting that after turning on the gauge-field the variation of the scalar current remains the same as in free theory, up to a $\tl$-dependent factor. Similarly, the scalar contact terms in the variation of the vector current only get rescaled, with no new terms appearing at finite $\tl$. Finally, note that the result for $F(\tl)$ matches the one derived in \cite{Giombi:2011kc}, and is valid to all orders in $\lambda=N/k$.

To further constrain the QF algebra, we write out the $\lg (\d\cdot J)_{\mu\nu}J_\alpha T_{\beta\gamma}\rg_{QF}$ Ward identity, using \eqref{5:dJ3} and \eqref{5:QFalgAnsatz}:
\setlength{\jot}{5pt}
\begin{flalign}\begin{aligned}\label{5.1:321qfwi}
	&\lg (\d\cdot J)_{\mu\nu}(x)J_\alpha(x_1) T_{\beta\gamma}(x_2) \rg_{QF} = \\[5pt]&
 
        \lg \delta J_\alpha^{QF}(x_1) T_{\beta\gamma}(x_2)\rg_{QF} + \lg J_\alpha(x_1) \delta T_{\beta\gamma}^{QF}(x_2)\rg_{QF} \\
	
	&+ \frac{1}{\tN}\frac{\tl}{\sqrt{1+\tl^2}}\frac{64i}{3}
	\lg \big(\frac{2}{5} O\d_{(\mu}J_{\nu)} - \frac{3}{5} J_{(\mu}\d_{\nu)}O + \frac{g_{\mu\nu}}{5}J^\sigma\d_\sigma O \big)(x)J_{\alpha}(x_1)T_{\beta\gamma}(x_2)\rg_{CB}
\end{aligned}\end{flalign}
\setlength{\jot}{0pt}

The contribution from the divergence of the spin-3 current is written in the second line, and at leading order in $\tN$ contributes as a product of two-point functions. It is easy to see that this term vanishes since the only contractions we can make in this case are $\lg JJ\rg_{QF}\lg TO\rg_{QF}$ and $\lg JT\rg_{QF}\lg JO\rg_{QF}$, however these terms vanish since two-point functions of different spin currents are zero. Therefore the right-hand side of \eqref{5.1:321qfwi} only contains contributions from the QF algebra. 

The first term is a sum of contact terms going as $\delta(x-x_1)$ while the second term is proportional to $\delta(x-x_2)$, hence we can be sure that the two contributions do not cancel between themselves for $x_1\neq x_2$. One can, for example, integrate the equation around a small region around $x_1$, not containing $x_2$. This means that the $\delta T^{QF}(x_2)$ terms can be put to zero. Furthermore, as $\lg J_s J_{s'}\rg_{QF}\propto \delta_{ss'}$, the remaining terms on the right-hand side are those proportional to $\delta_{T} J_\alpha^{QF}(x_1)$. The key point point is that the $\tl$-dependence on the left-hand side is given by:
\setlength{\jot}{5pt}
\begin{flalign}\begin{aligned}\label{5.1:321mz}
	&\lg (\d\cdot J)_{\mu\nu}(x)J_\alpha(x_1) T_{\beta\gamma}(x_2) \rg_{QF} =\\&
	
	\frac{1}{1+\tl^2}\lg (\d\cdot J)_{\mu\nu}(x)J_\alpha(x_1) T_{\beta\gamma}(x_2) \rg_{FF}  + \frac{\tl}{1+\tl^2}\lg (\d\cdot J)_{\mu\nu}(x)J_\alpha(x_1) T_{\beta\gamma}(x_2) \rg_{odd} \\&
	+ \frac{\tl^2}{1+\tl^2}\lg (\d\cdot J)_{\mu\nu}(x)J_\alpha(x_1) T_{\beta\gamma}(x_2) \rg_{CB} 
\end{aligned}\end{flalign}
\setlength{\jot}{0pt}
And comparing this to the ansatz \eqref{5:QFalgAnsatz} and \eqref{5:QFalgLim} leads to:
\begin{flalign}\begin{aligned}\label{5.1:321res}
	&f_2(\tl)=\frac{1}{1+\tl^2}, \qquad f_3(\tl)=\frac{\tl^2}{1+\tl^2}; \qquad h_2(\tl) \propto \frac{\tl}{1+\tl^2}
\end{aligned}\end{flalign}
The fact that $h_2(\tl)$ is an odd function leads to a strong constraint on the terms in \eqref{5:QFalgAnsatz} involving the stress-tensor, due to discrete symmetries of the theory. The action \eqref{FF+CS} is invariant under a parity transformation $\vec{x}\to -\vec{x}$ combined with $\lambda\to -\lambda$. As we are working in Euclidean signature, applying parity essentially contributes a minus sign in correlators for each Lorentz index\footnote{Note that the quasifermionic scalar transforms as $O\to -O$ under parity.}. Requiring that the Ward identity $\lg (\d\cdot J)_{\mu\nu}J_\alpha O O\rg_{QF}$ remains invariant under the combined transformation implies that $\delta_T J_\alpha^{int}$ terms  must be parity even, however conformal invariance and Lorentz symmetry restrict such terms from appearing, and therefore $\delta_T J_\alpha^{int}=0$. Similarly, the invariance of the $\lg (\d\cdot J)_{\mu\nu}J_\alpha J_\beta J_\gamma\rg_{QF}$ Ward identity constrains $h_3(\tl)$ to be even in $\tl$. We will revisit these Ward identities in Sections \ref{Sec5.2}.

Inserting these results back into \eqref{5:QFalgAnsatz}, we fix the local algebra of the QF theory, up to the constants $w_i^{FF, FB}$ appearing in the variations of operators, as well as a single unknown function $h_3(\tl)$ which vanishes in the free and critical limits. The action of the spin-3 pseudocharge on the scalar and vector currents now becomes:
\begin{flalign}\label{5.1:QFalg}
\begin{aligned}
	&[Q_{\mu\nu},O]_{QF}=
	\frac{1}{\sqrt{1+\tl^2}}\frac{8}{3}\eps_{\alpha\beta(\mu}\d_{\nu)}\d^\alpha J^\beta \\
	
	&[Q_{\mu\nu},J_\alpha]_{QF}=
	-\frac{1}{\sqrt{1+\tl^2}}\frac{4}{3}\eps_{\sigma\alpha(\mu}\d_{\nu)}\d^\sigma O 
	+ \frac{16i}{3}\frac{1-\tl^2}{1+\tl^2}\d_{(\mu}T_{\nu)\alpha}
	+\frac{h_3(\tl)}{\tN}\eps_{\alpha\rho(\mu}J_{\nu)}J^\rho
\end{aligned}
\end{flalign} 
Where $h_3(\tl)$ is an even function, which could be fixed by analyzing Ward identities of the QF theory involving higher correlators. The double-trace term in $[Q_{\mu\nu},J_\alpha]_{QF}$ does not contribute in Ward identities of three-point functions at large $\tN$. It appears when discussing Ward identities of higher-point functions. 

It is interesting to note that at the special point $\tl=1$, the stress-tensor terms vanish from the integrated equations \eqref{5.1:QFalg}. This does not necessarily imply that these terms vanish in local variations \eqref{5:QFalgAnsatz}, however it does suggest that that quasifermionic $n$-point functions $\lg(\d\cdot J_3)J_1..J_1O..O\rg_{QF}$ at $\tl=1$ may be expressed purely in terms of correlators of scalar and vector currents. It would be interesting to explore this further.

\subsection{Results for four-point functions}\label{Sec5.2}
After constraining the form of the higher-spin algebra, we can obtain non-trivial relations in correlators with an insertion of $(\d\cdot J_3)^{QF}$. Similarly to the case of three-point functions, higher-spin symmetry also constrains higher correlators to a tractable amount of different kinematic structures, each coming with a specific $\tl$ dependence. For correlators of QF Chern-Simons-matter theory, in the $\tl\to 0$ limit, we expect to recover only the FF structure, while in the $\tl\to\infty$ limit, only the CB structure survives. However, at finite $\tl$ several structures can appear. In the case of four-point functions, connections between different structures have been established in special kinematic regimes \cite{Kalloor:2019xjb} and it would be interesting to check if finite $\tl$ terms can generally be constrained in terms of free theory structures, as conjectured in \cite{Jain:2022ajd}. In this section we use weakly broken higher-spin symmetry to derive constraints on $\lg (\d\cdot J_3) J_1 O O\rg_{QF}$ and $\lg (\d\cdot J_3) J_1 J_1 J_1\rg_{QF}$.\\

$\underline{\bm{\lg [Q_{\mu\nu},J_\alpha O O]\rg_{QF}} \textbf{ Ward identity}}$
\vspace{3pt}

The higher-spin Ward identity for $\lg (i p\cdot J)_{\mu\nu}(p) J_\alpha(p_1) O(p_2) O(p_3)\rg_{QF}$ can be written out in momentum space using the divergence of the spin-3 current in the QF theory \eqref{5:dJ3} and the QF algebra \eqref{5:QFalgAnsatz}. We find:
\setlength{\jot}{5pt}
\begin{flalign}\begin{aligned}\label{5.2:3100qfwi}
	&\lg (ip\cdot J)_{\mu\nu}(p)J_\alpha(p_1)O(p_2)O(p_3)\rg_{QF} = \\& 
	
	\frac{\tN\tl}{\sqrt{1+\tl^2}}\frac{64i}{3}\bigg[
	
	-i\bigg(p_{1(\mu}+\frac{3}{5}p_{(\mu}-\text{Trace}\bigg)\lg J_{\nu)}(-p_1)J_\alpha(p_1)\rg_{QF}\lg O(p+p_1)O(p_2)O(p_3)\rg_{QF}\\& \qquad\qquad\quad
	
	+\bigg(i\bigg(p_{2(\mu}+\frac{2}{5}p_{(\mu}-\text{Trace}\bigg)\lg O(-p_2)O(p_2)\rg_{QF}\lg J_\alpha(p_1)J_{\nu)}(p+p_2)O(p_3)\rg_{QF} + 
	\begin{Bmatrix}2\leftrightarrow 3\end{Bmatrix}\bigg)\bigg] \\&
	
	+ \frac{1}{\sqrt{1+\tl^2}}\lg \delta_O J^{FF}_\alpha(p_1)O(p_2)O(p_3)\rg_{QF}
	+ \frac{1}{1+\tl^2}\lg \delta_T J^{FF}_\alpha(p_1)O(p_2)O(p_3)\rg_{QF} \\&
 
        + \frac{\tl^2}{1+\tl^2}\lg \delta_T J^{CB}_\alpha(p_1)O(p_2)O(p_3)\rg_{QF}
	
	
	+ \frac{1}{\sqrt{1+\tl^2}}\bigg(\lg J_\alpha(p_1)\delta O^{FF}(p_2)O(p_3)\rg_{QF}+ \begin{Bmatrix}2\leftrightarrow 3\end{Bmatrix}  \bigg)
\end{aligned}\end{flalign}
\setlength{\jot}{0pt}

We've disregarded the double-trace term in $\delta J_\alpha^{QF}$, since it doesn't contribute at leading order in $\tN$. To simplify the discussion, we also disregard terms proportional to $\lg OOO\rg_{QF}$ and $\lg \delta_O J_\alpha^{QF} OO\rg_{QF}$ as these terms do not depend on $p_{2,3}$ and hence should also correspond to contact terms in $\lg J_3 J_1 OO\rg_{QF}$. These contact terms are arbitrary and can be modified by adding local counterterms to the action.

Furthermore, we will separate out terms proportional to the Chern-Simons contact term of the vector current two-point function $\lg JJ\rg_{odd}$, as these will be analyzed separately in Section \ref{Sec5.3}. Using the form of three-point functions \eqref{5:3pt}, the four-point function can now be decomposed into free/critical theory pieces as:
\setlength{\jot}{5pt}
\begin{flalign}\begin{aligned}\label{5.2:3100qfwi2}
	&\lg (ip\cdot J)_{\mu\nu}(p)J_\alpha(p_1)O(p_2)O(p_3)\rg_{QF}\vert_{non-CS\; terms} = \\& \quad
	
	\frac{\tN}{1+\tl^2}\bigg[
		\bigg(\lg J_\alpha(p_1)\delta O^{FF}(p_2)O(p_3)\rg_{FF} 
		+\begin{Bmatrix}2\leftrightarrow 3\end{Bmatrix} \bigg) + 
		\lg \delta_T J^{FF}_\alpha(p_1)O(p_2)O(p_3)\rg_{FF} \bigg] \\& 
		
	+ \frac{\tN\tl}{1+\tl^2}\bigg[
		\bigg(\lg J_\alpha(p_1)\delta O^{FF}(p_2)O(p_3)\rg_{CB} 
		+\begin{Bmatrix}2\leftrightarrow 3\end{Bmatrix} \bigg) 
		\\& \qquad\quad 
		
		+\bigg(-\frac{64}{3}\bigg(p_{2(\mu}+\frac{2}{5}p_{(\mu}-\text{Trace}\bigg)
		\lg O(-p_2)O(p_2)\rg_{FF} \lg J_\alpha(p_1)J_{\nu)}(p+p_2)O(p_3)\rg_{FF}+
		\begin{Bmatrix}2\leftrightarrow 3\end{Bmatrix} \bigg)\bigg] \\& 
	
	+ \frac{\tN\tl^2}{1+\tl^2}\bigg[
		\lg \delta_T J^{CB}_\alpha(p_1)O(p_2)O(p_3)\rg_{CB} \\& \qquad\quad 
		+ \bigg(-\frac{64}{3}\bigg(p_{2(\mu}+\frac{2}{5}p_{(\mu}-\text{Trace}\bigg)
		\lg O(-p_2)O(p_2)\rg_{CB} \lg J_\alpha(p_1)J_{\nu)}(p+p_2)O(p_3)\rg_{CB}+
		\begin{Bmatrix}2\leftrightarrow 3\end{Bmatrix} \bigg) \bigg] \\& 
		  = 
		\frac{\tN}{1+\tl^2}\lg (ip\cdot J)_{\mu\nu} J_\alpha O O\rg_{FF} + 
		\frac{\tN\tl}{1+\tl^2}\lg (ip\cdot J)_{\mu\nu} J_\alpha O O \rg_{odd} + 
		\frac{\tN\tl^2}{1+\tl^2}\lg (ip\cdot J)_{\mu\nu} J_\alpha O O \rg_{CB}
\end{aligned}\end{flalign}
\setlength{\jot}{0pt}

In the $\tl\to 0, \infty$ limits, the Ward identity reduces to those of the free and critical theories, respectively. The four-point function $\lg (ip\cdot J)_{\mu\nu}(p)J_\alpha(p_1)O(p_2)O(p_3)\rg_{QF}$ is constrained by higher-spin symmetry to three structures, which can be written out only in terms of FF and CB correlators. We also note that three point functions appearing in \eqref{5.2:3100qfwi2} satsify simple relations in momentum space (see Appendix \ref{App2}, as well as \cite{Jain:2021gwa}):
\setlength{\jot}{5pt}
\begin{flalign}\label{5.2:OJJeps}
    \lg (\epsilon\cdot J)_\alpha(p_1)J_\beta(p_2) O(p_3)\rg_{FF} &= \lg (J_\alpha(p_1)J_\beta(p_2) O(p_3)\rg_{CB} \\[5pt]

     \lg T_{\mu\nu}OO\rg_{FF} &= \lg T_{\mu\nu} OO\rg_{CB}
\end{flalign}
\setlength{\jot}{0pt}

Given these simplifications, it is interesting to check if there exist additional relations between  the structures of $\lg(ip\cdot J)_{\mu\nu}J_\alpha OO\rg_{QF}$, similar to the ones observed \cite{Kalloor:2019xjb}. It was argued in \cite{Jain:2022ajd} that the correlator $\lg J_3 J_1 OO\rg_{QF}$ contains three structures appearing with the same $\tl$ dependence as in \eqref{5.2:3100qfwi2}, and conjectured that the intermediate structure satisfies an epsilon-transform relation in momentum space:
\begin{flalign}\label{5.2:3100conjecture}
	\lg J_{\mu\nu\rho}J_\alpha O O\rg_{odd} \stackrel{?}{\propto}
	  \lg (\eps\cdot J)_{\mu\nu\rho} J_\alpha O O \rg_{FF-CB} 
	
	\quad\text{ or }\quad 
	
	\lg J_{\mu\nu\rho}J_\alpha O O\rg_{odd} \stackrel{?}{\propto}	
	   \lg J_{\mu\nu\rho} (\eps\cdot J)_\alpha O O \rg_{FF-CB}
\end{flalign}

Using the Ward identity \eqref{5.2:3100qfwi2} we may check this conjecture indirectly. We find a combination of structures that does not contain any non-local terms:
\setlength{\jot}{5pt}
\begin{flalign}\begin{aligned}\label{5.2:3100check}
    &ip^\rho \cdot \bigg( \lg J_{\mu\nu\rho}(p) J_\alpha(p_1)O(p_2)O(p_3)\rg_{QF}-\lg J_{\mu\nu\rho}(p) (\epsilon\cdot J)_\alpha(p_1)O(p_2)O(p_3)\rg_{FF-CB}\bigg) = \\&
    \lg \delta_T (\eps\cdot J)^{CB}_\alpha(p_1)O(p_2)O(p_3)\rg_{CB} -
    \lg \delta_T (\eps\cdot J)^{FF}_\alpha(p_1)O(p_2)O(p_3)\rg_{FF} = \text{contact terms}
\end{aligned}\end{flalign}\setlength{\jot}{0pt}

The terms coming from the divergence of $J_3$ cancel out, and we find that the right-hand side is only a combination of contact terms coming from the higher-spin algebra\footnote{After writing out the local variations $\delta_T (\eps\cdot J)^{FF,CB}_{\alpha} (p_1)$, it is clear that the right-hand side is a function of $p+p_1$, which is equal to $-p_2-p_3$ by momentum conservation, and thus it only depends on $p_2$ and $p_3$.}. Although there is no unique way to invert the divergence in equation \eqref{5.2:3100check}, this analysis is consistent with the epsilon-transform \eqref{5.2:3100conjecture} with respect to the vector current, up to possible contact terms in $p, p_1$.\\

Furthermore, this analysis explicitly excludes the epsilon-transform \eqref{5.2:3100conjecture} with respect to the spin-3 current, as after acting on it by the divergence $ip^\rho$ and using the property $(ip\cdot (\eps\cdot J_3(p)))_{\mu\nu} = \frac{2}{3}(\eps\cdot(ip\cdot J_3))_{\mu\nu}$, we find that the non-contact terms of the two sides do not match.\\

$\underline{\bm{\lg [Q_{\mu\nu},J_\alpha J_\beta J_\gamma]\rg_{QF}} \textbf{ Ward identity}}$
\vspace{3pt}

Using the divergence of the spin-3 current \eqref{5:dJ3} and the higher-spin algebra \eqref{5:QFalgAnsatz} we analyze the Ward identity involving the correlator of a spin-3 current with three vector currents:
\setlength{\jot}{5pt}
\begin{flalign}\begin{aligned}\label{5.2:3111qfwi}
	&\lg (ip\cdot J)_{\mu\nu}(p)J_\alpha(p_1)J_\beta(p_2)J_\gamma(p_3)\rg_{QF} = 
	\\& 
	
	\tN\bigg[\frac{\tl}{\sqrt{1+\tl^2}}\frac{64}{3}
	\bigg(p_{1(\mu}+\frac{3}{5}p_{(\mu} - \text{Trace} \bigg)   
	\lg J_\alpha(p_1)J_{\nu)}(-p_1)\rg_{QF} 
	\lg O(p+p_1)J_\beta(p_2)J_\gamma(p_3) \rg_{QF} \\& \quad
	
	+ \frac{1}{\sqrt{1+\tl^2}}\lg \delta_O J^{FF}_\alpha(p_1)J_\beta(p_2)J_\gamma(p_3)\rg_{QF} + \frac{1}{1+\tl^2}\lg \delta_T J^{FF}_\alpha(p_1)J_\beta(p_2)J_\gamma(p_3)\rg_{QF} \\&\quad

	+ \frac{\tl}{1+\tl^2} \lg \delta_T J^{int}_\alpha(p_1)J_\beta(p_2)J_\gamma(p_3)\rg_{QF} + \frac{\tl^2}{1+\tl^2}\lg \delta_T J^{CB}_\alpha(p_1) J_\beta(p_2) J_\gamma(p_3)\rg_{QF} \\&\quad
	
	+ h_3(\tl)\bigg(\eps_{\alpha\rho(\mu}\lg J_{\nu)}(-p_2)J_\beta(p_2)\rg_{QF}\lg J^\rho(p+p_1+p_2) J_\gamma(p_3) \rg_{QF} + \begin{Bmatrix}\nu)\leftrightarrow \rho \end{Bmatrix}\bigg)\bigg] \\& 
        + \perm{1\leftrightarrow 2}{\alpha\leftrightarrow\beta} 
        + \perm{1\leftrightarrow 3}{\alpha\leftrightarrow\gamma} 
\end{aligned}\end{flalign}
\setlength{\jot}{0pt}

The correlators appearing on the right-hand side of \eqref{5.2:3111qfwi} are of the form $\lg JJ\rg_{QF}$, $\lg JJO\rg_{QF}$ and $\lg TJJ\rg_{QF}$. Each can be separated into FF and CB theory structures as in \eqref{5:2pt} and \eqref{5:3pt}. The three-point function $\lg TJJ\rg_{QF}$ contains an additional parity-odd structure which vanishes in the free/critical limits, however as discussed in Section \ref{Sec6.2}, we can relate it via an epsilon-transform to FF and CB pieces as \eqref{6.2:TJJepsFull}. Thus we can decompose \eqref{5.2:3111qfwi} into purely free theory structures.\\

We again separate out terms proportional to the Chern-Simons term $\lg JJ\rg_{odd}$, as these will be analyzed separately in Section \ref{Sec5.3}. The remaining terms can be grouped as follows:

\setlength{\jot}{5pt}
\begin{flalign}\begin{aligned}\label{5.2:3111nc}
	&\lg (ip\cdot J)_{\mu\nu}(p)J_\alpha(p_1)J_\beta(p_2)J_\gamma(p_3)\rg_{QF}\vert_{non-CS\; terms} = 
	\\& \quad
	
	\frac{1}{(1+\tl^2)^2}\bigg[ 
		\lg \delta_O J^{FF}_\alpha(p_1)J_\beta(p_2)J_\gamma(p_3)\rg_{FF} + 
		\lg \delta_T J^{FF}_\alpha(p_1)J_\beta(p_2)J_\gamma(p_3)\rg_{FF}
	\bigg] \\& 
	
	+ \frac{\tl}{(1+\tl^2)^2}\bigg[ 
		\lg \delta_O J^{FF}_\alpha(p_1)J_\beta(p_2)J_\gamma(p_3)\rg_{CB} + 
		\lg \delta_T J^{FF}_\alpha(p_1)J_\beta(p_2)J_\gamma(p_3)\rg_{odd} 
		\\& \qquad\qquad\qquad 
		+\frac{64}{3}\bigg(\frac{3}{5}p_{(\mu}+p_{1(\mu} - \text{Trace} \bigg)   
		\lg J_\alpha(p_1)J_{\nu)}(-p_1)\rg_{FF}
		\lg O(p+p_1)J_\beta(p_2)J_\gamma(p_3) \rg_{FF}
	\bigg] \\& 
	
	+ \frac{\tl^2}{(1+\tl^2)^2}\bigg[ 
		\lg \delta_O J^{FF}_\alpha(p_1)J_\beta(p_2)J_\gamma(p_3)\rg_{FF} +
		\lg \delta_T J^{FF}_\alpha(p_1)J_\beta(p_2)J_\gamma(p_3)\rg_{CB} 
        \\& \qquad\qquad\qquad
		+ \lg \delta_T J^{CB}_\alpha(p_1)J_\beta(p_2)J_\gamma(p_3)\rg_{FF} 
		\\[5pt]& \qquad\qquad\qquad
		+\frac{64}{3}\bigg(\frac{3}{5}p_{(\mu}+p_{1(\mu} - \text{Trace} \bigg)   
		\lg J_\alpha(p_1)J_{\nu)}(-p_1)\rg_{CB}
		\lg O(p+p_1)J_\beta(p_2)J_\gamma(p_3) \rg_{CB} 
	\bigg]\\&
	
	+ \frac{\tl^3}{(1+\tl^2)^2}\bigg[
		\lg \delta_O J^{FF}_\alpha(p_1)J_\beta(p_2)J_\gamma(p_3)\rg_{CB} +
		\lg \delta_T J^{CB}_\alpha(p_1)J_\beta(p_2)J_\gamma(p_3)\rg_{odd}
		\\& \qquad\qquad\qquad
		+\frac{64}{3}\bigg(\frac{3}{5}p_{(\mu}+p_{1(\mu} - \text{Trace} \bigg)   
		\lg J_\alpha(p_1)J_{\nu)}(-p_1)\rg_{FF}
		\lg O(p+p_1)J_\beta(p_2)J_\gamma(p_3) \rg_{FF}
	\bigg] \\& 
	
	+ \frac{\tl^4}{(1+\tl^2)^2}\bigg[
		\lg \delta_T J^{CB}_\alpha(p_1)J_\beta(p_2)J_\gamma(p_3)\rg_{CB} 
        \\&\qquad\qquad\qquad
		+\frac{64}{3}\bigg(\frac{3}{5}p_{(\mu}+p_{1(\mu} - \text{Trace} \bigg)   
		\lg J_\alpha(p_1)J_{\nu)}(-p_1)\rg_{CB}
		\lg O(p+p_1)J_\beta(p_2)J_\gamma(p_3) \rg_{CB}
	\bigg]\\& 
	
	+ \tN h_3(\tl)\bigg(\eps_{\alpha\rho(\mu}\lg J_{\nu)}(-p_2)J_\beta(p_2)\rg_{CB}\lg J^\rho(p+p_1+p_2) J_\gamma(p_3) \rg_{CB} + \begin{Bmatrix}\nu)\leftrightarrow \rho \end{Bmatrix}\bigg)\\&
        + \perm{1\leftrightarrow 2}{\alpha\leftrightarrow\beta} 
        + \perm{1\leftrightarrow 3}{\alpha\leftrightarrow\gamma} 
\end{aligned}\end{flalign}
\setlength{\jot}{0pt}

In the $\tl\to 0,\infty$ limits, we again recover free/critical theory Ward identities respectively, since $h_3(\tl)\to 0$.\\

An initial analysis of the $\lg J_{\mu\nu\rho} J_\alpha J_\beta J_\gamma\rg_{QF}$ four-point function was carried out in \cite{Jain:2022ajd}, suggesting that the correlator contains three or more structures at separated points. In \eqref{5.2:3111nc}, we observe five structures, and possibly an additional disconnected term, depending on the unfixed function $h_3(\tl)$. Curiously, not all the structures are independent, as we notice a specific combination that does not contain any non-contact terms. We name the five structures in \eqref{5.2:3111nc} appearing with factors $\tl^i(1+\tl^2)^{-2}$ for $i=0,1,2,3,4$ by subscripts $FF, X_1, X_2, X_3$, and $CB$, respectively. Using  $(ip\cdot (\eps\cdot J_3(p)))_{\mu\nu} = \frac{2}{3}(\eps\cdot(ip\cdot J_3))_{\mu\nu}$, we find:
\setlength{\jot}{5pt}
\begin{flalign}\begin{aligned}\label{5.2:3111check}
    &i p^\rho\cdot\bigg(\lg J_{\mu\nu\rho}(p) J_\alpha(p_1) J_\beta(p_2) J_\gamma(p_3)\rg_{FF+CB-X_2} 
    -\lg (\eps\cdot J)_{\mu\nu\rho}(p)J_\alpha(p_1) J_\beta(p_2) J_\gamma(p_3) \rg_{X_1-X_3} \bigg) =\\&
    \lg (\delta_T J_\alpha^{FF}-\delta_T J_\alpha^{CB})(p_1) J_\beta(p_2) J_\gamma(p_3)\rg_{FF-CB}
    -\frac{2}{3}\lg (\eps\cdot(\delta_T J_\alpha^{FF}-\delta_T J_\alpha^{CB}))(p_1) J_\beta (p_2) J_\gamma(p_3)\rg_{odd}\\& + \perm{1\leftrightarrow 2}{\alpha\leftrightarrow\beta} 
       + \perm{1\leftrightarrow 3}{\alpha\leftrightarrow\gamma} = \text{contact terms}
\end{aligned}\end{flalign}
\setlength{\jot}{0pt}

By expanding the local variations \eqref{4.1:dTJ}, we can confirm that the first term on the right-hand side is a function of $p+p_1=-p_2-p_3$ and thus a contact term, depending only on $p_2, p_3$, and similarly for the other two permutations. Thus it is possible, even when not acting by the divergence $ip^\rho$, that the left-hand side of \eqref{5.2:3111check} vanishes up to contact terms. Indeed, we confirm this in Section \ref{Sec:3111}, for collinear momenta and specific Lorentz indices, by direct perturbative calculation. It is worth noting that a similar (and in fact, slightly stronger) property was observed in the collinear limit in \cite{Kalloor:2019xjb} for the five linearly-dependent structures of $\lg J_1 J_1 J_1 J_1\rg_{QF}$.\\

The main results of this section are the momentum space decomposition of four-point functions \eqref{5.2:3100qfwi2} and \eqref{5.2:3111nc} into free theory structures, which further suggest possible epsilon-transform relations for $\lg J_3 J_1 OO\rg_{QF}$ and $\lg J_3 J_1 J_1 J_1 \rg_{QF}$:
\setlength{\jot}{10pt}
\begin{flalign}
    &\lg J_{\mu\nu\rho}J_\alpha O O\rg_{odd} = 
    \lg J_{\mu\nu\rho}(\eps\cdot J)_\alpha O O\rg_{FF-CB} + \text{contact terms} \label{4:3100epstr}\\&
    \lg J_{\mu\nu\rho}J_\alpha J_\beta J_\gamma \rg_{FF+CB-X_2}=
    \lg (\eps\cdot J)_{\mu\nu\rho}J_\alpha J_\beta J_\gamma \rg_{X_1-X_3} + \text{contact terms} \label{4:3111epstr}
\end{flalign}
\setlength{\jot}{0pt}

Our results are partially consistent with the conjectures of \cite{Jain:2022ajd}. Following the approach of \cite{Aharony:2024nqs,Jain:2024bza}, we emphasize that the form of $\lg J_3 J_1 OO\rg_{QF}$ and $\lg J_3 J_1 J_1 J_1\rg_{QF}$ further simplifies in spinor-helicity variables, as is often the case for two- and three-point functions of massless fields. 

\subsection{Relations between contact terms of QF (and QB) theories}\label{Sec5.3} 
Contact terms are terms appearing within correlation functions when two or more operator insertions coincide, reflecting the short distance physics of the theory. Most commonly, such terms are arbitrary, and can be changed by imposing a different regularization scheme or adding suitable counterterms to the action. In some cases however, contact terms hold physical significance, highlighting an underlying feature of the theory. A typical example of this is the seagull term of scalar QED, which is required by gauge invariance. In the case of Chern-Simons-matter theories, we encounter a third class of contact terms that is neither fully universal, nor arbitrary \cite{Closset:2012vp}. As mentioned in \eqref{5:2pt}, two-point functions of spinning currents develop conformally invariant contact terms at finite $\tl$. We encounter such a term in our analysis of higher-spin Ward identities, as $\lg J_\mu(p) J_\nu(-p)\rg_{QF}\propto \kappa_1(\tN,\tl) \eps_{\mu\nu\rho}p^\rho$. This term corresponds to a background Chern-Simons term in the generating functional of the theory. Following the discussion of \cite{Closset:2012vp}, one can only shift the value of this contact term by integer multiples, hence the fractional part of $\kappa_1(\tN,\tl)$ is a physical observable. Note that in \eqref{5:2pt}, we have defined the function $g_1(\tl)$ to be the leading contribution to $\kappa_1(\tN,\tl)$ in the large $\tN$ expansion, therefore the function $g_1(\tl)$ by itself does not define the observable $(\kappa_1(\tN,\tl) \mod 1)$. \\

To understand the appearance of new contact terms within QF correlators, it is instructive to work in the description \eqref{CB+CS}. Equivalently, we can reach this model by starting from the theory of a free boson coupled to Chern-Simons gauge interactions and Legendre transforming it with respect to the scalar current. The CFT pre-Legendre transforming is referred to as the quasibosonic (QB) Chern-Simons-matter theory, studied in \cite{Aharony:2012nh, Aharony:2011jz}. It is also constrained by approximate higher-spin symmetry, and its two-point functions of spinning currents are analogous to those of the QF theory\footnote{The kinematic form of these propagators is constrained by conformal invariance, however the appearing contact terms in principle have a general dependence $\tilde{g}_s(\tl_{qb})$ on some new higher-spin symmetry variable $\tl_{qb}$.} \eqref{5:2pt}.
\begin{figure}[htbp]
\centering
\includegraphics[width=0.3\textwidth]{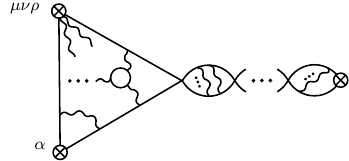}
\caption{A generic diagram contributing to $\lg J_{\mu\nu\rho}J_\alpha O\rg_{QF}$ in the CB+CS description \eqref{CB+CS}. The encircled crosses denote vertices in the QB theory, the straight lines denote full boson propagators and the squiggly lines denote gluon propagators. The scalar vertex receives corrections, similarly to the CB theory vertex. The diagrams are summed over all possible gluon line configurations and scalar bubbles, as well as permutations of external vertices. }
\label{fig:310diagQF}
\end{figure}
One can now ask how contact terms of associated QB and QF theories are related. The Legendre transform generally introduces new non-contact and contact terms through multiplication in momentum space \eqref{4:FBtoCB}. We start from the simplest Ward identity \eqref{5.1:310qfwi}, noting that the correlator $\lg(ip\cdot J)_{\mu\nu}(p)J_\alpha(p_1)O(p_2)\rg_{QF}$ receives new contact terms at finite $\tl$:
\begin{flalign}\begin{aligned}\label{5.3:QFct}
&\lg (ip\cdot J)_{\mu\nu}(p)J_\alpha(p_1)O(p_2)\rg_{QF}\vert_{CS\; terms} = 
\tN\frac{g_1(\tl)}{\sqrt{1+\tl^2}}\lg J_\alpha(p_1) \delta O^{FF}(p_2)\rg_{odd}\\& 
+\tN\frac{g_1(\tl)\tl}{\sqrt{1+\tl^2}}\frac{64i}{3}\bigg(-\frac{2}{5}ip_{1(\mu}+\frac{3}{5}ip_{2(\mu}-\text{Trace}\bigg)\lg J_\alpha(p_1)J_{\nu)}(p+p_2)\rg_{odd}\lg O(-p_2)O(p_2)\rg_{CB}
\end{aligned}\end{flalign} 
After expanding the variation of the scalar \eqref{5:ffvar} and the form of two-point functions, we find:
\setlength{\jot}{0pt} 
\begin{flalign}\begin{aligned}
    &\lg (ip\cdot J)_{\mu\nu}(p)J_\alpha(p_1)O(p_2)\rg_{QF}\vert_{CS\; terms} = \\&

    -\frac{\tN g_1(\tl)}{\sqrt{1+\tl^2}} \frac{8}{3}\bigg(

    p_{1\mu}p_{1\nu}p_{1\alpha}-w_1^{FF}p_{1(\mu}p_{\nu)}p_\alpha
    +(w_1^{FF}+w_2^{FF})(p_{1(\mu}p_{1\nu)}p_{\alpha}+p_{1(\mu}p_{\nu)}p_{1\alpha})\\&

    \qquad\qquad\qquad

    +\delta_{\alpha(\mu}\big( 
            - p_{1\nu)} p_1^2 
            - w_1^{FF} p_{\nu)} p \cdot p_1 
            - (w_1^{FF}+w_2^{FF})(p_{\nu)}p_1^2+p_{1\nu)}p\cdot p_1)
    \big)\bigg) \\&

    - \frac{\tN g_1(\tl)\tl}{\sqrt{1+\tl^2}}\frac{64i}{3}\eps_{\alpha\sigma(\mu}p_2
    \bigg(\frac{2i}{5}p_{1\nu)}p_1^\sigma - \frac{3i}{5}p_{2\nu)}p_1^\sigma-\text{Trace}\bigg)
\end{aligned}\end{flalign}
\setlength{\jot}{0pt} 

The corresponding QB correlator satisfies:
\begin{flalign}
    \lg (ip\cdot J)_{\mu\nu}(p)J_\alpha(p_1)O(p_2)\rg_{QF}=\lg (ip\cdot J)_{\mu\nu}(p)J_\alpha(p_1)O(p_2)\rg_{QB}\lg O(p_2)O(-p_2)\rg_{QB}^{-1}
\end{flalign}
As in Section \ref{Sec4}, this expression comes from summing Feynman diagrams, displayed in Figure \ref{fig:310diagQF}. We can invert the relation without ambiguity, to find  that the QB correlator also contains contact terms. We find\footnote{The scalar two-point function of the quasibosonic theory satisfies $\lg O(p)O(-p)\rg_{QB} = \tN \lg O(p)O(-p)\rg_{FB} = \tN \frac{8}{p}$.}:
\begin{flalign}\begin{aligned}\label{5.3:QBct}

 &\lg (ip\cdot J)_{\mu\nu}(p)J_\alpha(p_1)O(p_2)\rg_{QB}\vert_{CS\; terms} = \\&

    -\frac{\tN g_1(\tl)}{\sqrt{1+\tl^2}} \frac{64}{3}\frac{1}{p_2}\bigg(

    p_{1\mu}p_{1\nu}p_{1\alpha}-w_1^{FF}p_{1(\mu}p_{\nu)}p_\alpha
    +(w_1^{FF}+w_2^{FF})(p_{1(\mu}p_{1\nu)}p_{\alpha}+p_{1(\mu}p_{\nu)}p_{1\alpha})\\&

    \qquad\qquad\qquad

    +\delta_{\alpha(\mu}\big( 
            - p_{1\nu)} p_1^2 
            - w_1^{FF} p_{\nu)} p \cdot p_1 
            - (w_1^{FF}+w_2^{FF})(p_{\nu)}p_1^2+p_{1\nu)}p\cdot p_1)
    \big)\bigg) \\&

    - \frac{\tN g_1(\tl)\tl}{\sqrt{1+\tl^2}}\frac{64i}{3}\eps_{\alpha\sigma(\mu}p_2
    \bigg(\frac{16i}{5}p_{1\nu)}p_1^\sigma - \frac{24i}{5}p_{2\nu)}p_1^\sigma-\text{Trace}\bigg)
\end{aligned}\end{flalign} 
With the first term being a contact term in $p_1$ and the second term being a contact term in $p_2$. Going a step further, we argue that the contributions in \eqref{5.3:QFct} and \eqref{5.3:QBct} correspond to honest contact terms in $\lg J_3 J_1 O\rg_{QF,QB}$. It is obvious that these terms do not appear in the free and critical limits of QB and QF theories, where $\lg JJ\rg_{odd}=0$, and therefore do not arise as a result of the divergence acting on a non-contact term in the free/critical boson/fermion theory. However, since the non-contact pieces in $\lg J_{\mu\nu\rho} J_\alpha O\rg_{QF,QB}$ are fully determined by their free and critical structures \cite{Maldacena:2012sf}, the contact terms \eqref{5.3:QFct}, \eqref{5.3:QBct} must result from acting on a contact term of this correlator by $ip^\rho$. There is no unique way to invert this statement and obtain expressions for contact terms of $\lg J_{\mu\nu\rho} J_\alpha O\rg_{QF,QB}$, other than direct perturbative calculation. Nevertheless, we can present an example of how such contact terms could look like:
\setlength{\jot}{5pt}
\begin{flalign}\begin{aligned}\label{5.3:310ct}
&\lg J_{\mu\nu\rho}(p)J_\alpha(p_1)O(p_2)\rg_{QF}\vert_{CS\; terms} = \\&

\frac{\tN g_1(\tl)}{\sqrt{1+\tl^2}} \bigg[
 i\frac{8}{3}\frac{p_\rho}{3p^2}\bigg(

    p_{1\mu}p_{1\nu}p_{1\alpha}-w_1^{FF}p_{1(\mu}p_{\nu)}p_\alpha
    +(w_1^{FF}+w_2^{FF})(p_{1(\mu}p_{1\nu)}p_{\alpha}+p_{1(\mu}p_{\nu)}p_{1\alpha})\\&

    \qquad\qquad\qquad

    +i\frac{p_\rho}{3p^2}\delta_{\alpha(\mu}\big( 
            - p_{1\nu)} p_1^2 
            - w_1^{FF} p_{\nu)} p \cdot p_1 
            - (w_1^{FF}+w_2^{FF})(p_{\nu)}p_1^2+p_{1\nu)}p\cdot p_1)
    \big)\bigg)\bigg] \\&

    + \frac{\tN g_1(\tl)\tl}{\sqrt{1+\tl^2}}\frac{64i}{3}\frac{ip_\rho}{3p^2}\eps_{\alpha\sigma(\mu}p_2
    \bigg(\frac{2i}{5}p_{1\nu)}p_1^\sigma - \frac{3i}{5}p_{2\nu)}p_1^\sigma-\text{Trace}\bigg)

+(\text{Permutations over $\mu\leftrightarrow\nu\leftrightarrow\rho$})
\end{aligned}\end{flalign} 
\setlength{\jot}{0pt}

And similarly for the QB theory. We thus find that the three-point functions $\lg J_3 J_1 O\rg$ of Chern-Simons-matter theories must contain contact terms with specific dependences on the 't Hooft coupling for the theory to satisfy higher-spin symmetry locally. The function $g_1(\tl)$ was determined from perturbative calculations using \eqref{FF+CS} and \eqref{CB+CS} in \cite{Gur-Ari:2012lgt, Aharony:2012nh}. In the FF+CS theory, it's given by $g_1(\tl)\propto \tl$, while for the CB+CS theory it's given by $g_1(\tl)\propto \tl^{-1}$. The former description implies that the correlator $\lg J_3 J_1 O\rg_{QF}$ contains a contact term going like $\tl^2(1+\tl^2)^{-1/2}$, while in the latter description the same correlator contains a contact term proportional to $(\tl^2(1+\tl^2))^{-1/2}$. Although such terms appear diverging in the $\tl\to\infty, 0$ limits respectively, they are removed, similarly to $\lg JJ\rg_{odd}$. 

We find that the two descriptions are inequivalent at the level of contact terms to leading order in large $\tN$. It should be emphasized that still doesn't necessarily imply that \eqref{FF+CS} and \eqref{CB+CS} differ in the values of the observable $(\kappa_1(\tN,\tl) \mod 1)$, as this value is also dependent on $\mathcal{O}(1/\tN)$ corrections. Even so, unless the coefficient of $\lg JJ\rg_{odd}$ in \eqref{5:2pt} is integer valued, we find that the contact terms \eqref{5.3:310ct} cannot be naively removed by a different choice of regularization scheme. 

In the limits \cite{Aharony:2024nqs,Jain:2024bza}, three-point functions $\lg J_3 J_1 O\rg_{QF, QB}$ map to corresponding bulk vertices of chiral higher-spin gravity, when written in spinor-helicity variables. Although the $\tN$ and $\tl$ dependences of various contact terms in \eqref{5.3:310ct} causes these contributions to diverge, we note that we can always add arbitrary contact terms, with the same $\tN$, $\tl$ dependence, such that they cancel the Chern-Simons contact terms in the spinor-helicity basis. Thus we conclude that terms \eqref{5.3:310ct} are not visible in the chiral limits \cite{Aharony:2024nqs,Jain:2024bza}. \\

We can also track contact terms appearing in higher correlators through Ward identities of Section \ref{Sec5.2}, and determine if these terms are arbitrary or meaningful. For simplicity, we only consider contributions containing $\lg JJ\rg_{odd}$, as the remaining correlators do not contain physically meaningful contact terms. Separating out terms of \eqref{5.2:3100qfwi} leads to:
\setlength{\jot}{5pt}
\begin{flalign}\begin{aligned}\label{5.3:3100ct}
&\lg (ip\cdot J)_{\mu\nu}(p)J_\alpha(p_1)O(p_2)O(p_3)\rg_{QF}\vert_{CS\; terms} =\\& \frac{\tN g_1(\tl)\tl}{\sqrt{1+\tl^2}}\frac{64}{3}\bigg(p_{1(\mu}+\frac{3}{5}p_{(\mu}-\text{Trace}\bigg)\lg J_{\nu)}(-p_1)J_\alpha(p_1)\rg_{odd}\lg O(p+p_1)O(p_2)O(p_3)\rg_{QF}
\end{aligned}\end{flalign}
\setlength{\jot}{0pt}

The right hand side in $p_{2,3}$, however it is scheme-dependent, as it can be subtracted by adding counterterms corresponding to the scalar three-point function. It vanishes in perturbative calculations \cite{Gur-Ari:2012lgt} of the FF+CS theory. When Legendre transforming the QB theory \cite{Aharony:2012nh}, one finds a momentum space constant in $\lg OOO\rg_{QF}$, hence the correlator $\lg (ip\cdot J)_{\mu\nu}(p)J_\alpha(p_1)O(p_2)O(p_3)\rg_{QF}$ contains a contact term going as $g_1(\tl)\tl^2(1+\tl^2)^{1/2}$ in the CB+CS description. We analyze contact terms appearing in \eqref{5.2:3111qfwi} in similar fashion:
\setlength{\jot}{5pt}
\begin{flalign}\begin{aligned}\label{5.3:QF3111ct}
&\lg (ip\cdot J)_{\mu\nu}(p)J_\alpha(p_1)J_\beta(p_2)J_\gamma(p_3)\rg_{QF}\vert_{CS\; terms} =\\& \frac{\tN g_1(\tl)\tl}{\sqrt{1+\tl^2}}\frac{64}{3}\bigg(p_{1(\mu}+\frac{3}{5}p_{(\mu}-\text{Trace}\bigg)\lg J_{\nu)}(-p_1)J_\alpha(p_1)\rg_{odd}\lg O(p+p_1)J_\beta(p_2)J_\gamma(p_3)\rg_{QF}\\& 

+ \tN h_3(\tl)\eps_{\alpha\rho(\mu}\bigg(\lg J_{\nu)}(-p_2)J_\beta(p_2)\rg_{QF}\lg J^\rho(-p_3) J_\gamma(p_3) \rg_{QF} - \lg J_{\nu)}(-p_2)J_\beta(p_2)\rg_{FF}\lg J^\rho(-p_3) J_\gamma(p_3) \rg_{FF} \\&

\qquad \qquad \qquad \qquad 
+ \begin{Bmatrix}\nu)\leftrightarrow \rho \end{Bmatrix}\bigg) + \perm{1\leftrightarrow 2}{\alpha\leftrightarrow\beta} + 
	\perm{1\leftrightarrow 3}{\alpha\leftrightarrow\gamma}
\end{aligned}\end{flalign}
\setlength{\jot}{0pt}

We notice that this correlator contains contact terms that are scheme-independent, as removing them in a consistent way would require us to subtract the fractional contact term within the spin-1 two-point function. By expanding the correlators on the right-hand side of \eqref{5.3:QF3111ct} via \eqref{5:2pt} and \eqref{5:3pt}, we find the $\tl$ dependence of the contact terms. It would be interesting to try and reproduce this result perturbatively, and furthermore, it would provide constraints on the unfixed function $h_3(\tl)$.\\

In principle, one can decide to add various arbitrary contact terms to the theory by adding contributions in dynamical and background fields to the effective action. We may explicitly choose to add terms that break charge conjugation symmetry to a specific correlator. To maintain higher-spin symmetry in such cases, we must add similar terms to correlators related by Ward identities. For example, after adding a contact term to $\lg JOO\rg_{QF}$, leads to non-trivial Ward identities for $\lg (\d\cdot J_3) JJO\rg_{QF}$, and so forth.
\section{Perturbative Chern-Simons-matter theory calculations}\label{Sec6}

The results of the previous sections can be supplemented by direct computations using the action of QF Chern-Simons-matter theory \eqref{FF+CS}, at fixed, finite\footnote{The line of unitary CFT fixed points exists for $0\leq\lambda\leq 1$.} $\lambda=\frac{N}{k}$, while keeping both $N$ and $k$ large. We set up the computation in momentum space, as in \cite{Giombi:2011kc,Gur-Ari:2012lgt,Kalloor:2019xjb}. We employ light-cone coordinates $x^\pm=x_\mp=(x^1\pm ix^2)/\sqrt{2}$ and work in the light-cone gauge $A_-=0$. With this choice of gauge fixing, the self-interaction of gauge bosons vanishes, further simplifying perturbative calculations. The gluon propagator is given by:
\begin{flalign}\begin{aligned}\label{6:GluonProp}
	\lg A^a_\mu(-q)A^b_\nu(p)\rg_{QF} =& (2\pi)^3\delta^{(3)}(q-p)\delta^{ab}G_{\nu\mu}(p)\\
	G_{+3}(p)=&-G_{3+}(p) = \frac{4\pi i}{k}\frac{1}{p^+}
\end{aligned}\end{flalign}
With all other components vanishing. The fermion planar propagator was computed in \cite{Giombi:2011kc}, to all orders in $\lambda$:
\begin{flalign}\begin{aligned}\label{6:FermProp}
	&\lg\psi_i(p)\bps^j(-q)\rg_{QF} = (2\pi)^3\delta^{(3)}(p-q)\delta^j_i S(p)\\
	&S(p) = \frac{-i\g^\mu p_\mu + i\lambda^2\g^+p^-+\lambda p_s}{p^2}, \text{ where }p_s^2=2p^+p^-
\end{aligned}\end{flalign}
As in \cite{Aharony:2012nh,Gur-Ari:2012lgt}, we use cutoff regularization in the radial $p^{+-}$ plane, with cutoff $\Lambda$ and dimensional regularization along the $p^3$ direction. We will be working in the collinear limit, so that all external momenta point along the 3-direction, i.e. $p_i^\mu = p_i \delta^\mu_3$.

Correlators involving spin-0 and spin-1 currents in the QF theory were previously computed to all orders in 't Hooft coupling in the collinear limit in \cite{Gur-Ari:2012lgt,Kalloor:2019xjb}. The computation hinges upon computing the exact vertex functions, by solving the recursive Schwinger-Dyson equations at large $N$. The vertex operators are gauge-dependent and are also computed for external momenta $q$ along the 3-axis. We reference those results here:
\begin{flalign}
	\lg O(-q)\ps_i(k)\bps^j(-p)\rg_{QF} =& V_0(q,p)\delta^j_i(2\pi)^3\delta^{(3)}(q+p-k) \label{6:spin0vertex}\\	
	\lg J^\pm(-q)\ps_i(k)\bps^j(-p)\rg_{QF} =& V^\pm(q,p)\delta^j_i(2\pi)^3\delta^{(3)}(q+p-k) \label{6:spin1vertex}
\end{flalign}
As a shorthand, the vertex functions are written in terms of dimensionless variables. The loop momentum appears as $y=\frac{2p_s}{|q_3|}$, the cutoff in the $p^{+-}$ plane is given by $\Lambda'=\frac{2\Lambda}{|q_3|}$ and $\hat{\lambda}=\lambda\text{sign}(q_3)$ controls the coupling to the gauge field\footnote{Note that $\hat{\lambda}$ is invariant under a parity transformation combined with $\lambda\to - \lambda$}: 
\begin{flalign}\begin{aligned}\label{6:OJvertfunc}
	V_0(q,p)=& \frac{2\lambda p_+}{p_s}\frac{1-iy\hat{\lambda}-(1+iy\hat{\lambda})e^{-2i\hat{\lambda}\arctan(y)}}{\hat{\lambda}y(1+e^{-2i\hat{\lambda}\arctan(\Lambda')})}\g^+ + \frac{1+e^{-2i\hat{\lambda}\arctan(y)}}{1+e^{-2i\hat{\lambda}\arctan(\Lambda')}}\mathds{1} \\

	V^+(q,p)=& \frac{i}{2}(1-iy\hat{\lambda}+(1+iy\hat{\lambda})e^{2i\hat{\lambda}(\arctan(\Lambda')-\arctan(y))})\g^+ + \frac{ip^+}{q_3}(1-e^{2i\hat{\lambda}(\arctan(\Lambda')-\arctan(y))}) \mathds{1} \\
	
	V^-(q,p)=& i\g^- + \frac{i(p^-)^2}{p_s^2 y^2}((1+iy\hat{\lambda})e^{-2i\hat{\lambda}\arctan(y))}-(1-iy\hat{\lambda})) \g^+ + \frac{ip^-}{q_3 y^2}(1-2iy\hat{\lambda}-e^{-2i\hat{\lambda}\arctan(y)}) \mathds{1}
\end{aligned}\end{flalign}

In Appendix \ref{Sec6.1}, we compute the vertex corresponding to the plus-component of a general spin-$s$ current: 
\begin{flalign}\begin{aligned}\label{6.1:svertex}
  &\lg J_s^\pm(-q)\ps_i(k)\bps^j(-p)\rg_{QF} = V_s^{++...+}(q,p)\delta^j_i(2\pi)^3\delta^{(3)}(q+p-k)\\[3pt]&
   
   V_s^{++...+}(q,p)=\beta_s (p^+)^{s-1} V^+(q,p)
\end{aligned}\end{flalign}
Where $\beta_s$ is just an overall normalization factor\footnote{In our conventions \eqref{1FFCurrents} we have $\beta_2=1$ and $\beta_3=8i/3$.}. We now proceed to compute three- and four-point functions of the QF theory, involving the stress-energy tensor and spin-3 current.

\subsection{\texorpdfstring{$\lg T_{--} J_+ J_+\rg_{QF}$}{} calculation}\label{Sec6.2}
\begin{figure}[htbp]
\centering
\includegraphics[width=0.6\textwidth]{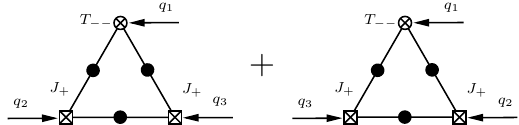}
\caption{Diagrams contributing to the three-point function $\lg T_{--}(-q_1)J_+(-q_2)J_+(-q_3)\rg_{QF}$. The exact spin-2 vertex is represented by an encircled cross, while the exact spin-1 vertices are represented by crossed squares. The exact fermion propagators are represented by filled circles.}
\label{fig:211qf}
\end{figure}

Using the vertex expressions of the previous section we can evaluate correlation functions of QF Chern-Simons-matter theory. We first compute the three-point function $\lg T_{--}(-q_1)J_+(-q_2)J_+(-q_3)\rg_{QF}$ for momenta $q_i$ pointing along the 3-axis. The one-loop contributions to the parity-odd structure of this correlator were analyzed in \cite{Giombi:2011kc}. At large $N$, the contributing diagrams are displayed in Figure \ref{fig:211qf}.
The expression reads:
\begin{flalign}\begin{aligned}\label{6.2:TJJint}
	&\lg T_{--}(-q_1)J_+(-q_2)J_+(-q_3)\rg_{QF} =\\[5pt]& 
 
        -N\int \frac{d^3 p}{(2\pi)^3} Tr\bigg[ S(p+q_1)V^{++}(q_1,p)S(p)V^-(q_2,p)S(p-q_2)V^-(q_3,p)\\&
	\qquad\qquad\qquad\qquad
	+ S(p+q_1)V^{++}(q_1,p)S(p)V^-(q_3,p)S(p-q_3)V^-(q_2,p)\bigg]
\end{aligned}\end{flalign}

Using the vertex expressions \eqref{6:spin1vertex}, \eqref{6.1:svertex} and the fermion propagator \eqref{6:FermProp} the integral can be evaluated explicitly. In the collinear limit, momentum conservation implies $q_1=-q_2-q_3$. After regularizing, we take the $p^{+-}$ plane cutoff $\Lambda$ to infinity, recovering a gauge-invariant expression for the correlator. Defining $\hl_i=\lambda \text{sign}(q_i)$, for $i=1,2,3$, we have:
\begin{flalign}\begin{aligned}
\lg T_{--}(-q_1)J_+(-q_2)J_+(-q_3)\rg_{QF} = 
&\frac{iN}{32\pi\lambda}
\frac{q_2q_3(2q_2^2+2q_3^2+5q_2q_3)-q_2^2q_3^2 e^{-i\pi(\hl_2+\hl_3)}}{q_2 q_3(q_2+q_3)} \\&
+\frac{iN}{32\pi\lambda}
\frac{q_2^2(q_2+q_3)^2e^{-i\pi\hl_2}+q_3^2(q_2+q_3)^2 e^{-i\pi\hl_3}-(q_2+q_3)^4 e^{i\pi\hl_1}}{q_2 q_3(q_2+q_3)}
\end{aligned}\end{flalign}
The result is $q_2\leftrightarrow q_3$ symmetric, as expected since the currents are bosonic operators. To rewrite the correlator in standard form, we expand the exponentials and use the relations between parameters of the theory and higher-spin variables $\tl$, $\tN$ \eqref{5:tNtldef}. We find:
\setlength{\jot}{5pt}
\begin{flalign}\begin{aligned}\label{6.2:TJJ}
&\lg T_{--}(-q_1)J_+(-q_2)J_+(-q_3)\rg_{QF} =\\& \quad
\frac{\tN}{1+\tl^2}\bigg(\frac{q_2^3(q_2+2q_3)\sign(q_2)+q_3^3(2q_2+q_3) \sign(q_3)-(q_2+q_3)^4\sign(q_2+q_3)}{64q_2q_3(q_2+q_3)}\bigg)\\& 

+\frac{\tN\tl}{1+\tl^2}\bigg(i\frac{q_2q_3(1+\sign(q_2)\sign(q_3))}{32(q_2+q_3)}\bigg)\\& 

+\frac{\tN\tl^2}{1+\tl^2}\bigg(\frac{(q_2+q_3)^2(q_2^2\sign(q_2)+q_3^2\sign(q_3))+q_2^2q_3^2(\sign(q_2)+\sign(q_3))-(q_2+q_3)^4\sign(q_2+q_3)}{64q_2q_3(q_2+q_3)}\bigg)\\& 

+\tN\tl\bigg(\frac{i}{32}(q_2+q_3)\bigg)
\end{aligned}\end{flalign}
\setlength{\jot}{0pt}

The correlator satisfies the form \eqref{5:3pt} predicted by weakly broken higher-spin symmetry. One can check that the three non-local terms in \eqref{6.2:TJJ} satisfy an epsilon-transform relation in the collinear limit:
\begin{flalign}\label{6.2:TJJepsColl}
	\lg T_{--}(-q_1)J_+(-q_2)J_+(-q_3)\rg_{odd} = -i\sign(q_2+q_3)\lg T_{--}(-q_1)J_+(-q_2)J_+(-q_3)\rg_{FF-CB}
\end{flalign}
This relation is just the collinear limit version of:
\begin{flalign}\label{6.2:TJJepsFull}
	\lg T_{\mu\nu}(-q_1)J_\alpha(-q_2)J_\beta(-q_3)\rg_{odd} =-\lg (\eps\cdot T)_{\mu\nu}(-q_1)J_\alpha(-q_2)J_\beta(-q_3)\rg_{FF-CB}
\end{flalign}
The perturbative result implies that the epsilon-transform holds only with respect to $T_{\mu\nu}$, and not with respect to the vector currents. These results further corroborate that the parity-odd structure can be represented in terms of correlators of the free and critical theories. Equations \eqref{6.2:TJJepsColl} and \eqref{6.2:TJJepsFull} are consistent with the relation observed in \cite{Jain:2021gwa}, obtained by analyzing higher-spin Ward identities of the spin-4 pseudocharge. Due to this property, it was argued in \cite{Aharony:2024nqs,Jain:2024bza} that in specific limits, the three-point function $\lg TJJ\rg_{QF}$ only contains components with a specific sign of overall helicity, when written in spinor-helicity variables. \\

The correlator $\lg TJJ\rg_{QF}$, computed in the FF+CS description, contains an additional parity-odd contact term which appears to diverge in the limit $\tl\to\infty$, however it is removed in the critical-bosonic theory due to parity. Given that our regularization scheme breaks conformal invariance, we should emphasize that the contact term in \eqref{6.2:TJJ} is not generally expected to be conformally-invariant. One may try to relate this term to the $\lg J_1 J_1\rg_{QF}$ parity-breaking contact term through the $\lg (\d\cdot T)_\nu J_\alpha J_\beta\rg_{QF}$ translation Ward identity, however its kinematic dependence suggests that it is a distinct contribution. This is consistent with the observations of \cite{Jain:2021vrv}. For general momenta and Lorentz indices, the term in \eqref{6.2:TJJ} could take the form:
\begin{flalign}\label{6.1:TJJctCov}
    \lg T_{\mu\nu}(-q_1)J_\alpha(-q_2)J_\beta(-q_3)\rg_{\text{contact term}} = \frac{\tN\tl}{64}g_{\alpha(\mu}\eps_{\nu)\beta\rho}q_1^\rho + (\alpha\leftrightarrow\beta)
\end{flalign}
This term is similar to the Chern-Simons contact term appearing in the two-point function of the spin-1 current. If we we denote the fields sourcing the stress tensor $T_{\mu\nu}$ and vector current $J_\alpha$ by $\mathcal{A}_2^{\mu\nu}$ and $\mathcal{A}_1^\alpha$, respectively, this contact term corresponds to the appearance of $q_1^\rho \mathcal{A}_2^{\sigma_1\sigma_2}(q_1)\eps_{\rho\sigma_2\sigma_3}\mathcal{A}_{1 \sigma_1}(q_2)\mathcal{A}_1^{\sigma_3}(q_3) + (\sigma_1\leftrightarrow\sigma_3)$ in the effective action of the theory. It would be interesting to further analyze the effects of such terms in the quasifermionic theory. Although the term \eqref{6.1:TJJctCov} diverges in the chiral limits discussed in \cite{Aharony:2024nqs,Jain:2024bza}, we note that one can add an arbitrary contact term with the same $\tN$, $\tl$ dependence, so that the contact terms cancel when written in spinor-helicity variables. Therefore, we argue that the term \eqref{6.1:TJJctCov} does not contribute in the chiral limit.\\

Finally, it is interesting to check the three-dimensional bosonization duality between QB and QF Chern-Simons-matter theories by comparing the correlator \eqref{6.2:TJJ} to the expression for $\lg T_{--}(-q_1)J_+(-q_2)J_+(-q_3)\rg_{QB}$ obtained in \cite{Aharony:2012nh}. The bosonic theory discussed therein has slightly broken higher-spin symmetry described by analogous variables $\tN_{QB}, \tl_{QB}$. The precise relation of these variables to the quasifermionic parameters used in this text was found in \cite{Gur-Ari:2012lgt} to be $\tN_{QB}=\tN$ and $\tl_{QB}=-\tl^{-1}$. Correlators of the QB theory can be decomposed in terms of $\tN_{QB}$ and $\tl_{QB}$ analogously\footnote{The $\tl_{QB}\to 0$ limit corresponds to the free boson (FB) theory, while the $\tl_{QB}\to \infty$ limit corresponds to the critical fermion (CF) theory.} to \eqref{5:3pt}. Expressing $\lg T_{--}J_+J_+\rg_{QB}$ via $\tN$ and $\tl$, we find:
\begin{flalign}\label{6.2:TJJqb}
	\lg T_{--}J_+J_+\rg_{QB} = \frac{\tN}{1+\tl^2}\lg  T_{--}J_+J_+\rg_{CF} - \frac{\tN\tl}{1+\tl^2}\lg  T_{--}J_+J_+\rg_{QB,odd} + \frac{\tN\tl^2}{1+\tl^2}\lg  T_{--}J_+J_+\rg_{FB}
\end{flalign}
Here we denote the odd structure appearing in the QB theory by $\lg TJJ\rg_{QB,odd}$, to emphasize its difference to the odd structure $\lg TJJ\rg_{odd}=\lg TJJ\rg_{QF,odd}$ appearing in the QF theory, in \eqref{6.2:TJJ}. The QB and QF theories are related by a Legendre transformation with respect to the scalar current. At large $N$, this does not affect correlators containing only currents with spin $s\geq 1$. Therefore, we expect the non-contact terms in \eqref{6.2:TJJ} and \eqref{6.2:TJJqb} to coincide. Comparing to the results of \cite{Aharony:2012nh}, we find:
\begin{flalign}\begin{aligned}
\lg T_{--}(-q_1)J_+(-q_2)J_+(-q_3)\rg_{CF}=\lg T_{--}(-q_1)J_+(-q_2)J_+(-q_3)\rg_{FF}-\frac{|q_2|}{32}-\frac{|q_3|}{32}\\

\lg T_{--}(-q_1)J_+(-q_2)J_+(-q_3)\rg_{FB}=\lg T_{--}(-q_1)J_+(-q_2)J_+(-q_3)\rg_{CB}-\frac{|q_2|}{32}-\frac{|q_3|}{32}\\

\lg T_{--}(-q_1)J_+(-q_2)J_+(-q_3)\rg_{QB,odd}=-\lg T_{--}(-q_1)J_+(-q_2)J_+(-q_3)\rg_{QF,odd}
\end{aligned}\end{flalign}
Which is consistent with the bosonization duality. In addition, we found that the QF correlator \eqref{6.2:TJJ} contains a contact term going as $\tl$ while $\lg T_{--}J_+J_+\rg_{QB}$ was found to have contact terms appearing with different odd functions of the quasibosonic 't Hooft coupling $\lambda_{QB}$. The difference in contact term coefficients reflects computations in different descriptions of the QF theory.

\subsection{\texorpdfstring{$\lg T_{--} J_+ J_+ O\rg_{QF}$}{} calculation}\label{Sec:2110}
Higher correlators in theories with weakly broken higher-spin symmetry have been a topic of interesting research. Computations of four-point functions using bootstrap arguments were carried out in \cite{Silva:2021ece,Li:2019twz,Turiaci:2018nua}. Furthermore, computations of scalar $n$-point functions were outlined in \cite{Yacoby:2018yvy}. To all orders in perturbation theory, four-point functions of spinning operators were determined in \cite{Kalloor:2019xjb}. We follow the procedure established therein to compute the four-point function $\lg T_{--}J_+J_+O\rg_{QF}$ in the collinear limit, first giving a brief overview of the computation.

\begin{figure}[htbp]
\centering
\includegraphics[width=\textwidth]{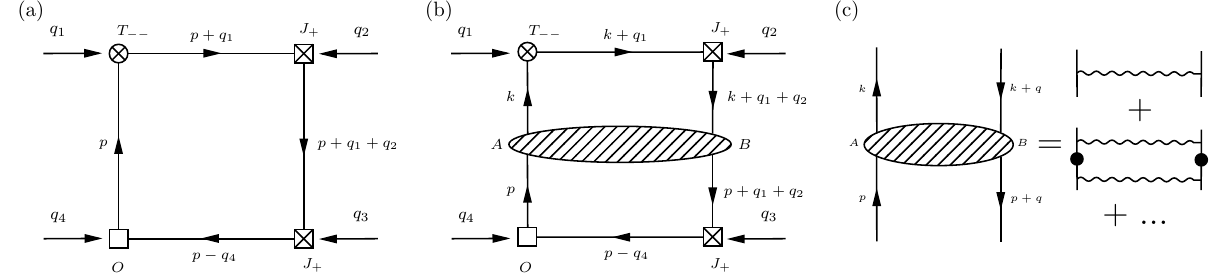}
\caption{Diagrams contributing to $\lg T_{--}(-q_1)J_+(-q_2)J_+(-q_3)O(-q_4)\rg_{QF}$: (a) Box-type diagram (b) Ladder-type diagram. The exact vertices of the spin-{0,1,2} currents are represented by encircled crosses, crossed squares and blank squares respectively. (c) The four-fermion vertex is obtained by summing all ladders made from gluon lines connected by exact fermion propagators.}
\label{fig:2110qf}
\end{figure}

The diagrams contributing at finite $\lambda$ can be split into two types: Box-type diagrams (Figure \ref{fig:2110qf}(a)) and Ladder-type diagrams (Figure \ref{fig:2110qf}(b)). The former are diagrams where non-adjacent fermion propagators are not connected. The latter involve the effective four-fermion vertex, as displayed in Figure \ref{fig:2110qf}(c), linking non-adjacent fermion lines. We write the vertex as $\Gamma^{jn}_{im}(k,q,p)=\Gamma_{AB}(k,q,p)(\gamma_A)^n_m(\gamma_B)^j_i$, for which the full expression can be found in \cite{Kalloor:2019xjb}. At leading order in large $N$, these are the only contributions to the four-point function. Accounting for all permutations of current insertions, there are 6 different Box diagrams and 12 different Ladder diagrams.

We denote the vertex functions as in \eqref{6:OJvertfunc} and \eqref{6.1:svertex}, but suppress the loop momentum variables. In the collinear limit, momentum conservation implies $q_1+q_2+q_3+q_4=0$. The box-type diagram in Figure \ref{fig:2110qf}(a) evaluates to:
\begin{flalign}\begin{aligned}\label{6.3:box}
	\text{Box}(p) = -N\int \frac{d^3 p}{(2\pi)^3}Tr\bigg[& 
 
        V^{++}(q_1)S(p)V_0(-q_4)S(p+q_1+q_2+q_3)\\&
        \cdot V^-(q_3)S(p+q_1+q_2)V^-(q_2)S(p+q_1)\bigg]
\end{aligned}\end{flalign}
And similarly, the ladder-type diagram in Figure \ref{fig:2110qf}(b) is expressed as:
\begin{flalign}\begin{aligned}
	\text{Ladder}(p,k) = -N\int \frac{d^3 p \; d^3 k}{(2\pi)^6}Tr\bigg[&S(k+q_1+q_2)V^-(q_2)S(k+q_1)V^{++}(q_1)S(k)\g^A \\
	
	&\cdot S(p)V_0(q_4)S(p-q_4)V^-(q_3)S(p-q_3-q_4)\g^B\bigg] \Gamma_{AB}(k,q_1+q_2,p)
\end{aligned}\end{flalign}
The integrals over loop momenta are now performed in specific order, to simplify the intermediate steps as much as possible. We first perform integrations along the 3-axis, by integrating over $dp_3$ and $dk_3$. The angular integrals $d\theta_p$, and $d\theta_k$ are performed by computing residues in the complex plane. The planar integral over $dk_s$ can now be performed without cutoffs in the $k^{+-}$ plane.

In the final step we sum up all the integrands, which leverages any exchange symmetries present in the correlator. Beforehand, one can check that the separate integrals do not diverge by themselves. Schematically, we write the final step as: 
\begin{flalign}
	\lg T_{--}(-q_1)J_+(-q_2)J_+(-q_3)O(q_1+q_2+q_3)\rg_{QF} = -N\int_0^\Lambda dp_s \sum_{
	\text{\renewcommand{\arraystretch}{0.5}
\begin{tabular}[t]{@{}c@{}} 
	Insertion \\ 
	permutations 
\end{tabular}}	
	} \bigg(\text{Box}(p_s) + \text{Ladder}(p_s)\bigg)
\end{flalign}
By sending $\Lambda\to\infty$, we recover the correlator in terms of $N$ and $\lambda$. When rewriting the expression in terms of higher-spin symmetry variables $\tN$ and $\tl$ it is important to note that the scalar vertex \eqref{6:spin0vertex} computed in \cite{Gur-Ari:2012lgt} is normalized so that the scalar two point function scales as $\lg OO\rg_{QF}=\tN(1+\tl^2)\lg OO\rg_{FF}$. In our normalization, the final result can be expressed as (suppresing the momentum arguments for brevity):
\setlength{\jot}{5pt}
\begin{flalign}\begin{aligned}\label{6.3:TJJO}
	\lg T_{--}(-q_1)J_+(-q_2)J_+(-q_3)O(-q_4)\rg_{QF} =&
	\frac{\tN}{\sqrt{1+\tl^2}^3}\lg T_{--}J_+J_+O\rg_{FF}  
	+ \frac{\tN\tl}{\sqrt{1+\tl^2}^3}\lg T_{--}J_+J_+O\rg_{X_1} \\& 
	+ \frac{\tN\tl^2}{\sqrt{1+\tl^2}^3}\lg T_{--}J_+J_+O\rg_{X_2} 
	+\frac{\tN\tl^3}{\sqrt{1+\tl^2}^3}\lg T_{--}J_+J_+O\rg_{CB}
\end{aligned}\end{flalign}
\setlength{\jot}{0pt}

The full expression for general values of collinear $q_{1,2,3}$ is given in Appendix \ref{App1}. As expected, the result is $q_2 \leftrightarrow q_3$ symmetric. In addition to free/critical theory results, we find two additional structures. For general signs of momenta, we do not find any relations between the four structures of $\lg T_{--}J_+J_+ O\rg_{QF}$ and they appear to be distinct, which is inconsistent with the form of $\lg J_{s_1}J_{s_2}J_{s_3}O\rg_{QF}$ conjectured in \cite{Jain:2022ajd}. It was pointed out\footnote{We thank Sachin Jain and Dhruva Sathyanarayanan for this remark.} that, when the momenta of the vector currents are of the same sign, and opposite the sign of the scalar and stress tensor, the four structures of $\lg T_{--}J_+J_+ O\rg_{QF}$ reduce to two structures, up to contact terms. It would be interesting to understand if this property holds more generally, beyond the collinear limit. 

\subsection{\texorpdfstring{$\lg J_{---} J_+ J_+ J_+\rg_{QF}$}{} calculation}\label{Sec:3111}
The computation of $\lg J_{---} J_+ J_+ J_+\rg_{QF}$ closely follows the steps of the previous subsection. The contributing diagrams are again displayed in Figure \ref{fig:2110qf}, however the scalar vertex is replaced by the spin-3 vertex using \eqref{6.1:svertex} and the stress tensor vertex is replaced by a third vector vertex \eqref{6:OJvertfunc}:
\setlength{\jot}{5pt}
\begin{flalign}\begin{aligned}
	&\text{Box}(p) = -N\int \frac{d^3 p}{(2\pi)^3}Tr\bigg[V^{-}(q_1)S(p)V^{+++}(-q_4)S(p+q_1+q_2+q_3) \\&
 
        \qquad\qquad\qquad\qquad\qquad\qquad \cdot V^-(q_3)S(p+q_1+q_2)V^-(q_2)S(p+q_1)\bigg]\\
	
	&\text{Ladder}(p,k) = -N\int \frac{d^3 p \; d^3 k}{(2\pi)^6}Tr\bigg[S(k+q_1+q_2)V^-(q_2)S(k+q_1)V^{-}(q_1)S(k)\g^A \\
	
	&\qquad\qquad\qquad\qquad\qquad
	\cdot S(p)V^{+++}(q_4)S(p-q_4)V^-(q_3)S(p-q_3-q_4)\g^B\bigg] \Gamma_{AB}(k,q_1+q_2,p)
\end{aligned}\end{flalign}
\setlength{\jot}{0pt}
And similarly for other diagrams which involve permutations of vertices. An additional subtlety arises as extra factors of $p^\pm$ in the denominator cause the integral to grow linearly with the cutoff $\Lambda$ in the $p^{+-}$ plane. Similarly to \cite{Kalloor:2019xjb}, one can check numerically that for general values of $q_i, \Lambda$ the separate integrals converge, hence we may sum the integrands before performing the final $dp_s$ integration. We find:
\begin{flalign}\label{6.3a:3111form}
	\lg J_{---}(q_1+q_2+q_3)J_+(-q_1)J_+(-q_2)J_+(-q_3)\rg_{QF} \propto 
	
	f\bigg(e^{2\pi i\lambda\arctan(\frac{2\Lambda}{P})} \bigg)+\frac{i\Lambda}{2\pi}
\end{flalign}
Where $P$ is a shorthand for different linear combinations of $q_1, q_2, q_3$. We may now take the $\Lambda\to\infty$ limit. Upon doing this, the first term on the right hand side of \eqref{6.3a:3111form} converges, while the linear divergence can be regularized by introducing a local counterterm. The final result can be written in terms of $\tN, \tl$ as ($q_4=-q_1-q_2-q_3$; again suppressing momentum arguments):
\begin{flalign}\begin{aligned}\label{6.3a:3111}
	&\lg J_{---}(-q_4)J_+(-q_1)J_+(-q_2)J_+(-q_3)\rg_{QF} =\\[5pt]& 
	
	\frac{\tN}{(1+\tl^2)^2}\lg J_{---}J_+J_+J_+\rg_{FF} 
        + \frac{\tN\tl}{(1+\tl^2)^2}\lg J_{---}J_+J_+J_+\rg_{X_1}      
	+ \frac{\tN\tl^2}{(1+\tl^2)^2}\lg J_{---}J_+J_+J_+\rg_{X_2} \\[5pt]&
 
	+ \frac{\tN\tl^3}{(1+\tl^2)^2}\lg J_{---}J_+J_+J_+\rg_{X_3} 
	+ \frac{\tN\tl^4}{(1+\tl^2)^2}\lg J_{---}J_+J_+J_+\rg_{CB}  
	+ \frac{\tN\tl^5}{(1+\tl^2)^2}\bigg(\frac{q_4}{6}\bigg)
\end{aligned}\end{flalign}
In addition to $FF$ and $CB$ structures, we find three additional pieces $X_{1,2,3}$ which vanish in both the free and critical limits. The full expression is given in Appendix \ref{App1a}. The $\tl$ dependence of non-contact terms in \eqref{6.3a:3111} nicely matches with the higher-spin Ward identity \eqref{5.2:3111nc}. The last term on the right hand side of \eqref{6.3a:3111} naively diverges as $\tl\to\infty$, however this should be understood as a contact term. Outside the collinear limit, we may guess the form of this term\footnote{Similarly to the $\lg TJJ\rg_{QF}$ case, we do not necessarily expect this term to be conformally invariant, as the regularization scheme we employ breaks conformal invariance.}, consistent with Lorentz covariance to be:
\begin{flalign}\label{6.3a:3111ct}
\lg J_{\mu\nu\rho}(-q_4)J_{\alpha}(-q_1)J_{\beta}(-q_2)J_{\gamma}(-q_3)\rg_{\text{contact}}\propto \frac{\tN\tl^5}{(1+\tl^2)^2}g_{\alpha\mu}g_{\beta\nu}\eps_{\sigma\gamma\rho}q_4^{\sigma} + \text{ index permutations} 
\end{flalign}
It would be natural to interpret this contact term as stemming from a slightly broken higher-spin Ward identity, and indeed there are various possibilities on how such a term may arise. For example, one may consider the action of the spin-3 pseudocharge on $\lg J_3 J_1 J_1 O\rg_{QF}$ and similar Ward identities involving currents of spin $s>3$. We will leave a comprehensive study of these four-point function contact terms to future work.

Here, we note that the $\tl$ dependence of the contact term \eqref{6.3a:3111ct}, obtained through perturbation theory, matches the expectation \eqref{5.3:QF3111ct} from equating local terms of the $\lg (\d\cdot J_3)J_1 J_1 J_1\rg_{QF}$ Ward identity, in the FF+CS description, i.e. when $g_1(\tl)\propto \tl$. Although this is not surprising, it is not precisely clear if the contact term appearing in \eqref{6.3a:3111} is apriori the same as the contact term in \eqref{5.3:QF3111ct}, since $\lg J_{---}J_+J_+J_+\rg_{QF}$ drops out of the Ward identity for collinear momenta along the 3-axis. It would be interesting to explore this further by computing the same correlator in the CB+CS picture, and ultimately checking if the computed terms come from Chern-Simons background terms.\\

The end result is written in Appendix \ref{App1a} in terms of $p_i=-q_i$. It satisfies the expected exchange symmetry $p_1\leftrightarrow p_2 \leftrightarrow p_3$. Interestingly, the five non-local structures also satisfy\footnote{We use the shorthand $\lg...\rg_{A\pm B}=\lg...\rg_{A}\pm\lg...\rg_{B}$}:
\begin{flalign}\label{6.3:3111rel}
\lg J_{---}(p_4)J_+(p_1)J_+(p_2)J_+(p_3)\rg_{FF+CB-X_2}=i \sign(p_4)\bigg(\lg J_{---}(p_4)J_+(p_1)J_+(p_2)J_+(p_3)\rg_{X_1-X_3}+\frac{p_4}{6}\bigg)
\end{flalign}
This relation appears as an epsilon-transform with respect to the spin-3 current in the collinear limit. In addition to the Ward identity analysis of $\lg (\d\cdot J_3)J_1 J_1 J_1\rg_{QF}$ (see discussion after \eqref{5.2:3111nc}), this strongly supports the claim that the general four-point function $\lg J_{\mu\nu\rho} J_\alpha J_\beta J_\gamma\rg_{QF}$ satisfies the following epsilon-transform relation up to contact terms:
\begin{flalign}
    \lg J_{\mu\nu\rho}(p)J_\alpha(p_1) J_\beta(p_2) J_\gamma(p_3)\rg_{FF+CB-X_2} = \lg (\eps\cdot J)_{\mu\nu\rho}(p)J_\alpha(p_1) J_\beta(p_2) J_\gamma(p_3)\rg_{X_1-X_3} + \text{contact terms}
\end{flalign}

Ultimately, a full check on this identity would require a direct calculation of the four-point function for general momenta and Lorentz indices, or possibly a different approach altogether, which we leave as a topic of future work.

\subsection{Checks of three- and four-point function Ward identities}

The perturbative results of the previous sections can be used to explicitly check higher-spin Ward identities. Such non-trivial checks provide further verification of the higher-spin algebra. However, in the collinear limit, identifying contact terms is not always obvious, which makes the equations difficult to track. Furthermore, for certain checks we require correlators with specific Lorentz indices, which can be difficult to compute in the full Chern-Simons-matter theory. We thus confine ourselves to verifying relations between correlators of free theories. When borrowing results from \cite{Gur-Ari:2012lgt, Aharony:2012nh, Kalloor:2019xjb}, we remind the reader that our FF (FB) correlators are normalized as $\frac{1}{2}$ those of a theory of a single Dirac fermion (complex boson)\footnote{In particular, the results for $\lg J_+J_-OO\rg$ and $\lg J_+J_-J_+J_-\rg$ in \cite{Kalloor:2019xjb} are corrected by a factor of $-\frac{1}{2}$, while the result for $\lg OOOO\rg$ is corrected by a minus sign. This can be checked by first taking the $\tl\to 0$ limit and then performing the calculation}. This corresponds to our $\tN\to 1, \tl\to 0$ and $\tN\to 1, \tl\to\infty$ limits, with the latter involving additional Legendre transformation discussed in Section \ref{Sec4}, to reach the critical fixed point. \\

$\underline{\textbf{Ward identities of three-point functions}}$
\vspace{3pt}

We will first discuss the $\lg [Q_{\mu\nu},J_\alpha OO]\rg_{FF,FB}$ Ward identities using the results of Appendix \ref{App2}. The FF Ward identity reads:
\begin{flalign}\begin{aligned}\label{6.4:JOOffwi}
&-a_1\eps_{\sigma\alpha(\mu}p_{1\nu)}p_1^\sigma\lg O(p_1)O(p_2)O(p_3)\rg_{FF}+
i a_2p_{1\mu}\lg T_{\nu)\alpha}(p_1)O(p_2)O(p_3)\rg_{FF}\\&

-a_0 \eps_{\rho\sigma(\mu}\bigg(p_{2\nu)}p_2^\rho \lg J_\alpha(p_1)J^\sigma(p_2)O(p_3)\rg_{FF} +\begin{Bmatrix}2\leftrightarrow 3\end{Bmatrix}\bigg) = 0
\end{aligned}\end{flalign}
The equation holds up to contact terms. The correlator $\lg OOO\rg_{FF}$, vanishes, as the scalar current is parity odd. Using \eqref{B:OJJff} and \eqref{B:TOOff}, we find that the right-hand side doesn't contain contact terms if and only if $a_2=2ia_0$, which is consistent with the relations found in Section \ref{Sec2.2}. After using this relation, we find that the left-hand side of \eqref{6.4:JOOffwi} becomes a sum of contact terms:
\setlength{\jot}{5pt}
\begin{flalign}
&\lg[Q_{\mu\nu},J_\alpha(p_1)O(p_2)O(p_3)]\rg_{FF}\vert_{\text{contact terms}} = \frac{a_0}{16}p_{1\alpha} p_{1\mu} p_{1\nu} + \bigg(\frac{1}{48}p_{2(\mu}\delta_{\nu)\alpha}(3p_1+2p_2-4p_3)+\begin{Bmatrix}2\leftrightarrow 3\end{Bmatrix}\bigg)
\end{flalign}
\setlength{\jot}{0pt}

Analogously, we check the $\lg [Q_{\mu\nu},J_\alpha(p_1)O(p_2)O(p_3)]\rg_{FB}$ Ward identity. This serves to confirm the FB algebra discussed in Section \ref{Sec2.3}, but also the associated algebra of the critical boson theory. The Ward identity was stated in \eqref{3.2:JOOwi}.

The expression for $\lg OOO\rg_{FB}$ can be obtained from a 1-loop calculation, and in our normalization reads $\lg O(p_1) O(p_2) O(p_3) \rg_{FB}=64(p_1p_2p_3)^{-1}$. The remaining correlators were computed similarly and are given by \eqref{B:OJJfb} and \eqref{B:TOOfb}. Inserting these results into \eqref{3.2:JOOwi}, we find that the non-contact terms on the left-hand side cancel if and only if the coefficients $b_{0,1,2,3}$ obey $b_1=-\frac{b_0}{128}$, $b_2=\frac{b_0}{128}$, $b_3=\frac{b_0}{4}$. This is consistent with the results of Section \ref{Sec2.3}. Given these constraints, the left-hand side simplifies to a sum of contact terms:
\setlength{\jot}{5pt}
\begin{flalign}\begin{aligned}
&\lg[Q_{\mu\nu},J_\alpha(p_1)O(p_2)O(p_3)]\rg_{FB}\vert_{\text{contact terms}} = -\frac{2i b_0 p_{1(\mu } \delta_{\nu)\alpha}}{3}\bigg(\frac{1}{p_2}+\frac{1}{p_3}\bigg)
\end{aligned}\end{flalign}
\setlength{\jot}{0pt}

We discuss the remaining Ward identities in the collinear limit, with all momenta pointing along the $3$-axis. 

The $\lg[Q_{\mu\nu},J_\alpha J_\beta J_\gamma]\rg_{FF}$ Ward identity can be written as:
\begin{flalign}\begin{aligned}
	&-a_1\eps_{\sigma\alpha(\mu}p_{1\nu)}p_1^\sigma\lg O(p_1)J_\beta(p_2)J_\gamma(p_3)\rg_{FF} 
        + ia_2p_{1(\mu}\lg T_{\nu)\alpha}(p_1)J_\beta(p_2) J_\gamma(p_3)\rg_{FF} \\[5pt]&

        + \perm{1\leftrightarrow 2}{\alpha\leftrightarrow \beta} 
        + \perm{1\leftrightarrow 3}{\alpha\leftrightarrow \gamma}=0 
\end{aligned}\end{flalign}
Choosing $p_i^\mu=p_i\delta_3^\mu$, and $\mu=3$, $\alpha=\beta=+$, $\gamma=\nu=-$, the equation becomes\footnote{Note that the coordinate system is right-handed, i.e. $\eps_{+-3}=i$}:
\begin{flalign}\begin{aligned}\label{6.4:JJJwi}
	&-\frac{ia_1}{2}p_1^2\lg O(p_1)J_+(p_2)J_-(p_3)\rg_{FF} -\frac{ia_1}{2}p_2^2\lg J_+(p_1)O(p_2)J_-(p_3)\rg_{FF} + \frac{ia_2}{2}p_3\lg J_+(p_1)J_+(p_2)T_{--}(p_3)\rg_{FF} \\&
	+\frac{ia_2}{2}p_1\lg T_{+-}(p_1)J_+(p_2)J_-(p_3)\rg_{FF} +\frac{ia_2}{2}p_1\lg T_{+-}(p_1)J_+(p_2)J_-(p_3)\rg_{FF} =0 
\end{aligned}\end{flalign}
Momentum conservation fixes $p_3=-p_1-p_2$. The correlator $\lg T_{+-}J_+J_-\rg$ is not determined by the preceeding analysis, however, one can find it analogous to \eqref{6.2:TJJint}, using the free fermion vertices $V^{+-}(q,p)=i p^{(+}\g^{-)}$ and $V^\pm(q,p)=i\g^\pm$ and taking the limit $\tl\to 0$. The final result is:
\begin{flalign}\label{6.4:TpmJJ}
	\lg T_{+-}(-p_2-p_3)J_+(p_2)J_-(p_3)\rg_{FF}=-\frac{p_2^2\sign(p_2)+p_3^2\sign(p_3)}{64 (p_2+p_3)}
\end{flalign}
Using \eqref{6.2:TJJ}, \eqref{6.4:TpmJJ} as well as the results of \cite{Gur-Ari:2012lgt}, the Ward identity \eqref{6.4:JJJwi} can be checked in the collinear limit. We find:
\setlength{\jot}{5pt}
\begin{flalign}\begin{aligned}\label{6.4:JJJwi2}
	
	&(4a_1-ia_2)\bigg( -\frac{p_1^4\sign(p_3)}{64p_2p_3}-\frac{p_1^4\sign(p_1)}{64p_2p_3} -\frac{p_1^3 \sign(p_1)}{32p_3} + \begin{Bmatrix}1\leftrightarrow 2\end{Bmatrix}\bigg) \\&
	
	+\frac{p_1p_2}{64p_3}\bigg(\sign(p_1)(4a_1p_2+ia_2p_1) + \begin{Bmatrix}1\leftrightarrow 2\end{Bmatrix}\bigg) \\&
	
	-\frac{1}{128}\sign(p_3)\bigg(p_3^2 (4a_1+ia_2)-5(4a_1-ia_2) \left(p_1^2+p_2^2\right)\bigg) = 0
\end{aligned}\end{flalign}
\setlength{\jot}{0pt}

The equation should hold up to contact terms, which can be removed by a suitable regularization scheme. It is clear that the only choice that makes all the non-contact terms on the right-hand side vanish is $a_2=-4ia_1$. This is consistent with the constraints obtained in Section \ref{Sec2}. After inserting this relation, equation \eqref{6.4:JJJwi2} simplifies, and we are left with:
\begin{flalign}
\lg[Q_{3-},J_+ J_+ J_-]\rg_{FF}\vert_{\text{contact terms}} = -\frac{a_1}{16}p_1\sign(p_1)p_2 -\frac{a_1}{16}p_1p_2\sign(p_2) + \frac{a_1}{16} p_3^2 \sign(p_3)
\end{flalign}
We argue that these terms are contact terms, as they all are polynomials in two of the three momenta. Therefore, we find that the correlators satisfy the higher-spin Ward identity, with the expected algebra coefficients. \\

$\underline{\textbf{Ward identities of four-point functions}}$
\vspace{3pt}

Next, we verify the $\lg[Q_3,J_1OOO]\rg_{FF}$ and $\lg[Q_3,J_1J_1J_1O]\rg_{FF}$ Ward identities in the collinear limit. We will make use of the results for $\lg OOOO\rg_{FF}$, $\lg J_+J_-OO\rg_{FF}$ and $\lg J_+J_+J_-J_-\rg_{FF}$ found in \cite{Kalloor:2019xjb} and our result of $\lg T_{--}J_+J_+O\rg_{FF}$ as in \eqref{TJJOfull}. Additionally, we will need correlators involving the $T_{+-}$ vertex. We only need the FF expressions, which can be obtained by combining free theory vertex $V^{+-}(q,p)=i p^{(+}\g^{-)}$ with expressions \eqref{6:FermProp}, \eqref{6:spin0vertex}, \eqref{6:spin1vertex} in the $\tl\to 0$ limit. Each correlator is obtained by summing 6 box diagrams, given by \eqref{6.3:box}. We find that $\lg T_{+-}(p_1)O(p_2)O(p_3)O(p_4)\rg_{FF}$ vanishes, while:
\setlength{\jot}{5pt}
\begin{flalign}\begin{aligned}\label{6.4:TpmJJO}
	\lg T_{+-}(p_1)J_+(p_2)J_-(p_3)O(-p_1-p_2-p_3)\rg_{FF}= \frac{i}{32p_1p_2p_3}
	
	\bigg(\frac{p_2(p_1+p_2)^3}{|p_1+p_2|} - \frac{p_3(p_1+p_3)^3}{|p_1+p_3|} 
	- \frac{p_2^4}{|p_2|} + \frac{p_3^4}{|p_3|}\\
	
	\qquad\qquad\qquad
	- \frac{(p_2-p_3)\left(|p_2+p_3|(p_1+p_2+p_3)^3-(p_2+p_3)^3 | p_1+p_2+p_3| \right)}{|p_2+p_3||p_1+p_2+p_3|} \bigg)
\end{aligned}\end{flalign}
\setlength{\jot}{0pt}

As a consistency check on this result, we can confirm that it aligns with the translation Ward identity:
\setlength{\jot}{5pt}
\begin{flalign}\begin{aligned}
&\lg (\d\cdot T)_\nu(p_1) J_\alpha(p_2)J_\beta(p_3)O(p_4)\rg_{FF}=
\delta(x_1-x) \lg \d_\nu J_{\alpha}(x)J_\beta(x_2)O(x_3)\rg_{FF} \\&
+ \delta(x_2-x) \lg J_{\alpha}(x_1)\d_\nu J_\beta(x)O(x_3)\rg_{FF} 
+ \delta(x_3-x) \lg J_{\alpha}(x_1)J_\beta(x_2)\d_\nu O(x)\rg_{FF}
\end{aligned}\end{flalign}
\setlength{\jot}{0pt}

Tracelessness of the stress-tensor implies $T_{33}+2T_{+-}=0$. As all momenta point along the 3-axis, we have:
\setlength{\jot}{5pt}
\begin{flalign}\begin{aligned}
ip_1\lg T_{+-}(p_1)J_+(p_2)J_-(p_3)O(p_4)\rg_{FF}=&-\frac{1}{2}\bigg(
i(p_1+p_2)\lg J_+(p_1+p_2)J_-(p_3)O(p_4)\rg_{FF}\\&\qquad\quad
+i(p_1+p_3)\lg J_+(p_1)J_-(p_1+p_3)O(p_4)\rg_{FF}\\&\qquad\quad
-i(p_2+p_3)\lg J_+(p_2)J_-(p_3)O(p_1+p_4)\rg_{FF}\bigg)
\end{aligned}\end{flalign}
\setlength{\jot}{0pt}

Where $p_4=-p_1-p_2-p_3$. Borrowing the result for $\lg J_+J_-O\rg_{FF}$ from \cite{Gur-Ari:2012lgt}, we verify that this relation indeed holds. \\

We first check the $\lg[Q_{\mu\nu},J_\alpha OOO]\rg_{FF}$ higher-spin Ward identity. The equation reads:
\begin{flalign}\begin{aligned}
	
	&-a_1\eps_{\sigma\alpha(\mu}p_{1\nu)}p_1^\sigma \lg O(p_1)O(p_2)O(p_3)O(p_4)\rg_{FF} + ic_2p_{1(\mu}\lg T_{\nu)\alpha}(p_1)O(p_2)O(p_3)O(p_4)\rg_{FF} \\
	
	&-a_0\bigg(\eps_{\sigma\rho(\mu}p_{2\nu)}p_2^\sigma \lg J_\alpha(p_1)J_\rho(p_2)O(p_3)O(p_4)\rg_{FF} + \{2 \leftrightarrow 3\} + \{2 \leftrightarrow 4\} \bigg) = 0
\end{aligned}\end{flalign}
In the collinear limit, with $\mu=3$, $\nu=-$, $\alpha=+$, we find:
 \begin{flalign}\begin{aligned}\label{6.4:JOOOwi}
 
	&-\frac{ia_1}{2}p_1^2\lg O(p_1)O(p_2)O(p_3)O(p_4)\rg_{FF}	
	+\cancel{\frac{ia_2}{2}p_1\lg T_{+-}(p_1)O(p_2)O(p_3)O(p_4)\rg_{FF}}\\&
	-\frac{ia_0}{2}\bigg(p_2^2\lg J_+(p_1)J_-(p_2)O(p_3)O(p_4)\rg_{FF} + \{2 \leftrightarrow 3\} + \{2 \leftrightarrow 4\}\bigg) = 0
	
\end{aligned}\end{flalign}

Momentum conservation implies $p_4=-p_1-p_2-p_3$, therefore we will discuss the equation in terms of $p_{1,2,3}$. The Ward identities can be discussed for different values of momenta. A way to check all cases is to write out \eqref{6.4:JOOOwi} for all possible values of $\{\sign(p_1),\sign(p_2),\sign(p_3),\sign(p_1+p_2),\sign(p_1+p_3),\sign(p_2+p_3),\sign(p_1+p_2+p_3)\}$. Excluding contradictory assumptions, there are 32 possible regimes. We will analyze the equation for $p_1>0, p_2>0, p_3>0$ and leave to the reader to check that the same conclusions follow in all cases. The terms of \eqref{6.4:JOOOwi} become:
 \begin{flalign}\begin{aligned}\label{6.4:JOOOterms}
 
	&p_1^2\lg O(p_1)O(p_2)O(p_3)O(p_4)\rg_{FF} = \frac{p_1^2(p_1p_2+p_2p_3+p_3p_1)}{2(p_1+p_2)(p_2+p_3)(p_3+p_1)}\\
		
	&p_2^2\lg J_+(p_1)J_-(p_2)O(p_3)O(p_4)\rg_{FF} =\frac{p_2^2}{4p_4}+\frac{p_1p_2^3}{4(p_1+p_2)(p_2+p_3)(p_3+p_1)}
	
\end{aligned}\end{flalign}
Inserting $a_0=-2a_1$ and \eqref{6.4:JOOOterms} into \eqref{6.4:JOOOwi}, we find that the equation is satisfied up to contact terms. Furthermore, it is easy to verify that any other choice of constants $a_{0,1}$ results in terms that are not polynomials in three of the four momenta, and thus cannot be interpreted as contact terms. Therefore, $a_0=-2a_1$ is the unique constraint obtained from the Ward identity and is consistent with the FF algebra \eqref{2.2:QJ}. The remaining terms in \eqref{6.4:JOOOwi} after enforcing the constraint are:
 \begin{flalign}\begin{aligned}\label{6.4:JOOOct}
 	\lg[Q_{3-},J_+(p_1)O(p_2)O(p_3)O(p_4)]\rg_{FF}\vert_{\text{contact terms}} 
 	= \frac{i a_1p_1^2}{4p_4}- \frac{ia_1 p_2p_3}{2p_4}
\end{aligned}\end{flalign}
These would need to be absorbed into some of the correlators involved, for the Ward identity to be consistent up to contact terms. Different expressions for \eqref{6.4:JOOOct} are obtained in different kinematic regimes, however we argue that these should come from contact terms of the covariant equation, since all the terms appearing are polynomials in at least two of the four momenta. It would be interesting to reconstruct these contact terms, by comparing the results with general momentum expressions.\\

Next, we check the Ward identity $\lg[Q_{\mu\nu},J_\alpha J_\beta J_\gamma O]\rg_{FF}$. The equation reads:
\begin{flalign}\begin{aligned}
&\bigg(-a_1\eps_{\sigma\alpha(\mu}p_{1\nu)}p_1^\sigma\lg O(p_1)J_\beta(p_2)J_\gamma(p_3)O(p_4)\rg_{FF} 
+ ia_2p_{1(\mu}\lg T_{\nu)\alpha}(p_1)J_\beta(p_2) J_\gamma(p_3)O(p_4)\rg_{FF}\\ 

&+ \perm{1\leftrightarrow 2}{\alpha\leftrightarrow \beta} + \perm{1\leftrightarrow 3}{\alpha\leftrightarrow \gamma}\bigg)
 - a_0\eps_{\sigma\rho(\mu}p_{4\nu)}p_4^\sigma\lg J_\alpha(p_1)J_\beta(p_2)J_\gamma(p_3)J^\rho(p_4)\rg_{FF} = 0
\end{aligned}\end{flalign}
To test this relation with correlators at our disposal, we choose $\mu=3$, $\alpha=\beta=+$, $\gamma=\nu=-$ and work in the collinear limit. The equation then simplifies to:
\begin{flalign}\begin{aligned}\label{6.4:JJJOwi}
	&-\frac{ia_1}{2}\bigg(p_1^2\lg O(p_1)J_+(p_2)J_-(p_3)O(p_4)\rg_{FF}+p_2^2\lg J_+(p_1)O(p_2)J_-(p_3)O(p_4)\rg_{FF}\bigg) \\&
	
	+\frac{ia_2}{2}p_1\lg T_{+-}(p_1)J_+(p_2)J_-(p_3)O(p_4)\rg_{FF}+\frac{ia_2}{2}p_2\lg J_+(p_1)T_{+-}(p_2)J_-(p_3)O(p_4)\rg_{FF}\\&
	
	+\frac{ia_2}{2}p_3\lg J_+(p_1)J_+(p_2)T_{--}(p_3)O(p_4)\rg_{FF}-\frac{ia_0}{2}p_4^2\lg J_+(p_1)J_+(p_2)J_-(p_3)J_-(p_4)\rg_{FF} =0
\end{aligned}\end{flalign}
Similarly to $\lg[Q_3,J_1OOO]\rg_{FF}$, we can analyze the \eqref{6.4:JJJOwi} for different values of $p_{1,2,3}$. We confine ourselves to analyzing the $p_1>0$, $p_2>0$, $p_3>0$. It's straightforward to check that the conclusions carry over to the more general case. The terms of the Ward identity become:
\begin{flalign}\begin{aligned}
	p_1^2\lg O(p_1)J_+(p_2)J_-(p_3)O(p_4)\rg_{FF} =& 
	\frac{p_1^2}{4p_4}+\frac{p_1^2p_2p_3}{4(p_1+p_2)(p_1+p_3)(p_2+p_3)}\\
	
	p_3\lg J_+(p_1)J_+(p_2)T_{--}(p_3)O(p_4)\rg_{FF} =&
	-\frac{ip_1p_2p_3^2}{16(p_1+p_2)(p_1+p_3)(p_2+p_3)}\\

	p_4^2\lg J_+(p_1)J_+(p_2)J_-(p_3)J_-(p_4)\rg_{FF} =& 
	\frac{p_1p_2p_3(p_1+p_2+p_3)}{8(p_1+p_2)(p_1+p_3)(p_2+p_3)}\\
	
	p_1\lg T_{+-}(p_1)J_+(p_2)J_-(p_3)O(p_4)\rg_{FF} =& 0
\end{aligned}\end{flalign}
The non-contact terms of the Ward identity \eqref{6.4:JJJOwi} then amount to:
\begin{flalign}
-\frac{p_1p_2p_3(2ia_0(p_1+p_2+p_3)+4ia_1(p_1+p_2)-a_2p_3)}{32(p_1+p_2)(p_1+p_3)(p_2+p_3)}= 0
\end{flalign}
Which is uniquely satisfied for $a_2=2ia_0$ and $a_0=-2a_1$. These constraints are consistent with those of Sections \ref{Sec2}. The remaining terms are all polynomials in three of the four momenta:
\begin{flalign}\begin{aligned}\label{6.4:JJJOwi-ct}
	\lg[Q_{3-},J_+J_+J_-O]\rg_{FF}\vert_{\text{contact terms}}=-\frac{ia_1}{8}\bigg(\frac{p_1^2}{p_4}+\frac{p_2^2}{p_4}\bigg)
\end{aligned}\end{flalign}
We argue that these terms originate from contact terms of the full covariant expression, and ought to be absorbed into some of the correlators, for the Ward identity to be fully consistent. After imposing the FF algebra constraints, one can verify that \eqref{6.4:JJJOwi} is consistent in all other kinematic regimes, once appropriate contact terms are subtracted from the correlators. Writing these contact terms in closed form just from the collinear limit expressions is tedious, however it is obvious that higher-spin symmetry imposes constraints on contact terms of different four-point functions.
\section{Summary and Discussion}\label{Sec7}
We summarize the main results of previous sections and detail the possibilities for follow-ups of this work:

\subsection{Local higher-spin algebra}
The higher-spin algebra of QF Chern-Simons-matter theory is constrained by analyzing local Ward idenitities of three-point functions, for operators of spin $s<4$. By construction, it reproduces the higher-spin Ward identities of singlet sectors of the free-fermionic $U(N)$ model and of the critical-bosonic $U(N)$ model, in the limits $\tl\to 0$ and $\tl \to \infty$, respectively. The former is verified by making free field contractions (see Appendix \ref{Sec6.1}), while the latter is verified by a resummation of Feynman diagrams at large $N$ (see Section \ref{Sec4}). Under the insertion of the spin-3 current $(\d\cdot J)^{QF}_{\mu\nu}(x)$, the local variations of the scalar and vector current are given by:
\begin{flalign}
    &\delta O^{QF}(x_1) = \frac{1}{\sqrt{1+\tl^2}}\delta O^{FF}(x_1)\\[5pt]&
    \delta J^{QF}_\alpha(x_2) = \frac{1}{\sqrt{1+\tl^2}}\delta_O J^{FF}_\alpha(x_2) + \frac{\delta_T J^{FF}(x_2) + \tl^2\delta_T J^{CB}(x_2)}{1+\tl^2} + \frac{h_3(\tl)}{\tN}\delta(x_2-x)\eps_{\alpha\rho(\mu}J_{\nu)}J^\rho(x)
\end{flalign}
Where $\delta O^{FF}, \delta_O J^{FF}, \delta_T J^{FF}$ and $\delta_T J^{CB}=\delta_T J^{FB}$ stand in for linear combinations of contact terms, defined in \eqref{5:ffvar} and \eqref{4.1:dTJ}, respectively. The divergence of the spin-3 current is given by \eqref{4.2:CBalgebra} in our normalization, and is consistent with \cite{Giombi:2011kc}. $h_3(\tl)$ is an even function of the 't Hooft coupling, which vanishes in the free/critical limits and is unfixed by our analysis. The action of the spin-3 pseudocharge $Q_{\mu\nu}$ on spin-0 and spin-1 currents is now given by:
\begin{flalign}\begin{aligned}\label{6:[Q,J]}
    &[Q_{\mu\nu},O]_{QF} = \int_x \delta O^{QF} = 
    \frac{1}{\sqrt{1+\tl^2}}\frac{8}{3}\eps_{\alpha\beta(\mu}\d_{\nu)}\d^\alpha J^\beta \\[5pt]&
     [Q_{\mu\nu},J_\alpha]_{QF} = \int_x \delta J^{QF} = -\frac{1}{\sqrt{1+\tl^2}}\frac{4}{3}\eps_{\sigma\alpha(\mu}\d_{\nu)}\d^\sigma O 
	+ \frac{16i}{3}\frac{1-\tl^2}{1+\tl^2}\d_{(\mu}T_{\nu)\alpha}
	+\frac{h_3(\tl)}{\tN}\eps_{\alpha\rho(\mu}J_{\nu)}J^\rho
\end{aligned}\end{flalign}

Our result \eqref{6:[Q,J]} is similar to the form of commutators guessed in \cite{Jain:2022ajd} and builds on it by fixing the exact coefficients of each term. Furthermore, as we've shown, the analysis at this level still allows for the existence of a double-trace term in $[Q_3,J]_{QF}$, which was not originally present in \cite{Jain:2022ajd}. \\

Pertaining to the QF theory discussed in this text, the algebra \eqref{5.1:QFalg} can be used to investigate $n$-point functions of currents with spin $s\leq 3$, for $n>4$. Combining this approach with bootstrap techniques \cite{Aharony:2018npf} may lead to stronger constraints on higher correlators of the theory. Further algebraic constraints can also be obtained. One may discuss Ward identities which contain the action of the spin-3 pseudocharge on operators of spin $s>1$. Moreover, it would also be interesting to consider Ward identities of $Q_{s>3}$ pseudocharges. This was already done for integrated spin-4 identities in several instances \cite{Maldacena:2012sf,Jain:2022ajd, Jain:2021gwa}. Indeed, the compelling premise of higher-spin symmetry is that an arbitrary number of constraints on a specific correlator can be obtained. A further understanding of the interplay of different higher-spin Ward identities is in principle limited only by computational difficulty. In the case of free theories, extending the approach of Section \ref{Sec2} for higher-spin currents should be straightforward. The Ward identities may also be used to constrain correlation functions of non-local operators \cite{Gabai:2022vri,Gabai:2022mya,Gabai:2023lax}.\\

Another natural direction is to investigate the constraints of higher-spin symmetry in quasibosonic Chern-Simons-matter theories \cite{Aharony:2012nh,Aharony:2011jz}, as well as generalizations with bosons and fermions gauged by Chern-Simons interactions \cite{Aharony:2012ns, Jain:2013gza}. Such theories are closely related to the QF model discussed here, hence establishing their higher-spin algebras could lead to further checks on correlators, both at separated and coincident points. Furthermore, it would be interesting to see if such current algebra methods can be extended to theories with $\mathcal{N}=2$ supersymmetry \cite{Giveon:2008zn,Benini:2011mf}. It would also be insightful to study higher-spin Ward identities along the full flow between free-bosonic and critical-bosonic vector models, at large $N$. This would require extending the analysis of Sections \ref{Sec2} and \ref{Sec4} to theories where conformal symmetry is broken.

Finally, it would be interesting to examine whether higher-spin symmetry imposes constraints on subleading terms in the large $N$ expansion. This would involve the computation of non-planar corrections to correlation functions, as well as the analysis of Ward identities at order $\mathcal{O}(1/N)$. 

\subsection{Results for quasifermionic four-point functions}
Four-point functions involving a single insertion of $(\d\cdot J_3)^{QF}$ are decomposed into two- and three-point correlators of free/critical theories using higher-spin Ward identities, as described in Section \ref{Sec5.2}. We write the form of these correlators schematically, excluding Chern-Simons contact terms:
\begin{flalign}
    &\lg (\d\cdot J_3) J_1 O O \rg_{QF} = 
    \frac{\tN}{1+\tl^2}\bigg( \lg ...\rg_{FF} + \tl\lg ...\rg_{odd} + \tl^2\lg ...\rg_{CB} \bigg) \label{7.2:Wi4pt3100}\\[5pt]&

    \lg (\d\cdot J_3) J_1 J_1 J_1 \rg_{QF} = 
    \frac{\tN}{(1+\tl^2)^2}\bigg( \lg ...\rg_{FF} + \tl\lg ...\rg_{X_1} + \tl^2\lg ...\rg_{X_2} + \tl^3\lg ...\rg_{X_3} + \tl^4\lg ...\rg_{CB} + h_3(\tl)\lg ...\rg_{X_4} \bigg)   \label{7.2:Wi4pt3111}
\end{flalign}
Four-point functions with specific Lorentz indices were also computed directly in the FF+CS description, in the limit of collinear momenta (see Sections \ref{Sec:2110} and \ref{Sec:3111}). The non-contact terms of these correlators are:
\begin{flalign}
    &\lg T_{--} J_+ J_+ O \rg_{QF} = 
    \frac{\tN}{\sqrt{1+\tl^2}^3}\bigg( \lg ...\rg_{FF} + \tl\lg ...\rg_{X_1} + \tl^2\lg ...\rg_{X_2} + \tl^3\lg ...\rg_{CB} \bigg) \label{7.2:Pert4pt2110}\\[5pt]&

    \lg J_{---} J_+ J_+ J_+ \rg_{QF} = 
    \frac{\tN}{(1+\tl^2)^2}\bigg( \lg ...\rg_{FF} + \tl\lg ...\rg_{X_1} + \tl^2\lg ...\rg_{X_2} + \tl^3\lg ...\rg_{X_3} + \tl^4\lg ...\rg_{CB}\bigg) \label{7.2:Pert4pt3111}

\end{flalign}
With $\lg J_{---} J_+ J_+ J_+ \rg_{QF}$ containing an additional contact term, as observed in \eqref{6.3a:3111}. We highlight some features of these results:
\begin{itemize}
    \item The $\tl$-dependence of $\lg T_{--}J_+J_+O\rg_{QF}$ is inconsistent with the form conjectured in \cite{Jain:2022ajd}, while the $\tl$-dependence of $\lg (\d\cdot J_3) J_1 O O \rg_{QF}$ is consistent with the form conjectured therein.

    \item We do not observe any relations between the four terms of \eqref{7.2:Pert4pt2110}. This suggests that the full correlator $\lg T_{\mu\nu}J_\alpha J_\beta O\rg_{QF}$ contains four independent non-local structures, outside the collinear limit. The Ward identity analysis \eqref{5.2:3100qfwi} suggests that the four-point function $\lg J_{\mu\nu\rho} J_\alpha O O \rg_{QF}$ may satisfy an epsilon-transform relation with respect to the vector current \eqref{4:3100epstr}, consistent with the conjecture of \cite{Jain:2022ajd}. It would be interesting to check these statements directly, which requires either a more complicated perturbative calculation, or an alternate approach.

    \item The form of $\lg J_3 J_1 J_1 J_1\rg_{QF}$ is consistent between the spin-3 Ward identity \eqref{7.2:Wi4pt3111} and direct perturbative result \eqref{7.2:Pert4pt3111}. The Ward identity contains an additional disconnected piece, multiplied by an even function\footnote{$h_3(\tl)$ vanishes as $\tl\to 0,\infty$, so it does not affect the free/critical theory results.} $h_3(\tl)$, which was not fixed by our analysis. It would be interesting to check if this function vanishes, as anticipated in \cite{Jain:2022ajd}. In the collinear limit, the five structures of $\lg J_3 J_1 J_1 J_1\rg_{QF}$ satisfy an epsilon-transform relation \eqref{6.3:3111rel}, and the Ward identity analysis \eqref{5.2:3111nc} suggests that this property may hold for general momenta. We note that the same (and, in fact, slightly stronger) property was observed in \cite{Kalloor:2019xjb}, for the five structures of $\lg J_1 J_1 J_1 J_1\rg_{QF}$, in the collinear limit. This could imply that such a property holds for general correlators of the form $\lg J_s J_1 J_1 J_1\rg_{QF}$, and perhaps for more general four-point functions of spinning currents.
    
\end{itemize}

Although the momentum space relations for the structures of four-point functions \eqref{7.2:Wi4pt3100} and \eqref{7.2:Wi4pt3111} appear intricate, we emphasize that these relations simplify when written in spinor-helicity variables, as is usually the case for correlators of massless fields. A better understanding of the structure of QF four-point functions would allow for direct comparisons to dual higher-spin gravity theories \cite{Skvortsov:2018uru, Skvortsov:2022wzo} and further investigation of the chiral limits discussed in \cite{Aharony:2024nqs,Jain:2024bza}. 

\subsection{Background Chern-Simons contact terms}

Imposing higher-spin symmetry at coincident points implies that higher correlators develop scheme-independent contact terms related to the presence of the background Chern-Simons term in the effective action of the theory, as discussed in Section \ref{Sec5.3}. Interestingly, these contact terms have different dependences on the 't Hooft coupling, depending on whether we are working in the FF+CS description \eqref{FF+CS}, or the CB+CS description \eqref{CB+CS}. By analyzing higher-spin Ward identities, we argue that the following correlators develop scheme-independent contact terms. Schematically:
\begin{flalign}
    &\lg J_{\mu\nu\rho} J_\alpha O\rg_{QF}\vert_{CS\text{ contact terms}} = 
    \frac{\tN g_1(\tl)}{\sqrt{1+\tl^2}}\lg ...\rg_{ct_1} + 
    \frac{\tN g_1(\tl)\tl}{\sqrt{1+\tl^2}}\lg ...\rg_{ct_2} \label{7.3:wi310}\\[5pt]&

    \lg (\d \cdot J)_{\mu\nu} J_\alpha J_\beta J_\gamma\rg_{QF}\vert_{CS\text{ contact terms}} = 
    \frac{\tN g_1(\tl)\tl}{1+\tl^2}\lg ...\rg_{ct_1} + 
    \frac{\tN g_1(\tl)\tl^2}{1+\tl^2}\lg ...\rg_{ct_2} + 
    \tN h_3(\tl)\lg ...\rg_{ct_3}  \label{7.3:wi3111ct}
\end{flalign}

Where $g_1(\tl)=\tl$ in the FF+CS description and $g_1(\tl)=-\tl^{-1}$ in the CB+CS description and $h_3(\tl)$ is again, an unfixed even function. Although some of these terms appear divergent in the $\tl\to 0,\infty$ limits, they are not present in FF and CB theories. We also note that the correlator $\lg (\d\cdot J_3)J_1 OO\rg_{QF}$ contains a higher-spin contact term in the CB+CS description \eqref{5.3:3100ct}, but not in the FF+CS description. Similar statements can be made for correlators of the quasibosonic Chern-Simons-matter theory (see Section \ref{Sec5.3}). 

The perturbative results of Section \ref{Sec6} imply the presence of additional parity-breaking contact terms. We infer the Lorentz covariant form of these terms from collinear limit results\footnote{The underlines and overlines denote index symmetrizations.}:
\setlength{\jot}{5pt}
\begin{flalign}
    &\lg T_{\mu\nu}(-q_1) J_\alpha(-q_2) J_\beta(-q_3) \rg_{QF}\vert_{\text{contact terms}} = \tN \tl \frac{1}{64}g_{\alpha(\mu}\eps_{\nu)\beta\rho} q_1^\rho + (\alpha\leftrightarrow\beta) \label{7:TJJct}\\&

    \lg J_{\mu\nu\rho}(-q_4) J_\alpha(-q_1) J_\beta(-q_2) J_\gamma(-q_3) \rg_{QF}\vert_{\text{contact terms}} = \frac{\tN\tl^5}{(1+\tl^2)^2} g_{\underline{\alpha}\overline{\mu}}g_{\underline{\beta}\overline{\nu}}\eps_{\sigma\underline{\gamma}\overline{\rho}}\frac{q_4^\sigma}{6} \label{7.3:3111pert}
\end{flalign}
\setlength{\jot}{0pt}

We emphasize that this computation is done in the FF+CS description. These terms are similar to the $U(1)$ Chern-Simons term appearing in $\lg J_\mu J_\nu\rg_{QF}$ and it would be useful to better understand how they complement contact terms found in \eqref{7.3:wi310} and \eqref{7.3:wi3111ct} and if they can be interpreted as arising from other Ward identities. In particular, it is not obvious if the contact terms present in \eqref{7.3:3111pert} are the same terms appearing in \eqref{7.3:wi3111ct}. \\

Chern-Simons contact terms of various QF correlators correspond to generalized $\theta$-terms in the action of the bulk dual higher-spin gravity. In this work, we have only considered contact terms of the vector current two-point function, which corresponds to a specific tuning of the bulk $F\wedge F$ term, however the analysis is by no means complete\footnote{Similarly, one can analyze relations imposed by contact terms in $\lg TT\rg_{QF}$, associated to a gravitational Chern-Simons term $R\wedge R$ in the bulk action \cite{Closset:2012vp}, as well as more complicated cases.}. A better understanding of the map between contact terms \eqref{7.3:wi310}-\eqref{7.3:3111pert} and bulk $\theta$-terms would serve to further probe higher-spin gravity duals of Chern-Simons vector models. It would be interesting to analyze the effects of these terms on higher-point correlators in gravity-side computations, as well as to understand to what extent they affect higher-point correlators in the chiral limits discussed in \cite{Aharony:2024nqs,Jain:2024bza}. In the main text, it was argued that the three-point function contact terms \eqref{7.3:wi310} and \eqref{7:TJJct} do not contribute in these limits, which is consistent with the expected matching between CFT three-point functions and cubic vertices in the bulk.\\

\textbf{Acknowledgements.} We thank Ofer Aharony and Rohit R. Kalloor for many helpful discussions and comments on the manuscript. We thank for Sachin Jain and Dhruva Sathyanarayanan for useful discussions. This work is an extension of the M.Sc. thesis \cite{Kukolj:2024}, submitted to the Weizmann Institute. This work was supported in part by an Israel Science Foundation (ISF) grant number 2159/22, by Simons Foundation grant 994296 (Simons Collaboration on Confinement and QCD Strings), by the Minerva foundation with funding from the Federal German Ministry for Education and Research, by the German Research Foundation through a German-Israeli Project Cooperation (DIP) grant ``Holography and the Swampland'', and by a research grant from Martin Eisenstein. 

\appendix

\section{\texorpdfstring{$\lg T_{--}J_+J_+O\rg_{QF}$}{}: The full expression}\label{App1}
We find four distinct structures of this correlator, as defined in \eqref{6.3:TJJO}. We write the result in terms of collinear momenta $p_i=-q_i$:
\setlength{\jot}{5pt}
\begin{flalign*}\begin{aligned}
&\lg T_{--}(p_1)J_+(p_2)J_+(p_3)O(-p_1-p_2-p_3)\rg_{FF} = \\&

-\frac{ip_1^3\sign(p_1)(p_1 (p_2+p_3)+p_2^2+p_2 p_3+p_3^2)}{32 p_2 p_3(p_1+p_2)(p_1+p_3)(p_2+p_3)} 
+\frac{i (p_1-p_3) (p_1+p_3)^2 \sign(p_1+p_3)}{32 p_1 p_2 p_3}\\&

+ \frac{ip_2^2\sign(p_2)(p_1^2 (p_2+2 p_3)+p_1(p_2^2+3 p_2 p_3+2 p_3^2)+p_2 p_3(p_2+p_3))}{32 p_1 p_3(p_1+p_2)(p_1+p_3)(p_2+p_3)}
-\frac{i (p_2+p_3)^3 \sign(p_2+p_3)}{32 p_1 p_2 p_3} \\&

+\frac{ip_3^2\sign(p_3)(p_1^2 (2 p_2+p_3)+p_1(2 p_2^2+3 p_2 p_3+p_3^2)+ p_2 p_3 (p_2+p_3))}{32 p_1 p_2 (p_1+p_2)(p_1+p_3)(p_2+p_3)}\\&
+\frac{i(p_1-p_2)(p_1+p_2)^2\sign(p_1+p_2)}{32 p_1 p_2 p_3} -\frac{i(p_1+p_2+p_3)^2(p_1^3 (p_2+p_3)+p_1^2 p_2 p_3) \sign(p_1+p_2+p_3)}{32 p_1 p_2 p_3(p_1+p_2)(p_1+p_3) (p_2+p_3)}\\&

-\frac{i(p_1+p_2+p_3)^2(-p_1(p_2^3+p_2^2 p_3+p_2 p_3^2+p_3^3)-p_2 p_3 (p_2+p_3)^2) \sign(p_1+p_2+p_3)}{32 p_1 p_2 p_3(p_1+p_2)(p_1+p_3) (p_2+p_3)}\\\\

\cline{1-2}\\

&\lg T_{--}(p_1)J_+(p_2)J_+(p_3)O(-p_1-p_2-p_3)\rg_{X_1} = \\&

\frac{p_3^3 \left(2 p_1^2+p_1 (p_2+p_3)+p_2 (p_3-2 p_2)\right)}{32 p_1 p_2 (p_1+p_2) (p_1+p_3) (p_2+p_3)}
-\frac{4 p_1^2 p_3^2+(p_1-p_2) (p_1+p_2)^3}{32 p_3 (p_1+p_2) (p_1+p_3) (p_2+p_3)}\\&
+\frac{-p_1^4-4 p_1^3 p_2+p_2^4}{32 p_1 p_2 (p_1+p_3) (p_2+p_3)}

-\frac{p_3 \left(p_1^4+3 p_1^2 p_2^2+3 p_1 p_2^2 (p_2+p_3)+p_2^3 (p_2+3 p_3)\right)}{16 p_1 p_2 (p_1+p_2) (p_1+p_3) (p_2+p_3)}\\&
- \frac{p_1^3 \sign(p_1) \left(p_1 (p_2+p_3)+p_2^2+p_2 p_3+p_3^2\right) \sign(p_1+p_2+p_3)}{32 p_2 p_3 (p_1+p_2) (p_1+p_3) (p_2+p_3)} \\&

+\frac{p_2 \sign(p_2) \left(2 p_1^3 (p_2+p_3)+p_1^2 \left(3 p_2^2+4 p_2 p_3+2 p_3^2\right)\right) \sign(p_1+p_2+p_3)}{32 p_1 p_3 (p_1+p_2) (p_1+p_3) (p_2+p_3)} \\&

+\frac{p_2 \sign(p_2) \left(p_1 p_2 \left(p_2^2+3 p_2 p_3+2 p_3^2\right)+p_2^2 p_3 (p_2+p_3)\right) \sign(p_1+p_2+p_3)}{32 p_1 p_3 (p_1+p_2) (p_1+p_3) (p_2+p_3)} \\&

+\frac{p_3 \sign(p_3) \left(p_1 p_3 \left(2 p_2^2+3 p_2 p_3+p_3^2\right)+p_2 p_3^2 (p_2+p_3)\right) \sign(p_1+p_2+p_3)}{32 p_1 p_2 (p_1+p_2) (p_1+p_3) (p_2+p_3)} \\&

+\frac{p_3 \sign(p_3) \left(2 p_1^3 (p_2+p_3)+p_1^2 \left(2 p_2^2+4 p_2 p_3+3 p_3^2\right)\right) \sign(p_1+p_2+p_3)}{32 p_1 p_2 (p_1+p_2) (p_1+p_3) (p_2+p_3)} \\&

-\frac{p_1 p_2 p_3 \sign(p_2) \sign(p_3)}{16 (p_1+p_2) (p_1+p_3) (p_2+p_3)} 
-\frac{p_2 (p_1+p_2) \sign(p_2) \sign(p_1+p_2)}{16 p_1 p_3} \\&

-\frac{p_3 (p_1+p_3) \sign(p_3) \sign(p_1+p_3)}{16 p_1 p_2}
+\frac{(p_1-p_2) (p_1+p_2)^2 \sign(p_1+p_2) \sign(p_1+p_2+p_3)}{32 p_1 p_2 p_3}\\&

+\frac{(p_1-p_3) (p_1+p_3)^2 \sign(p_1+p_3) \sign(p_1+p_2+p_3)}{32 p_1 p_2 p_3} 
+\frac{(p_2+p_3)^3 \sign(p_2+p_3) \sign(p_1+p_2+p_3)}{32 p_1 p_2 p_3}\\\\

\cline{1-2}\\

\end{aligned}\end{flalign*}
\begin{flalign}\begin{aligned}\label{TJJOfull}

&\lg T_{--}(p_1)J_+(p_2)J_+(p_3)O(-p_1-p_2-p_3)\rg_{X_2} = \\&

-\frac{i p_1^3 \sign(p_1) \left(p_1 (p_2+p_3)+p_2^2+p_2 p_3+p_3^2\right)}{32 p_2 p_3 (p_1+p_2) (p_1+p_3) (p_2+p_3)}
+\frac{i (p_1+p_2) \left(p_1^2+p_2^2\right) \sign(p_1+p_2)}{32 p_1 p_2 p_3}\\&

-\frac{i p_3^3 \left(2 p_1^2+p_1 (p_2+p_3)+p_2 (p_3-2 p_2)\right) \sign(p_1+p_2+p_3)}{32 p_1 p_2 (p_1+p_2) (p_1+p_3) (p_2+p_3)}
+\frac{i (p_1+p_3) \left(p_1^2+p_3^2\right) \sign(p_1+p_3)}{32 p_1 p_2 p_3}\\&

-\frac{i p_3 \sign(p_3) \left(2 p_1^3 (p_2+p_3)+p_1^2 p_3 (4 p_2+3 p_3)+p_1 p_3 \left(2 p_2^2+3 p_2 p_3+p_3^2\right)+p_2 p_3^2 (p_2+p_3)\right)}{32 p_1 p_2 (p_1+p_2) (p_1+p_3) (p_2+p_3)}\\&

-\frac{i p_2 \sign(p_2) \left(2 p_1^3 (p_2+p_3)+p_1^2 p_2 (3 p_2+4 p_3)+p_1 p_2 \left(p_2^2+3 p_2 p_3+2 p_3^2\right)+p_2^2 p_3 (p_2+p_3)\right)}{32 p_1 p_3 (p_1+p_2) (p_1+p_3) (p_2+p_3)}\\&

-\frac{i \left(p_3 (p_1+p_2)^2 \left(p_1^3+3 p_1^2 p_2-p_1 p_2^2+p_2^3\right)+p_1^2 p_2 \left(p_1^3+2 p_1^2 p_2+2 p_1 p_2^2+2 p_2^3\right)\right) \sign(p_1+p_2+p_3)}{32 p_1 p_2 p_3 (p_1+p_2) (p_1+p_3) (p_2+p_3)}\\&

-\frac{i \left(2 p_3^3 (p_1+p_2) \left(p_1^2-3 p_2^2\right)+2 p_3^2 \left(p_1^4+3 p_1^3 p_2+p_1^2 p_2^2-3 p_1 p_2^3-p_2^4\right)+p_1 p_2^5\right) \sign(p_1+p_2+p_3)}{32 p_1 p_2 p_3 (p_1+p_2) (p_1+p_3) (p_2+p_3)}\\&

-\frac{i p_1 p_2 p_3 \sign(p_2) \sign(p_3) \sign(p_1+p_2+p_3)}{16 (p_1+p_2) (p_1+p_3) (p_2+p_3)}

+\frac{i p_2 (p_1+p_2) \sign(p_2) \sign(p_1+p_2) \sign(p_1+p_2+p_3)}{16 p_1 p_3}\\&
+\frac{i p_3 (p_1+p_3) \sign(p_3) \sign(p_1+p_3) \sign(p_1+p_2+p_3)}{16 p_1 p_2}
-\frac{i (p_2+p_3)^3 \sign(p_2+p_3)}{32 p_1 p_2 p_3}\\\\

\cline{1-2}\\

&\lg T_{--}(p_1)J_+(p_2)J_+(p_3)O(-p_1-p_2-p_3)\rg_{CB} = \\&

-\frac{p_3 \sign(p_3) \left(p_1^2 \left(2 p_2^2+2 p_2 p_3+p_3^2\right)+p_1 p_3 \left(2 p_2^2+3 p_2 p_3+p_3^2\right)+p_2 p_3^2 (p_2+p_3)\right) \sign(p_1+p_2+p_3)}{32 p_1 p_2 (p_1+p_2) (p_1+p_3) (p_2+p_3)}\\&

-\frac{p_2 \sign(p_2) \left(p_1^2 \left(p_2^2+2 p_2 p_3+2 p_3^2\right)+p_1 p_2 \left(p_2^2+3 p_2 p_3+2 p_3^2\right)+p_2^2 p_3 (p_2+p_3)\right) \sign(p_1+p_2+p_3)}{32 p_1 p_3 (p_1+p_2) (p_1+p_3) (p_2+p_3)}\\&

+\frac{(p_1+p_2) \left(p_1^2+p_2^2\right) \sign(p_1+p_2) \sign(p_1+p_2+p_3)}{32 p_1 p_2 p_3} 
\\&

+\frac{(p_1+p_3) \left(p_1^2+p_3^2\right) \sign(p_1+p_3) \sign(p_1+p_2+p_3)}{32 p_1 p_2 p_3} - \frac{p_3^2 \left(2 p_1^3+2 p_1^2 p_3+5 p_1 p_2 (2 p_2+p_3)+6 p_2^3\right)}{32 p_1 p_2 (p_1+p_2) (p_1+p_3) (p_2+p_3)}\\&

-\frac{p_3^2 \left(2 \left(p_1^3+5 p_1 p_2^2+3 p_2^3\right)+p_3 \left(2 p_1^2+5 p_1 p_2+4 p_2^2\right)+p_3^2 (p_1+p_2)\right)}{32 p_1 p_2 (p_1+p_2) (p_1+p_3) (p_2+p_3)}\\&

+\frac{(p_2+p_3)^3 \sign(p_2+p_3) \sign(p_1+p_2+p_3)}{32 p_1 p_2 p_3}-\frac{\left(p_1^2+p_2^2\right) \left(p_3 \left(p_1^2+4 p_1 p_2+p_2^2\right)+p_1 p_2 (p_1+p_2)\right)}{32 p_1 p_2 p_3 (p_1+p_3) (p_2+p_3)}\\&

-\frac{p_1^3 \sign(p_1) \left(p_1 (p_2+p_3)+p_2^2+p_2 p_3+p_3^2\right) \sign(p_1+p_2+p_3)}{32 p_2 p_3 (p_1+p_2) (p_1+p_3) (p_2+p_3)}

\end{aligned}\end{flalign}
\section{\texorpdfstring{$\lg J_{---}J_+J_+J_+\rg_{QF}$}{}: The full expression}\label{App1a}

We find five structures of $\lg J_{---}J_+J_+J_+\rg_{QF}$, as defined in \eqref{6.3a:3111}, excluding the contact term which diverges as $\tl\to\infty$. In terms of collinear momenta $p_i=-q_i$, we have:
\setlength{\jot}{5pt}
\begin{flalign*}\begin{aligned}
&\lg J_{---}(-p_1-p_2-p_3)J_+(p_1)J_+(p_2)J_+(p_3)\rg_{FF} = \\&

\frac{i p_1^3 \sign(p_1) \left(p_1^3 (p_2+p_3)+p_1^2 \left(3 p_2^2+5 p_2 p_3+3 p_3^2\right)+2 p_1 (p_2+p_3)^3+2 p_2 p_3 (p_2+p_3)^2\right)}{48 p_2 p_3 (p_1+p_2) (p_1+p_3) (p_2+p_3) (p_1+p_2+p_3)}\\&

+\frac{i p_2^3 \sign(p_2) \left(2 p_1^3 (p_2+p_3)+p_1^2 \left(3 p_2^2+6 p_2 p_3+4 p_3^2\right)+p_1 \left(p_2^3+5 p_2^2 p_3+6 p_2 p_3^2+2 p_3^3\right)\right)}{48 p_1 p_3 (p_1+p_2) (p_1+p_3) (p_2+p_3) (p_1+p_2+p_3)}\\&

+\frac{i p_3^3 \sign(p_3) \left(2 p_1^3 (p_2+p_3)+p_1^2 \left(4 p_2^2+6 p_2 p_3+3 p_3^2\right)+p_1 \left(2 p_2^3+6 p_2^2 p_3+5 p_2 p_3^2+p_3^3\right)\right)}{48 p_1 p_2 (p_1+p_2) (p_1+p_3) (p_2+p_3) (p_1+p_2+p_3)}\\&

+\frac{i p_2^3 \sign(p_2) \left(p_2 p_3 \left(p_2^2+3 p_2 p_3+2 p_3^2\right)\right)}{48 p_1 p_3 (p_1+p_2) (p_1+p_3) (p_2+p_3) (p_1+p_2+p_3)}
+\frac{i (p_1+p_2)^4 (p_1+p_2+2 p_3) \sign(-p_1-p_2)}{48 p_1 p_2 p_3 (p_1+p_2+p_3)}\\&

+\frac{i (p_1+p_3)^4 (p_1+2 p_2+p_3) \sign(-p_1-p_3)}{48 p_1 p_2 p_3 (p_1+p_2+p_3)}
+\frac{i (p_2+p_3)^4 (2 p_1+p_2+p_3) \sign(-p_2-p_3)}{48 p_1 p_2 p_3 (p_1+p_2+p_3)}\\&

-\frac{i (p_1 (p_2+p_3)+p_2 p_3) (p_1+p_2+p_3)^5 \sign(-p_1-p_2-p_3)}{48 p_1 p_2 p_3 (p_1+p_2) (p_1+p_3) (p_2+p_3)}\\&
+\frac{i p_3^3 \sign(p_3) \left(p_2 p_3 \left(2 p_2^2+3 p_2 p_3+p_3^2\right)\right)}{48 p_1 p_2 (p_1+p_2) (p_1+p_3) (p_2+p_3) (p_1+p_2+p_3)}\\\\

\cline{1-2}\\

&\lg J_{---}(-p_1-p_2-p_3)J_+(p_1)J_+(p_2)J_+(p_3)\rg_{X_1} = \\&

\frac{p_1^2 \left(16 p_2^3+55 p_2^2 p_3+55 p_2 p_3^2+16 p_3^3\right)+p_1 \left(4 p_2^4+32 p_2^3 p_3+55 p_2^2 p_3^2+32 p_2 p_3^3+4 p_3^4\right)}{24 (p_1+p_2) (p_1+p_3) (p_2+p_3) (p_1+p_2+p_3)}\\&

+\frac{4 p_1^4 (p_2+p_3)+16 p_1^3 (p_2+p_3)^2+4 p_2 p_3 \left(p_2^3+4 p_2^2 p_3+4 p_2 p_3^2+p_3^3\right)}{24 (p_1+p_2) (p_1+p_3) (p_2+p_3) (p_1+p_2+p_3)}\\&

-\frac{p_2 p_3 \sign(p_2) \sign(p_3) (p_1+p_2+p_3) (p_1 (p_2+p_3)+p_2 p_3)}{24 p_1 (p_1+p_2) (p_1+p_3) (p_2+p_3)}-\frac{p_3 (p_1+p_2)^3 \sign(p_3) \sign(-p_1-p_2)}{24 p_1 p_2 (p_1+p_2+p_3)}\\&

-\frac{p_1 p_2 \sign(p_1) \sign(p_2) (p_1+p_2+p_3) (p_1 (p_2+p_3)+p_2 p_3)}{24 p_3 (p_1+p_2) (p_1+p_3) (p_2+p_3)}-\frac{p_2 (p_1+p_3)^3 \sign(p_2) \sign(-p_1-p_3)}{24 p_1 p_3 (p_1+p_2+p_3)}\\&

-\frac{p_1 p_3 \sign(p_1) \sign(p_3) (p_1+p_2+p_3) (p_1 (p_2+p_3)+p_2 p_3)}{24 p_2 (p_1+p_2) (p_1+p_3) (p_2+p_3)}-\frac{p_1 \sign(p_1) (p_2+p_3)^3 \sign(-p_2-p_3)}{24 p_2 p_3 (p_1+p_2+p_3)}\\\\

\cline{1-2}\\

&\lg J_{---}(-p_1-p_2-p_3)J_+(p_1)J_+(p_2)J_+(p_3)\rg_{X_2} = \\&

\frac{i p_2 \sign(p_2) \left(p_1^2 \left(p_2^2-2 p_3^2\right)+p_1 p_2^2 (p_2+p_3)+p_2^2 p_3 (p_2+p_3)\right) (p_1+p_2+p_3)}{24 p_1 p_3 (p_1+p_2) (p_1+p_3) (p_2+p_3)}\\&

+\frac{i p_3 \sign(p_3) \left(p_1^2 \left(p_3^2-2 p_2^2\right)+p_1 p_3^2 (p_2+p_3)+p_2 p_3^2 (p_2+p_3)\right) (p_1+p_2+p_3)}{24 p_1 p_2 (p_1+p_2) (p_1+p_3) (p_2+p_3)}\\&

\end{aligned}\end{flalign*}
\begin{flalign}\begin{aligned}\label{B2:3111full}

&+\frac{i p_1 \sign(p_1) \left(p_1^3 (p_2+p_3)+p_1^2 \left(p_2^2+p_2 p_3+p_3^2\right)-2 p_2^2 p_3^2\right) (p_1+p_2+p_3)}{24 p_2 p_3 (p_1+p_2) (p_1+p_3) (p_2+p_3)}\\&

-\frac{i (p_1 (p_2+p_3)+p_2 p_3) (p_1+p_2+p_3)^5 \sign(-p_1-p_2-p_3)}{24 p_1 p_2 p_3 (p_1+p_2) (p_1+p_3) (p_2+p_3)}\\&

+\frac{i (p_1+p_3)^3 (p_1+p_2+p_3) \sign(-p_1-p_3)}{24 p_1 p_2 p_3}
+\frac{i (p_2+p_3)^3 (p_1+p_2+p_3) \sign(-p_2-p_3)}{24 p_1 p_2 p_3}\\&

+\frac{i (p_1+p_2)^3 (p_1+p_2+p_3) \sign(-p_1-p_2)}{24 p_1 p_2 p_3}-\frac{i p_1 p_2 p_3 \sign(p_1) \sign(p_2) \sign(p_3) (p_1+p_2+p_3)}{12 (p_1+p_2) (p_1+p_3) (p_2+p_3)}\\&

\cline{1-2}\\

&\lg J_{---}(-p_1-p_2-p_3)J_+(p_1)J_+(p_2)J_+(p_3)\rg_{X_3} = \\&

\frac{7 p_1^2 \left(4 p_2^3+13 p_2^2 p_3+13 p_2 p_3^2+4 p_3^3\right)+p_1 \left(8 p_2^4+54 p_2^3 p_3+91 p_2^2 p_3^2+54 p_2 p_3^3+8 p_3^4\right)}{24 (p_1+p_2) (p_1+p_3) (p_2+p_3) (p_1+p_2+p_3)}\\&

-\frac{p_2 p_3 \sign(p_2) \sign(p_3) (p_1+p_2+p_3) \left(2 p_1^2+p_1 (p_2+p_3)+p_2 p_3\right)}{24 p_1 (p_1+p_2) (p_1+p_3) (p_2+p_3)}\\&

-\frac{p_1 p_3 \sign(p_1) \sign(p_3) (p_1+p_2+p_3) (p_1 (p_2+p_3)+p_2 (2 p_2+p_3))}{24 p_2 (p_1+p_2) (p_1+p_3) (p_2+p_3)}\\&

-\frac{p_1 p_2 \sign(p_1) \sign(p_2) (p_1+p_2+p_3) (p_1 (p_2+p_3)+p_3 (p_2+2 p_3))}{24 p_3 (p_1+p_2) (p_1+p_3) (p_2+p_3)}\\&

+\frac{8 p_1^4 (p_2+p_3)+p_1^3 \left(28 p_2^2+54 p_2 p_3+28 p_3^2\right)+4 p_2 p_3 \left(2 p_2^3+7 p_2^2 p_3+7 p_2 p_3^2+2 p_3^3\right)}{24 (p_1+p_2) (p_1+p_3) (p_2+p_3) (p_1+p_2+p_3)}\\&

-\frac{p_1 \sign(p_1) (p_2+p_3)^3 \sign(-p_2-p_3)}{24 p_2 p_3 (p_1+p_2+p_3)}
-\frac{p_3 (p_1+p_2)^3 \sign(p_3) \sign(-p_1-p_2)}{24 p_1 p_2 (p_1+p_2+p_3)}\\&

-\frac{p_2 (p_1+p_3)^3 \sign(p_2) \sign(-p_1-p_3)}{24 p_1 p_3 (p_1+p_2+p_3)}\\\\

\cline{1-2}\\

&\lg J_{---}(-p_1-p_2-p_3)J_+(p_1)J_+(p_2)J_+(p_3)\rg_{CB} =\\&

\frac{i (p_1+p_2)^3 \left(p_1^2+2 p_1 (p_2+p_3)+p_2^2+2 p_2 p_3+2 p_3^2\right) \sign(-p_1-p_2)}{48 p_1 p_2 p_3 (p_1+p_2+p_3)}\\&

+\frac{i (p_1+p_3)^3 \left(p_1^2+2 p_1 (p_2+p_3)+2 p_2^2+2 p_2 p_3+p_3^2\right) \sign(-p_1-p_3)}{48 p_1 p_2 p_3 (p_1+p_2+p_3)}\\&

+\frac{i (p_2+p_3)^3 \left(2 p_1^2+2 p_1 (p_2+p_3)+(p_2+p_3)^2\right) \sign(-p_2-p_3)}{48 p_1 p_2 p_3 (p_1+p_2+p_3)}\\&

+\frac{i p_1^3 \sign(p_1) \left(p_1^3 (p_2+p_3)+p_1^2 \left(3 p_2^2+5 p_2 p_3+3 p_3^2\right)+4 p_1 \left(p_2^3+2 p_2^2 p_3+2 p_2 p_3^2+p_3^3\right)\right)}{48 p_2 p_3 (p_1+p_2) (p_1+p_3) (p_2+p_3) (p_1+p_2+p_3)}\\&

+\frac{i p_2^3 \sign(p_2) \left(2 p_1^4+4 p_1^3 (p_2+p_3)+p_1^2 \left(3 p_2^2+8 p_2 p_3+4 p_3^2\right)+p_1 (p_2+p_3) (p_2+2 p_3)^2\right)}{48 p_1 p_3 (p_1+p_2) (p_1+p_3) (p_2+p_3) (p_1+p_2+p_3)}\\&

+\frac{i p_3^3 \sign(p_3) \left(2 p_1^4+4 p_1^3 (p_2+p_3)+p_1^2 \left(4 p_2^2+8 p_2 p_3+3 p_3^2\right)+p_1 (p_2+p_3) (2 p_2+p_3)^2\right)}{48 p_1 p_2 (p_1+p_2) (p_1+p_3) (p_2+p_3) (p_1+p_2+p_3)}\\

\end{aligned}\end{flalign}
\begin{flalign*}\begin{aligned}

&+\frac{i p_3^3 \sign(p_3) \left(p_2 \left(2 p_2^3+4 p_2^2 p_3+3 p_2 p_3^2+p_3^3\right)\right)}{48 p_1 p_2 (p_1+p_2) (p_1+p_3) (p_2+p_3) (p_1+p_2+p_3)}
+\frac{i (p_1+p_2+p_3)^5 \sign(p_1+p_2+p_3)}{48 p_1 (p_1+p_2) (p_1+p_3) (p_2+p_3)}\\&

+\frac{i p_1^3 \sign(p_1) \left(2 (p_2+p_3)^2 \left(p_2^2+p_3^2\right)\right)}{48 p_2 p_3 (p_1+p_2) (p_1+p_3) (p_2+p_3) (p_1+p_2+p_3)}
+\frac{i (p_1+p_2+p_3)^5 \sign(p_1+p_2+p_3)}{48 p_3 (p_1+p_2) (p_1+p_3) (p_2+p_3)}\\&

+\frac{i p_2^3 \sign(p_2) \left(p_3 \left(p_2^3+3 p_2^2 p_3+4 p_2 p_3^2+2 p_3^3\right)\right)}{48 p_1 p_3 (p_1+p_2) (p_1+p_3) (p_2+p_3) (p_1+p_2+p_3)}

+\frac{i(p_1+p_2+p_3)^5 \sign(p_1+p_2+p_3)}{48 p_2 (p_1+p_2) (p_1+p_3) (p_2+p_3)} 
\end{aligned}\end{flalign*}
\section{Supplementary computations}\label{App2}

We present the more technical computations referenced throughout the main text. 

\subsection{Integrated FF Ward identities}\label{C:FFWI}

Continuing subsection \ref{Sec3.1}, we verify that the higher-spin algebra \eqref{2.2:QO}, \eqref{2.2:QJ} obtained through the free-fermionic OPE leads to consistent higher-spin Ward identities. We consider the action of the spin-3 charge $Q_{\mu\nu}$ on a general $n$-point function of $m-1$ vector currents and $n-m+1$ scalar currents, denoted by $\lg [Q_{\mu\nu},J_{\alpha_1}J_{\alpha_2}...J_{\alpha_{m-1}}O...O]\rg_{FF}$.

Specially, the result also holds when the number of scalar or vector currents is zero. Furthermore, when $m$ is odd, the Ward identity is trivially satisfied as all correlators vanish due to charge conjugation.\\

For simplicity, we will assume the algebra only contains terms appearing in \eqref{2.2:QO}, \eqref{2.2:QJ} with unknown coefficients, and then find constraints for the Ward identity to be satisfied. The reader is welcome to start from the more general ansatz \eqref{2.1:[Q,J]qf}, and verify that it leads to the same result. The Ward identity \eqref{3.1:FFWI} reads:
\begin{flalign}\begin{aligned}\label{3.1:JOWI}

	&\bigg(	a_1
		\epsilon_{\sigma\alpha_1(\mu}\d_{1\nu)}\d_1^\sigma \lg O(x_1) J_{\alpha_2}(x_2) .. J_{\alpha_{m-1}}(x_{m-1}) O(x_m) ... O(x_n)\rg_{FF} \\[5pt]& 
	+a_2 	
		\d_{1(\mu}\lg T_{\nu)\alpha_1}(x_1) J_{\alpha_2}(x_2) .. J_{\alpha_{m-1}}(x_{m-1}) O(x_m) ... O(x_n)\rg_{FF} 
		+ \perm{1 \leftrightarrow 2}{\alpha_1 \leftrightarrow \alpha_2} + ..
		+ \perm{1 \leftrightarrow m-1}{\alpha_1 \leftrightarrow \alpha_{m-1}} \bigg) \\[5pt]& +
	a_0\bigg(	
		\epsilon_{\rho\sigma(\mu}\d_{m\nu)}\d_m^\rho \lg J_{\alpha_1}(x_1) J_{\alpha_2}(x_2) .. J_{\alpha_{m-1}}(x_{m-1}) J^\sigma(x_m) O(x_{m+1}) ... O(x_n)\rg_{FF} 
		+ \perm{m \leftrightarrow m+1}{\alpha_m \leftrightarrow \alpha_{m+1}} +\\[5pt]&
        ... + \perm{m \leftrightarrow n}{\alpha_m \leftrightarrow \alpha_n} \bigg) = 0
\end{aligned}\end{flalign}
We will define $\sum_{(i)}$ to be the sum over all permutations $\{p_1,p_2,..,p_n\}$ of $\{1,2,...,n\}$ where $p_i=1$ and $\{p_2,..,p_n\}$ is a permutation of $\{1,2,...,n\}/\{i\}$. We also define $\G_{p_i}=\g_{\alpha_{p_i}}$ if $p_i< m$ and $\G_{p_i}=1$ if $p_i\geq m$. Finally, the fermionic propagator from $x_i$ to $x_j$ is denoted by $S_{ij}$. The first term in \eqref{3.1:JOWI} can now be written as:
\begin{flalign}\begin{aligned}

	&\epsilon_{\sigma\alpha_1(\mu}\d_{1\nu)}\d_1^\sigma \lg O(x_1)J_{\alpha_2}(x_2).. J_{\alpha_{m-1}}(x_{m-1})O(x_m).. O(x_n) \rg_{FF} \\[5pt]& =
	
	-i^{m-2}\sum_{(1)}\epsilon_{\sigma\alpha_{p_1}(\mu}\d_{p_1\nu)}\d_{p_1}^\sigma Tr\big( S_{p_np_1}S_{p_1p_2}\G_{p_2}S_{p_2p_3}...\G_{p_n}\big) \\& =
	
	-i^{m-1}\sum_{(1)}\epsilon_{\sigma\alpha_{p_1}(\mu}\d_{p_1\nu)}\d_{p_1\rho} 
	Tr\bigg( S_{p_np_1} \frac{1}{2}\{\g^\rho,\g^\sigma\} S_{p_1p_2}\G_{p_2}S_{p_2p_3}...\G_{p_n}\bigg)
\end{aligned}\end{flalign}
It is simple to show that this can be rewritten as 

$\epsilon_{\sigma\alpha_1(\mu}\d_{1\nu)}\d_1^\sigma \lg O(x_1)J_{\alpha_2}(x_2).. J_{\alpha_{m-1}}(x_{m-1})O(x_m).. O(x_n) \rg_{FF} = A_1 - B_1$, where:
\begin{flalign}\begin{aligned}

	A_k &= i^{m-1}\sum_{(k)}	Tr\big(
		\d_{p_1\mu}\d_{p_1\nu}S_{p_np_1}\G_{p_1}S_{p_1p_2}...\G_{p_n} - S_{p_np_1}\G_{p_1}\d_{p_1\mu}\d_{p_1\nu}S_{p_1p_2}...\G_{p_n}
	\big) \\
	
	B_k &= -i^{m-1}\sum_{(k)} \d_{p_1(\nu} Tr\big(
		\d_{p_1\alpha_{p_1}}S_{p_np_1}\g_{\mu)}S_{p_1p_2}\G_{p_2}...\G_{p_n} - S_{p_np_1}\g_{\mu)}\d_{p_1\alpha_{p_1}}S_{p_1p_2}\G_{p_2}...\G_{p_n}
	\big)
	
\end{aligned}\end{flalign}
The stress tensor terms can also be rewritten, by expanding the operators \eqref{1FFCurrents} and making contractions:
\begin{flalign}\begin{aligned}

	&\d_{1(\mu}\lg T_{\nu)\alpha_1}(x_1) J_{\alpha_2}(x_2) .. J_{\alpha_{m-1}}(x_{m-1}) O(x_m) ... O(x_n)\rg_{FF} \\[5pt]&=
	
	\frac{i^{m-2}}{4}\sum_{(1)}\d_{p_1(\mu}Tr\big(
			S_{p_np_1}\g_{\nu)}\d_{p_1\alpha_{p_1}}S_{p_1p_2}\G_{p_2}...\G_{p_n} - \d_{p_1\alpha_{p_1}}S_{p_np_1}\g_{\nu}S_{p_1p_2}\G_{p_2}... \G_{p_n}  
		\big) \\& \quad
		
	+\frac{i^{m-2}}{4}\sum_{(1)}\d_{p_1(\mu}Tr\big(
		S_{p_np_1}\G_{p_1}\d_{p_1\nu)}S_{p_1p_2}\G_{p_1}...\G_{p_n} - \d_{p_1\nu)}S_{p_np_1}\G_{p_1}S_{p_1p_2}...\G_{p_n}
	
	\big) = \frac{i}{4}A_1 + \frac{i}{4}B_1
					
\end{aligned}\end{flalign}
The remaining terms in \eqref{3.1:JOWI} are those coming from $[Q_{\mu\nu},O]_{FF}$.
\begin{flalign}\begin{aligned}\label{app:FFWI1}

	&\epsilon_{\rho\sigma(\mu}\d_{m\nu)}\d_m^\rho \lg J_{\alpha_1}(x_1) J_{\alpha_2}(x_2) .. J_{\alpha_{m-1}}(x_{m-1}) J^\sigma(x_m) O(x_{m+1}) ... O(x_n)\rg_{FF} \\[5pt]&=
	
	-i^m \sum_{(m)}\epsilon_{\rho\sigma(\mu}\d_{p_1\nu)}\d_{p_1}^\rho Tr\big( S_{p_np_1}\g^\sigma S_{p_1p_2}\G_{p_2} ... \G_{p_n} \big) \\& =
	
	-i^{m-1} \sum_{(m)} \d_{p_1(\nu}\d_{p_1}^\rho Tr\bigg(S_{p_np_1}\frac{1}{2}[\g_{\mu)},\g_\rho]S_{p_1p_2}\G_{p_2}...\G_{p_n} \bigg) \\& =
	
	-i^{m-1} \sum_{(m)} \d_{p_1(\nu} Tr\bigg(
		\d_{p_1}^\rho S_{p_np_1}\frac{1}{2}[\g_{\mu)},\g_\rho]S_{p_1p_2}\G_{p_2}...\G_{p_n} +  
		S_{p_np_1}\frac{1}{2}[\g_{\mu)},\g_\rho]\d_{p_1}^\rho S_{p_1p_2}\G_{p_2}...\G_{p_n} 
	\bigg)
\end{aligned}\end{flalign}
We can now replace commutators by anticommutators. This modifies \eqref{app:FFWI1} by adding terms proportional to the equations of motion, and thus does not change non-local terms\footnote{Here we are only interested in matching non-contact terms. Contact terms can also be included tracking them throughout the derivation.}. We find:
\begin{flalign*}

	&\epsilon_{\rho\sigma(\mu}\d_{m\nu)}\d_m^\rho \lg J_{\alpha_1}(x_1) J_{\alpha_2}(x_2) .. J_{\alpha_{m-1}}(x_{m-1}) J^\sigma(x_m) O(x_{m+1}) ... O(x_n)\rg_{FF} \\[5pt]&=
	
	-i^{m-1}\sum_{(m)}\d_{p_1(\nu} Tr\big(
		\d_{p_1\mu)} S_{p_np_1}S_{p_1p_2}\G_{p_2}...\G_{p_n} - S_{p_np_1}\d_{p_1\mu)}S_{p_1p_2}\G_{p_2}...\G_{p_n}
	\big) \\& =
	
	-i^{m-1}\sum_{(m)} Tr\big(
		\d_{p_1\mu}\d_{p_1\nu} S_{p_np_1}S_{p_1p_2}\G_{p_2}...\G_{p_n} - S_{p_np_1}\d_{p_1\mu}\d_{p_1\nu}S_{p_1p_2}\G_{p_2}...\G_{p_n}
	\big) = -A_m

\end{flalign*}

Finally, one can verify that $A_k$'s satisfy the following relation:
\begin{flalign}
\sum_{k=1}^{m-1} A_k + \sum_{k=m}^{n}A_k = 0 \implies \sum_{k=m}^{n}A_k = - \sum_{k=1}^{m-1} A_k
\end{flalign}
We insert these expressions into the Ward identity \eqref{3.1:JOWI}, to obtain:
\begin{flalign}\begin{aligned}

	&\bigg(a_1
		\epsilon_{\sigma\alpha_1(\mu}\d_{1\nu)}\d_1^\sigma \lg O(x_1) J_{\alpha_2}(x_2) .. J_{\alpha_{m-1}}(x_{m-1}) O(x_m) ... O(x_n)\rg_{FF}\\[5pt]& 
	+a_2 \d_{1(\mu}\lg T_{\nu)\alpha_1}(x_1) J_{\alpha_2}(x_2) .. J_{\alpha_{m-1}}(x_{m-1}) O(x_m) ... O(x_n)\rg_{FF} 
		+ \perm{1 \leftrightarrow 2}{\alpha_1 \leftrightarrow \alpha_2}  + 
        ... + \perm{1 \leftrightarrow m-1}{\alpha_1 \leftrightarrow \alpha_{m-1}} \bigg) \\[5pt]& + 
	a_0\bigg(	
		\epsilon_{\rho\sigma(\mu}\d_{m\nu)}\d_m^\rho \lg J_{\alpha_1}(x_1) J_{\alpha_2}(x_2) .. J_{\alpha_{m-1}}(x_{m-1}) J^\sigma(x_m) O(x_{m+1}) ... O(x_n)\rg_{FF} 
		+ \perm{m \leftrightarrow m+1}{\alpha_m \leftrightarrow \alpha_{m+1}} + \\[5pt]&
  
  ... + \perm{m \leftrightarrow n}{\alpha_m \leftrightarrow \alpha_n} \bigg) =
	
	\bigg(a_1+\frac{i a_2}{4} + a_0 \bigg)\sum_{k=1}^{m-1} A_k + \bigg(-a_1+\frac{i a_2}{4} \bigg)\sum_{k=1}^{m-1} B_k = 0

\end{aligned}\end{flalign}
Requiring that the equation is consistent for all $n$ and $m$ and for generic $x_i$ gives us a system of two linear homogenous equations. We solve this to find $a_1=-\frac{a_0}{2}=\frac{a}{2}$ and $a_2=-4ia_1=-2ia$ which is consistent with the coefficients obtained in Section \ref{Sec2.2}.

\subsection{Integrated FB Ward identities}\label{C:FBWI}

In this subsection we verify that the algebra coefficients in \eqref{2.3:QO} and \eqref{2.3:QJ} lead to consistent free-bosonic higher-spin Ward identities. We confine ourselves to Ward identities of three-point functions, however extending the analysis to general $n$-point functions should be straightforward.

There are four different Ward identities one can consider. However, two of these, specifically $\lg [Q_{\mu\nu},OOO]\rg_{FB}$ and $\lg [Q_{\mu\nu},J_\alpha J_\beta O]\rg_{FB}$ are trivially satisfied, as all appearing correlators are zero, due to charge conjugation. Assuming general coefficients in \eqref{2.3:QO} and \eqref{2.3:QJ}, we obtain constraints on the algebra from the two non-trivial Ward identities:\\

$\underline{\bm{\lg [Q_{\mu\nu},J_\alpha OO]\rg_{FB}} \textbf{ Ward identity}}$
\vspace{3pt}

The equation reads:
\begin{flalign}\begin{aligned}\label{3.2:JOOwi}
	&b_1\d_{1\mu}\d_{1\nu}\d_{1\alpha}\lg O(x_1)O(x_2)O(x_3)\rg_{FB} 
	+ b_2 g_{\alpha(\mu}\d_{1\nu)}\d_1^2 \lg O(x_1)O(x_2)O(x_3)\rg_{FB}	\\[5pt]	&
	+ b_3 \d_{1(\mu}\lg T_{\nu)\alpha}(x_1)O(x_2)O(x_3)\rg_{FB} 
	  + b_0 \big(\d_{2(\mu}\lg J_\alpha(x_1) J_{\nu)}(x_2) O(x_3)\rg_{FB} \\[5pt] &
        + \d_{3(\mu}\lg J_\alpha(x_1) O(x_2) J_{\nu)}(x_3) \rg_{FB} \big) = 0
\end{aligned}\end{flalign}
We denote the free scalar propagator from $x_i$ to $x_j$ by $F_{ij}$. Terms involving the scalar three-point function $\lg OOO\rg_{FB}=8^3\cdot 2 F_{12}F_{23}F_{31}$ can be written as:
\setlength{\jot}{5pt}
\begin{flalign}\label{3.2:JOOe1}
	\d_{1\mu}\d_{1\nu}\d_{1\alpha}\lg O(x_1)O(x_2)O(x_3)\rg_{FB}=
	
	& 8^3\cdot 2 \big(\d_{1\mu}\d_{1\nu}\d_{1\alpha}F_{12}F_{23}F_{31} + F_{12}F_{23}\d_{1\mu}\d_{1\nu}\d_{1\alpha}F_{31} \big)  \nonumber\\&
	
	+ 8^3\cdot 6 \big(\d_{1(\alpha}F_{12}F_{23}\d_{1\mu}\d_{1\nu)}F_{31} + \d_{1(\mu}\d_{1\nu}F_{12}F_{23}\d_{1\alpha)}F_{31} \big)\\
	
	g_{\alpha(\mu}\d_{1\nu)}\d_1^2 \lg O(x_1)O(x_2)O(x_3)\rg_{FB} =& 8^3\cdot 4g_{\alpha(\mu}(\d_{1\nu)}\d_{1\sigma}F_{12}F_{23}\d_1^\sigma F_{31} + \d_{1\sigma}F_{12}F_{23}\d_{1\nu)}\d_1^\sigma F_{31})
\end{flalign}
\setlength{\jot}{0pt}

Using the definition of the stress-tensor \eqref{1FBCurrents}, we similarly write out the term involving $\lg TOO\rg_{FB}$, while neglecting contact terms that come from the equations of motion:
\setlength{\jot}{5pt}
\begin{flalign}\begin{aligned}
	&\d_{1(\mu}\lg T_{\nu)\alpha}(x_1)O(x_2)O(x_3)\rg_{FB} =\\&
	
	8^2\bigg(-\frac{1}{2}(\d_{1\alpha}\d_{1\mu}\d_{1\nu}F_{12}F_{23}F_{31} + F_{12}F_{23}\d_{1\mu}\d_{1\nu}\d_{1\alpha}F_{31}) 
	
	+\frac{3}{2}(\d_{1\mu}\d_{1\nu}F_{12}F_{23}\d_{1\alpha}F_{31} + \d_{1\alpha}F_{12}F_{23}\d_{1\mu}\d_{1\nu}F_{31}) \\&
	
	+ (\d_{1\alpha}\d_{1(\nu}F_{12}F_{23}\d_{1\mu)}F_{31} + \d_{1(\mu}F_{12}F_{23}\d_{1\nu)}\d_{1\alpha}F_{31}) 
	
	- g_{\alpha(\mu}(\d_{1\nu)}\d_{1\sigma}F_{12}F_{23}\d_1^\sigma F_{31} + \d_{1\sigma}F_{12}F_{23}\d_{1\nu)}\d_1^\sigma F_{31})\bigg)
\end{aligned}\end{flalign}
\setlength{\jot}{0pt}
And analogously, we write out the terms coming from $[Q_{\mu\nu},O]_{FB}$ as:
%
\setlength{\jot}{5pt}
\begin{flalign}\begin{aligned}\label{3.2:JOOe2}
	&\d_{2(\mu}\lg J_\alpha(x_1)J_{\nu)}(x_2)O(x_3) \rg_{FB} + \begin{Bmatrix}2\leftrightarrow 3\end{Bmatrix} =\\&
	
	8\bigg(2(\d_{1\alpha}\d_{1\mu}\d_{1\nu}F_{12}F_{23}F_{31} + F_{12}F_{23}\d_{1\mu}\d_{1\nu}\d_{1\alpha}F_{31})
	-2(\d_{1\mu}\d_{1\nu}F_{12}F_{23}\d_{1\alpha}F_{31} + \d_{1\alpha}F_{12}F_{23}\d_{1\mu}\d_{1\nu}F_{31})\bigg)
\end{aligned}\end{flalign}
\setlength{\jot}{0pt}
Inserting the relations \eqref{3.2:JOOe1} - \eqref{3.2:JOOe2} into the Ward identity \eqref{3.2:JOOwi}, we obtain a sum of linearly independent terms:
\begin{flalign}\begin{aligned}
	&(2b_0+128b_1-4b_3)\bigg(\d_{1\alpha}\d_{1\mu}\d_{1\nu}F_{12}F_{23}F_{31} + F_{12}F_{23}\d_{1\mu}\d_{1\nu}\d_{1\alpha}F_{31} \bigg) +\\&
	
	 (-2b_0+128b_1+12b_3)\bigg(\d_{1\mu}\d_{1\nu}F_{12}F_{23}\d_{1\alpha}F_{31} + \d_{1\alpha}F_{12}F_{23}\d_{1\mu}\d_{1\nu}F_{31} \bigg) \\&

	+(256b_1+8b_3)\bigg(\d_{1\mu}\d_{1\nu}F_{12}F_{23}\d_{1\alpha}F_{31} + \d_{1\alpha}F_{12}F_{23}\d_{1\mu}\d_{1\nu}F_{31}\bigg) \\&
	 
	+(256b_2-8b_3)g_{\alpha(\mu}\bigg(\d_{1\nu)}\d_{1\sigma}F_{12}F_{23}\d_1^\sigma F_{31} + \d_{1\sigma}F_{23}\d_{1\nu)}\d_1^\sigma F_{31}\bigg) = 0
\end{aligned}\end{flalign}
Since the terms are linearly independent, the consistency of this equation implies $b_1=-\frac{b_0}{128}$, $b_2 = -b_1 =\frac{b_0}{128}$, $b_3=-32b_1=\frac{b_0}{4}$, which are the same constraints obtained through the free-bosonic OPE \eqref{2.3:QJ}.\\
 
$\underline{\bm{\lg [Q_{\mu\nu},J_\alpha J_\beta J_\gamma]\rg_{FB}} \textbf{ Ward identity}}$
\vspace{3pt}

Inserting \eqref{2.3:QJ} into \eqref{3.2:FBWI}, we have the integrated Ward identity: 
\begin{flalign}\begin{aligned}\label{3.2:JJJwi}
	&b_1 \d_{1\mu}\d_{1\nu}\d_{1\alpha}\lg O(x_1)J_\beta(x_2)J_\gamma(x_3)\rg_{FB}  + 
	b_2 g_{\alpha(\mu}\d_{1\nu)}\d_1^2 \lg O(x_1)J_\beta(x_2)J_\gamma(x_3)\rg_{FB} \\
	
	&+ b_3\d_{1(\mu}\lg T_{\nu)\alpha}(x_1)J_\beta(x_2)J_\gamma(x_3)\rg_{FB} 
	+ \perm{1\leftrightarrow 2}{\alpha\leftrightarrow\beta} + \perm{1\leftrightarrow 3}{\alpha\leftrightarrow\gamma} = 0
\end{aligned}\end{flalign}
The quickest way to verify this is to expand \eqref{2.3:QJ} using the definitions of current operators \eqref{1FBCurrents}:
\begin{flalign}\begin{aligned}\label{3.2:JJJwi1}
	&[Q_{\mu\nu},J_\alpha] =\\[5pt]&
	
	\bigg(8b_1-\frac{1}{4}b_3\bigg)(\d_\mu\d_\nu\d_\alpha\phd\ph + \phd\d_\mu\d_\nu\d_\alpha\ph)
	 +\bigg(8b_1+\frac{3}{4}b_3\bigg)(\d_\mu\d_\nu\phd\d_\alpha\ph + \d_\alpha\phd\d_\mu\d_\nu\ph) \\&
	 
	  +\bigg(8b_1+\frac{1}{4}b_3\bigg)(\d_\alpha\d_{(\mu}\phd\d_{\nu)}\ph + \d_{(\nu}\phd\d_{\mu)}\d_\alpha\ph) 
	  +\bigg(16b_2-\frac{1}{2}b_3\bigg)g_{\alpha(\mu}(\d_\sigma\d_{\nu)}\phd\d^\sigma\ph + \d_\sigma\phd\d_{\nu)}\d^\sigma\ph)
\end{aligned}\end{flalign}

Let us now look at contractions that come from each term separately, and then sum over permutations. Doing so, we find that the first two terms in \eqref{3.2:JJJwi1} lead to the same contributions, which we will label $A_1$:
\begin{flalign}\begin{aligned}
	&A_1 = \lg(\d_{1\mu}\d_{1\nu}\d_{1\alpha}\phd\ph + \phd\d_{1\mu}\d_{1\nu}\d_{1\alpha}\ph)(x_1) J_\beta(x_2)J_\gamma(x_3)\rg_{FB} 
	+ \perm{1\leftrightarrow 2}{\alpha\leftrightarrow\beta} + \perm{1\leftrightarrow 3}{\alpha\leftrightarrow\gamma} = \\
	
	&\quad = \lg(\d_{1\mu}\d_{1\nu}\phd  \d_{1\alpha}\ph + \d_{1\alpha}\phd\d_{1\mu}\d_{1\nu}\ph)(x_1) J_\beta(x_2)J_\gamma(x_3)\rg_{FB} 
	+ \perm{1\leftrightarrow 2}{\alpha\leftrightarrow\beta} + \perm{1\leftrightarrow 3}{\alpha\leftrightarrow\gamma} = \\
	
	& = -2\bigg(
	\d_{1\alpha}\d_{1\mu}\d_{1\nu}F_{12}(\d_{2\beta}F_{23}\d_{3\gamma}F_{31}-F_{31}\d_{2\beta}\d_{3\gamma}F_{23}) +\d_{1\mu}\d_{1\nu}\d_{1\alpha}F_{13}(\d_{3\gamma}F_{32}\d_{2\beta}F_{21}-F_{21}\d_{2\beta}\d_{3\gamma}F_{32}) \\&\qquad 
	+\d_{2\mu}\d_{2\nu}\d_{2\beta}F_{21}(\d_{1\alpha}F_{13}\d_{3\gamma}F_{32}-F_{32}\d_{1\alpha}\d_{3\gamma}F_{13}) +\d_{2\mu}\d_{2\nu}\d_{2\beta}F_{23}(\d_{3\gamma}F_{31}\d_{1\alpha}F_{12}-F_{12}\d_{3\gamma}\d_{1\alpha}F_{31}) \\&\qquad 
	+\d_{3\mu}\d_{3\nu}\d_{3\gamma}F_{32}(\d_{2\beta}F_{21}\d_{1\alpha}F_{13}-F_{13}\d_{2\beta}\d_{1\alpha}F_{21}) +\d_{3\mu}\d_{3\nu}\d_{3\gamma}F_{31}(\d_{1\alpha}F_{12}\d_{2\beta}F_{23}-F_{23}\d_{1\alpha}\d_{2\beta}F_{12})	
	\bigg)
\end{aligned}\end{flalign}
The remaining terms in \eqref{3.2:JJJwi1} give different contributions, which we label $A_2$ and $A_3$:
\begin{flalign}
	&A_2 = \lg(\d_{1\alpha}\d_{1(\mu}\phd\d_{1\nu)}\ph + \d_{1(\mu}\phd\d_{1\nu)}\d_{1\alpha}\ph)(x_1) J_\beta(x_2)J_\gamma(x_3)\rg_{FB} = \nonumber\\[5pt]&
	
	\qquad -2\d_{1\alpha}\big(
	\d_{1(\mu}F_{12}\d_{2\beta}F_{23}\d_{3\gamma}\d_{1\nu)}F_{31} +    
        \d_{1(\mu}F_{13}\d_{3\gamma}F_{32}\d_{2\beta}\d_{1\nu)}F_{21}\\[5pt]&
	
	\qquad\qquad\quad -\d_{1(\mu}F_{12}\d_{2\beta}\d_{3\gamma}F_{23}\d_{1\nu)}F_{31} - \d_{1(\mu}\d_{3\gamma}F_{13}F_{32}\d_{2\beta}\d_{1\nu)}F_{21}
	\big)  \nonumber\\ \nonumber\\

	&A_3 =\lg(\d_{1\sigma}\d_{1\alpha}\phd\d_1^\sigma\ph + \d_{1\sigma}\phd\d_{1\alpha}\d_1^\sigma\ph)(x_1) J_\beta(x_2)J_\gamma(x_3)\rg_{FB} =\nonumber\\[5pt]&
	
	\qquad -2\d_{1\alpha}(
	\d_{1\sigma}F_{12}\d_{2\beta}F_{23}\d_{3\gamma}\d_1^\sigma F_{31} + \d_{1\sigma}F_{13}\d_{3\gamma}F_{32}\d_{2\beta}\d_1^\sigma F_{21} \\[5pt]&
	
	\qquad\qquad\quad -\d_{1\sigma}F_{12}\d_{2\beta}\d_{3\gamma}F_{23}\d_1^\sigma F_{31} - \d_{1\sigma}\d_{3\gamma}F_{13}F_{32}\d_{2\beta}\d_1^\sigma F_{21} \nonumber
	)
\end{flalign}
The Ward identity \eqref{3.2:JJJwi} now becomes:
\begin{flalign}
	\bigg(8b_1-\frac{1}{4}b_3+8b_1+\frac{3}{4}b_3\bigg)A_1 + 
	\bigg(8b_1+\frac{1}{4}b_3\bigg) A_2 +
	\bigg(16b_2-\frac{1}{2}b_3\bigg) A_3 =0
\end{flalign}
Since $A_{1,2,3}$ are linearly independent for generic $x_{1,2,3}$, the equation can only be satisfied if all the coefficients are zero. This leads to $b_2=-b_1$ and $b_3=32b_2=-32b_1$, which is consistent with constraints obtained in \eqref{2.3:FBOPEfinal} and other FB Ward identities.

\subsection{Free/critical theory three-point functions}\label{sec:AppC3}

The correlators depend on three momenta $p_{1,2,3}^\mu$ satisfying $p_1^\mu+p_2^\mu+p_3^\mu=0$. We denote the modulus of each vector by $p_i = \sqrt{p_i^\mu p_{i,\mu}}$. By using dimensional regularization, one can compute the following integrals:

\setlength{\jot}{5pt}
\begin{flalign}\begin{aligned}\label{B:loopint}
		&\lg J_{s_1}(p_1)J_{s_2}(p_2)J_{s_3}(p_3)\rg_{FB} =\\&
  
        \frac{N}{2} \int \frac{d^3 \vec{k}}{(2\pi)^3} 
		\bigg(\frac{V^{FB}_{s_1}(p_1,k-p_2)V^{FB}_{s_2}(p_2,k)V^{FB}_{s_3}(p_3,k+p_3)}{(\vec{k}-\vec{p}_2)^2\vec{k}^2(\vec{k}+\vec{p}_3)^2} +\perm{p_2\leftrightarrow p_3}{s_2\leftrightarrow s_3}\bigg)\\\\
		
		&\lg J_{s_1}(p_1)J_{s_2}(p_2)J_{s_3}(p_3)\rg_{FF} =\\& -\frac{iN}{2} \int \frac{d^3 \vec{k}}{(2\pi)^3} 
		Tr\bigg[\frac{(\slashed{k}-\slashed{p_2})V^{FF}_{s_1}(p_1,k-p_2)(\slashed{k}+\slashed{p_3})V^{FF}_{s_2}(p_2,k)\slashed{k}V^{FF}_{s_3}(p_3,k+p_3)}{(\vec{k}-\vec{p}_2)^2\vec{k}^2(\vec{k}+\vec{p}_3)^2} 
        +\perm{p_2\leftrightarrow p_3}{s_2\leftrightarrow s_3}
  \bigg]
\end{aligned}\end{flalign}
\setlength{\jot}{0pt}

The vertex functions $V_s(p,k)$ of currents in the FB, FF theories are obtained by Fourier transforming the position space operators in \eqref{1FBCurrents} and \eqref{1FFCurrents}, respectively. $p$ is the momentum of the insertion, while $k$ is the outgoing loop momentum. We've included a $\frac{1}{2}$ factor to reflect the fact that the correlators are normalized as in one-half of those in the theories of a single complex boson/Dirac fermion. We use this normalization in Sections \ref{Sec5} and \ref{Sec6}. Correlators of the CB theory can be obtained via FB correlators using the Legendre transformation \eqref{4:FBtoCB}. \\

Of special interest are three-point functions with one or two scalar currents, as in this case the full QF correlators can be obtained from free/critical theory computations. For the three-point function of one scalar current and two vector currents, we find:

\begin{flalign}
&\lg O(p_1)J^\mu(p_2)J^\nu(p_3)\rg_{FF} = \nonumber\\[5pt]&
\frac{p_2^{\mu } \eps^{\nu  p_2 p_3}}{8 p_2 (p_1+p_2+p_3)^2} +\frac{p_2^{\nu } \eps^{\mu  p_2 p_3}}{8 p_2 (p_1+p_2+p_3)^2}+\frac{p_2 \eps^{\mu  \nu  p_3}}{8 (p_1+p_2+p_3)^2}-\frac{p_3 \eps^{\mu\nu p_2}}{8(p_1+p_2+p_3)^2}\nonumber\\& 

-\frac{p_3^{\nu}\eps^{\mu p_2 p_3}}{8 p_3 (p_1+p_2+p_3)^2}-\frac{p_3^{\mu}\eps^{\nu p_2 p_3}}{8 p_3 (p_1+p_2+p_3)^2}+\frac{\left(\vec{p_2}\cdot\vec{p_3}\right) \eps^{\mu  \nu  p_2}}{8 p_2 (p_1+p_2+p_3)^2}-\frac{\left(\vec{p_2}\cdot\vec{p_3}\right) \eps^{\mu  \nu  p_3}}{8 p_3 (p_1+p_2+p_3)^2} \label{B:OJJff} \\[20pt]

&\lg O(p_1)J^\mu(p_2)J^\nu(p_3)\rg_{FB} = \label{B:OJJfb} \nonumber\\[5pt]&

\frac{2 \delta^{\mu\nu }}{p_1+p_2+p_3}+\frac{2 p_3 p_2^{\mu } p_2^{\nu }}{p_1 p_2 (p_1+p_2+p_3)^2}+\frac{p_3 p_2^{\mu } p_3^{\nu }}{p_1 p_2 (p_1+p_2+p_3)^2}+\frac{p_2 p_2^{\mu } p_3^{\nu }}{p_1 p_3 (p_1+p_2+p_3)^2} \nonumber\\&

-\frac{p_1 p_2^{\mu } p_3^{\nu }}{p_2 p_3 (p_1+p_2+p_3)^2}-\frac{2 p_2^{\nu } p_3^{\mu }}{p_1 (p_1+p_2+p_3)^2}+\frac{2 p_2 p_3^{\mu } p_3^{\nu }}{p_1 p_3 (p_1+p_2+p_3)^2}

\end{flalign}
The CB correlator is obtained by \eqref{4:FBtoCB} as $\lg O(p_1)J^\mu(p_2)J^\nu(p_3)\rg_{CB} = \frac{p_1}{8}\lg O(p_1)J^\mu(p_2)J^\nu(p_3)\rg_{FB}$. It is interesting to analyze the $U(1)$ Ward identities corresponding to these correlators. We have:
\begin{flalign}
	\lg O(p_1)(ip_2\cdot J)(p_2)J^\nu(p_3)\rg_{FF} = 0; \qquad
	\lg O(p_1)(ip_2\cdot J)(p_2)J^\nu(p_3)\rg_{FB} = \frac{ip_2^{\nu}}{p_1}=\frac{ip_2^{\nu}}{4}\lg O(p_1)O(-p_1)\rg_{FB}
\end{flalign}
Naively, we expect both expressions to be equal to zero, as the scalar and vector currents are not charged under the global $U(1)$ symmetry. However the Ward identity of the FB theory contains an additional contact term, corresponding to an appearance of the nonlinear source term $\propto\mathcal{A}_{1\mu}\mathcal{A}^{1\mu}O$ in the generating functional of the theory, where $\mathcal{A}_{1\mu}$ is the background field coupled to the vector current. The $\lg JJO\rg$ correlators of FF and CB theories satisfy satisfy the epsilon-transform relation as shown in \cite{Jain:2021gwa}, up to a contact term that can be removed by a suitable regularization scheme. We find:
\begin{flalign}
	\lg O(p_1)J^\mu(p_2)J^\nu(p_3)\rg_{CB} = \lg O(p_1)(\eps\cdot J)^\mu(p_2)J^\nu(p_3)\rg_{FF}-\frac{\delta^{\mu\nu}}{8}
\end{flalign}
After subtracting this contact term, one can contract the Lorentz indices with polarization vectors $(z^\pm_{p_1})^\mu (z^\pm_{p_2})^\nu$ to write the QF correlator in spinor-helicity variables:
\begin{flalign}\begin{aligned}
\lg J^+J^-O\rg_{QF}=0; \qquad\qquad
\lg J^\pm J^\pm O\rg_{QF}=\tN\sqrt{\frac{1\mp i\tl}{1\pm i\tl}}\lg J^\pm J^\pm O\rg_{FF}
\end{aligned}\end{flalign}
For a more comprehensive overview of the spinor-helicity basis, we refer the reader to \cite{Baumann:2020dch, Jain:2021vrv}.

Similarly, we find three-point functions involving the stress tensor and two scalars. The computation yields:
\setlength{\jot}{5pt}
\begin{flalign}
&\lg T^{\mu\nu}(p_1)O(p_2)O(p_3)\rg_{FF} =\nonumber\\[5pt]&

\frac{-3 p_1^2-2 p_1 (p_2+p_3)+(p_2-p_3)^2}{32 p_1 (p_1+p_2+p_3)^2}\left(p_2^{\mu } p_3^{\nu }+p_2^{\nu } p_3^{\mu }\right)+ p_2^{\mu } p_2^{\nu }\frac{ (p_1+p_2-p_3)^2}{32 p_1 (p_1+p_2+p_3)^2} \nonumber\\&

+p_3^{\mu } p_3^{\nu } \frac{(p_1-p_2+p_3)^2}{32 p_1 (p_1+p_2+p_3)^2}+\left(\frac{p_2 p_3}{8 (p_1+p_2+p_3)}+\frac{p_1}{32}-\frac{p_2}{24}-\frac{p_3}{24}\right) \delta^{\mu\nu } \label{B:TOOff} \\[20pt]

&\lg T^{\mu\nu}(p_1)O(p_2)O(p_3)\rg_{FB} =\nonumber \\[5pt]& 

\frac{8}{p_2}\frac{8}{p_3}\bigg[
\frac{-3 p_1^2-2 p_1 (p_2+p_3)+(p_2-p_3)^2}{32 p_1 (p_1+p_2+p_3)^2}\left(p_2^{\mu } p_3^{\nu }+p_2^{\nu } p_3^{\mu }\right)+ p_2^{\mu } p_2^{\nu }\frac{ (p_1+p_2-p_3)^2}{32 p_1 (p_1+p_2+p_3)^2} \nonumber \\&

+p_3^{\mu } p_3^{\nu } \frac{(p_1-p_2+p_3)^2}{32 p_1 (p_1+p_2+p_3)^2}+\left(\frac{p_2 p_3}{8 (p_1+p_2+p_3)}+\frac{p_1}{32}-\frac{p_2}{24}-\frac{p_3}{24}\right) \delta^{\mu\nu }\bigg]
\label{B:TOOfb}

\end{flalign}
\setlength{\jot}{0pt}

Conformal symmetry constrains the correlator $\lg TOO\rg_{QF}$ to a single, parity-even structure \eqref{5:3pt}, fully determined by the number of dimensions and the scaling dimension of the scalar. This structure should be the same in both the $\tl\to 0$ and $\tl\to\infty$ limits, and indeed after Legendre-transforming the FB result \eqref{B:TOOfb}, we recover the FF result \eqref{B:TOOff}, even up to contact terms.  By contracting the result with polarization vectors $(z^\pm_{p_1})^\mu (z^\pm_{p_1})^\nu$, we can write $\lg TOO\rg_{QF}$ in spinor-helicity variables:
\begin{flalign}
	\lg T^{\pm\pm}OO\rg_{QF} = A \pm iB
\end{flalign}
The positive- and negative-helicity components are consistently related by complex conjugation. Similarly, three-point correlators with four or more Lorentz indices can be computed by solving more complicated Feynman integrals.

\subsection{Spin-\texorpdfstring{$s^+$}{} vertex in the collinear limit}\label{Sec6.1}

The computation of vertex factors for $(J_s)^{++...+}$ in the collinear limit $q^\pm=0$ follows the analogous computation for the spin-0 and spin-1 vertices of \cite{Aharony:2012nh}. For simplicity, we will first compute the stress-tensor vertex, defined as:
\begin{flalign}
	\lg T^{++}(-q)\psi_i(k)\bps^j(-p)\rg = V^{++}(q,p)\delta^j_i(2\pi)^3\delta^{(3)}(q+p-k)
\end{flalign}
The result can then easily be generalized to arbitrary $s$.
For operators with $s\geq 2$, one has to worry about external gluon lines connecting to the vertex. However, our choice of gauge $A_-=0$ ensures that the vertex functions $V^{++}(q,p)=V_{--}(q,p)$ do not receive any such corrections, making the calculation more tractable. In similar fashion, correlators involving only $O$, $J_1^\pm$ and $J_s^{--...-}$ (for $s\geq 2$) vertices are simplest to compute in the $A_+=0$ gauge. 
\begin{figure}[htbp]
\centering
\includegraphics[width=0.6\textwidth]{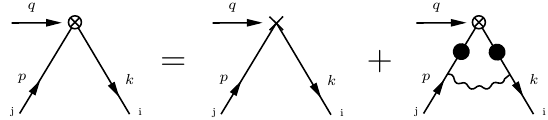}
\caption{Schematic Schwinger-Dyson equation for spin-2 vertex. The circled cross denotes the insertion of the exact $T^{++}$ vertex, while the un-circled cross represents the free theory vertex. The filled circle denotes the exact fermion propagator.}
\label{fig:spin2vertex}
\end{figure}

For the stress-tensor normalized as in \eqref{1FFCurrents}, the bootstrap equation, schematically represented in Figure \ref{fig:spin2vertex}, is given by:
\begin{flalign}\label{6.1:vertexEq}
	V^{++}(q,p)=ip^+\g^+ - 2\pi i\lambda\int\frac{d^3 k}{(2\pi)^3}\g^{[3}S(k+q)V^{++}(q,k)S(k)\g^{+]}\frac{1}{(k-p)^+}
\end{flalign}
For a general $A=a_\mu\g^\mu+a_0\mathds{1}$ we have $\g^{[3}A\g^{+]}=2a_0\g^+-2a_-\mathds{1}$, so for the solution to be self-consistent, we see that $V^{++}(q,p)$ does not have terms proportional to $\g^-$ and $\g^3$. We assume a general ansatz for the vertex, based on dimensional analysis and rotational invariance in the $p^{+-}$ plane:
\begin{flalign}\label{6.1:vertex}
	V^{++}(q,p)= p^+ v_+\bigg(\frac{2p_s}{|q_3|}\bigg)\g^+ + v_1\bigg(\frac{2p_s}{|q_3|}\bigg)\frac{(p^+)^2}{q_3}\mathds{1}
\end{flalign}
With $v_+\big(\frac{2p_s}{|q_3|}\big)$ and $v_1\big(\frac{2p_s}{|q_3|}\big)$ being dimensionless functions. The vertex equation \eqref{6.1:vertexEq} can now be expanded using the ansatz and the fermion propagator \eqref{6:FermProp}. We find:
\setlength{\jot}{5pt}
\begin{flalign}
	p^+ v_+-ip^+ =& 	
	 4\pi i\lambda\int\frac{d^3 k}{(2\pi)^3} 
	\frac{(k^+)^2}{(k+q)^2k^2(k-p)^+}\bigg[(q_3+2i\lambda k_s)v_+ - \bigg((2\lambda^2-1)\frac{k_s^2}{q_3}-\frac{k_3}{q_3}(k_3+q_3)\bigg) v_1\bigg] \\
	
	\frac{(p^+)^2}{q_3}v_i =& -4\pi i\lambda \int\frac{d^3 k}{(2\pi)^3} 
	\frac{(k^+)^3}{(k+q)^2k^2(k-p)^+}\bigg[2 v_+ +\bigg(2i\lambda \frac{k_s}{q_3}-1\bigg)v_1\bigg]
\end{flalign}
\setlength{\jot}{0pt}

We've suppressed the arguments of $v_1$, $v_+$ for brevity. The integrals over $k_3$ and the polar angle $\theta$ in the $p^{+-}$ plane can now be performed straightforwardly, since the functions $v_1$, $v_+$ on the right-hand side of the equations depend only on $\big(\frac{2k_s}{|q_3|}\big)$. The remaining integral over $k_s$ introduces a UV cutoff $\Lambda$. It is convenient to rewrite the resulting equations in terms of "parity-invariant" variables:
$x=\frac{2k_s}{|q_3|}$, $y=\frac{2p_s}{|q_3|}$, $\Lambda'=\frac{2\Lambda}{|q_3|}$ and $\hat{\lambda}=\lambda \text{sign}(q_3)$:
\begin{flalign}
	v_+(y)=&i-i\hat{\lambda}\int_y^{\Lambda'} dx \frac{1}{1+x^2}\bigg(\frac{\hat{\lambda}^2-1}{2}x^2 v_1(x) - (1+i\hat{\lambda}x)v_+(x) \bigg)\\
	
	v_1(y)=& -i\hat{\lambda}\int_y^{\Lambda'} dx \frac{2v_+(x)+(i\hat{\lambda}x-1)v_1(x)}{1+x^2}
\end{flalign}
These equations can be solved to give:
\begin{flalign}
	v_1(y)=&i(1-e^{-2i\hat{\lambda}(\arctan(y)-\arctan(\Lambda'))})\\
	v_+(y)=&\frac{i}{2}(1-iy\hat{\lambda} + e^{-2i\hat{\lambda}(\arctan(y)-\arctan(\Lambda'))}(1+iy\hat{\lambda}))
\end{flalign}

Inserting this back into \eqref{6.1:vertex} the vertex function $V^{++}(q,p)$. Notably, the relation between spin-1 and spin-2 vertices in the collinear limit is given by $V^{++}(q,p)=p^+ V^+(q,p)$. To generalize the result to spin-$s$ currents, note that the free theory vertex in the bootstrap equation \eqref{6.1:vertexEq} in the collinear limit is modified by $(p^+)^{s-1}$ and similarly, Lorentz invariance requires us to modify the ansatz \eqref{6.1:vertex} by the same factor. The computation carries through identically, and we find:
\begin{flalign}
  V_s^{++...+}(q,p)=\beta_s (p^+)^{s-1} V^+(q,p)
\end{flalign}
Where $\beta_s$ is just an overall normalization factor. In our conventions \eqref{1FFCurrents} we have $\beta_2=1$ and $\beta_3=8i/3$.

An analogous relation was observed in \cite{Aharony:2012nh} when computing vertex functions in the QB Chern-Simons-Matter theory.

\printbibliography

@article{Maldacena:2011jn,
    author = "Maldacena, Juan and Zhiboedov, Alexander",
    title = "{Constraining Conformal Field Theories with A Higher Spin Symmetry}",
    eprint = "1112.1016",
    archivePrefix = "arXiv",
    primaryClass = "hep-th",
    doi = "10.1088/1751-8113/46/21/214011",
    journal = "J. Phys. A",
    volume = "46",
    pages = "214011",
    year = "2013"
}

@article{Maldacena:2012sf,
    author = "Maldacena, Juan and Zhiboedov, Alexander",
    title = "{Constraining conformal field theories with a slightly broken higher spin symmetry}",
    eprint = "1204.3882",
    archivePrefix = "arXiv",
    primaryClass = "hep-th",
    reportNumber = "PUPT-2410",
    doi = "10.1088/0264-9381/30/10/104003",
    journal = "Class. Quant. Grav.",
    volume = "30",
    pages = "104003",
    year = "2013"
}

@article{Klebanov:2002ja,
    author = "Klebanov, I. R. and Polyakov, A. M.",
    title = "{AdS dual of the critical O(N) vector model}",
    eprint = "hep-th/0210114",
    archivePrefix = "arXiv",
    reportNumber = "PUPT-2053",
    doi = "10.1016/S0370-2693(02)02980-5",
    journal = "Phys. Lett. B",
    volume = "550",
    pages = "213--219",
    year = "2002"
}

@inproceedings{Giombi:2016ejx,
    author = "Giombi, Simone",
    title = "{Higher Spin \textemdash{} CFT Duality}",
    booktitle = "{Theoretical Advanced Study Institute in Elementary Particle Physics}: {New Frontiers in Fields and Strings}",
    eprint = "1607.02967",
    archivePrefix = "arXiv",
    primaryClass = "hep-th",
    doi = "10.1142/9789813149441_0003",
    pages = "137--214",
    year = "2017"
}

@article{Giombi:2016zwa,
    author = "Giombi, S. and Gurucharan, V. and Kirilin, V. and Prakash, S. and Skvortsov, E.",
    title = "{On the Higher-Spin Spectrum in Large N Chern-Simons Vector Models}",
    eprint = "1610.08472",
    archivePrefix = "arXiv",
    primaryClass = "hep-th",
    reportNumber = "PUPT-2512, LMU-ASC-52-16",
    doi = "10.1007/JHEP01(2017)058",
    journal = "JHEP",
    volume = "01",
    pages = "058",
    year = "2017"
}

@article{Giombi:2011kc,
    author = "Giombi, Simone and Minwalla, Shiraz and Prakash, Shiroman and Trivedi, Sandip P. and Wadia, Spenta R. and Yin, Xi",
    title = "{Chern-Simons Theory with Vector Fermion Matter}",
    eprint = "1110.4386",
    archivePrefix = "arXiv",
    primaryClass = "hep-th",
    doi = "10.1140/epjc/s10052-012-2112-0",
    journal = "Eur. Phys. J. C",
    volume = "72",
    pages = "2112",
    year = "2012"
}

@article{Giombi:2011rz,
    author = "Giombi, Simone and Prakash, Shiroman and Yin, Xi",
    title = "{A Note on CFT Correlators in Three Dimensions}",
    eprint = "1104.4317",
    archivePrefix = "arXiv",
    primaryClass = "hep-th",
    doi = "10.1007/JHEP07(2013)105",
    journal = "JHEP",
    volume = "07",
    pages = "105",
    year = "2013"
}

@article{Jain:2022ajd,
    author = "Jain, Prabhav and Jain, Sachin and Sahoo, Bibhut and Dhruva, K. S. and Zade, Aashna",
    title = "{Mapping Large N Slightly Broken Higher Spin (SBHS) theory correlators to free theory correlators}",
    eprint = "2207.05101",
    archivePrefix = "arXiv",
    primaryClass = "hep-th",
    doi = "10.1007/JHEP12(2023)173",
    journal = "JHEP",
    volume = "12",
    pages = "173",
    year = "2023"
}

@article{Aharony:2018npf,
    author = "Aharony, Ofer and Alday, Luis F. and Bissi, Agnese and Yacoby, Ran",
    title = "{The Analytic Bootstrap for Large $N$ Chern-Simons Vector Models}",
    eprint = "1805.04377",
    archivePrefix = "arXiv",
    primaryClass = "hep-th",
    doi = "10.1007/JHEP08(2018)166",
    journal = "JHEP",
    volume = "08",
    pages = "166",
    year = "2018"
}

@article{Closset:2012vp,
    author = "Closset, Cyril and Dumitrescu, Thomas T. and Festuccia, Guido and Komargodski, Zohar and Seiberg, Nathan",
    title = "{Comments on Chern-Simons Contact Terms in Three Dimensions}",
    eprint = "1206.5218",
    archivePrefix = "arXiv",
    primaryClass = "hep-th",
    reportNumber = "PUTP-2417, WIS-10-12-JUNE-DPPA",
    doi = "10.1007/JHEP09(2012)091",
    journal = "JHEP",
    volume = "09",
    pages = "091",
    year = "2012"
}

@article{Gur-Ari:2012lgt,
    author = "Gur-Ari, Guy and Yacoby, Ran",
    title = "{Correlators of Large N Fermionic Chern-Simons Vector Models}",
    eprint = "1211.1866",
    archivePrefix = "arXiv",
    primaryClass = "hep-th",
    doi = "10.1007/JHEP02(2013)150",
    journal = "JHEP",
    volume = "02",
    pages = "150",
    year = "2013"
}

@article{Aharony:2012nh,
    author = "Aharony, Ofer and Gur-Ari, Guy and Yacoby, Ran",
    title = "{Correlation Functions of Large N Chern-Simons-Matter Theories and Bosonization in Three Dimensions}",
    eprint = "1207.4593",
    archivePrefix = "arXiv",
    primaryClass = "hep-th",
    reportNumber = "WIS-13-12-JUL-DPPA",
    doi = "10.1007/JHEP12(2012)028",
    journal = "JHEP",
    volume = "12",
    pages = "028",
    year = "2012"
}

@article{Aharony:2011jz,
    author = "Aharony, Ofer and Gur-Ari, Guy and Yacoby, Ran",
    title = "{d=3 Bosonic Vector Models Coupled to Chern-Simons Gauge Theories}",
    eprint = "1110.4382",
    archivePrefix = "arXiv",
    primaryClass = "hep-th",
    doi = "10.1007/JHEP03(2012)037",
    journal = "JHEP",
    volume = "03",
    pages = "037",
    year = "2012"
}

@article{Aharony:2015mjs,
    author = "Aharony, Ofer",
    title = "{Baryons, monopoles and dualities in Chern-Simons-matter theories}",
    eprint = "1512.00161",
    archivePrefix = "arXiv",
    primaryClass = "hep-th",
    reportNumber = "WIS-12-15-NOV-DPPA",
    doi = "10.1007/JHEP02(2016)093",
    journal = "JHEP",
    volume = "02",
    pages = "093",
    year = "2016"
}

@article{Hsin:2016blu,
    author = "Hsin, Po-Shen and Seiberg, Nathan",
    title = "{Level/rank Duality and Chern-Simons-Matter Theories}",
    eprint = "1607.07457",
    archivePrefix = "arXiv",
    primaryClass = "hep-th",
    doi = "10.1007/JHEP09(2016)095",
    journal = "JHEP",
    volume = "09",
    pages = "095",
    year = "2016"
}

@article{Aharony:2016jvv,
    author = "Aharony, Ofer and Benini, Francesco and Hsin, Po-Shen and Seiberg, Nathan",
    title = "{Chern-Simons-matter dualities with $SO$ and $USp$ gauge groups}",
    eprint = "1611.07874",
    archivePrefix = "arXiv",
    primaryClass = "cond-mat.str-el",
    reportNumber = "SISSA-62-2016-FISI",
    doi = "10.1007/JHEP02(2017)072",
    journal = "JHEP",
    volume = "02",
    pages = "072",
    year = "2017"
}

@article{Aharony:2012ns,
    author = "Aharony, Ofer and Giombi, Simone and Gur-Ari, Guy and Maldacena, Juan and Yacoby, Ran",
    title = "{The Thermal Free Energy in Large N Chern-Simons-Matter Theories}",
    eprint = "1211.4843",
    archivePrefix = "arXiv",
    primaryClass = "hep-th",
    reportNumber = "WIS-18-12-NOV-DPPA",
    doi = "10.1007/JHEP03(2013)121",
    journal = "JHEP",
    volume = "03",
    pages = "121",
    year = "2013"
}

@article{Aharony:2018pjn,
    author = "Aharony, Ofer and Jain, Sachin and Minwalla, Shiraz",
    title = "{Flows, Fixed Points and Duality in Chern-Simons-matter theories}",
    eprint = "1808.03317",
    archivePrefix = "arXiv",
    primaryClass = "hep-th",
    doi = "10.1007/JHEP12(2018)058",
    journal = "JHEP",
    volume = "12",
    pages = "058",
    year = "2018"
}

@article{Jain:2013gza,
    author = "Jain, Sachin and Minwalla, Shiraz and Yokoyama, Shuichi",
    title = "{Chern Simons duality with a fundamental boson and fermion}",
    eprint = "1305.7235",
    archivePrefix = "arXiv",
    primaryClass = "hep-th",
    reportNumber = "TIFR-TH-13-17",
    doi = "10.1007/JHEP11(2013)037",
    journal = "JHEP",
    volume = "11",
    pages = "037",
    year = "2013"
}

@article{Gabai:2022vri,
    author = "Gabai, Barak and Sever, Amit and Zhong, De-liang",
    title = "{Line Operators in Chern-Simons\textendash{}Matter Theories and Bosonization in Three Dimensions}",
    eprint = "2204.05262",
    archivePrefix = "arXiv",
    primaryClass = "hep-th",
    doi = "10.1103/PhysRevLett.129.121604",
    journal = "Phys. Rev. Lett.",
    volume = "129",
    number = "12",
    pages = "121604",
    year = "2022"
}

@article{Gabai:2022mya,
    author = "Gabai, Barak and Sever, Amit and Zhong, De-liang",
    title = "{Line operators in Chern-Simons-Matter theories and Bosonization in Three Dimensions II: Perturbative analysis and all-loop resummation}",
    eprint = "2212.02518",
    archivePrefix = "arXiv",
    primaryClass = "hep-th",
    doi = "10.1007/JHEP04(2023)070",
    journal = "JHEP",
    volume = "04",
    pages = "070",
    year = "2023"
}

@article{Gabai:2023lax,
    author = "Gabai, Barak and Sever, Amit and Zhong, De-liang",
    title = "{Bootstrapping smooth conformal defects in Chern-Simons-matter theories}",
    eprint = "2312.17132",
    archivePrefix = "arXiv",
    primaryClass = "hep-th",
    doi = "10.1007/JHEP03(2024)055",
    journal = "JHEP",
    volume = "03",
    pages = "055",
    year = "2024"
}

@article{Giveon:2008zn,
    author = "Giveon, Amit and Kutasov, David",
    title = "{Seiberg Duality in Chern-Simons Theory}",
    eprint = "0808.0360",
    archivePrefix = "arXiv",
    primaryClass = "hep-th",
    doi = "10.1016/j.nuclphysb.2008.09.045",
    journal = "Nucl. Phys. B",
    volume = "812",
    pages = "1--11",
    year = "2009"
}

@article{Benini:2011mf,
    author = "Benini, Francesco and Closset, Cyril and Cremonesi, Stefano",
    title = "{Comments on 3d Seiberg-like dualities}",
    eprint = "1108.5373",
    archivePrefix = "arXiv",
    primaryClass = "hep-th",
    reportNumber = "PU-2391, WIS-07-11-AUG-DPPA, TAUP-2931-11",
    doi = "10.1007/JHEP10(2011)075",
    journal = "JHEP",
    volume = "10",
    pages = "075",
    year = "2011"
}

@article{Kalloor:2019xjb,
    author = "Kalloor, Rohit R.",
    title = "{Four-point functions in large $N$ Chern-Simons fermionic theories}",
    eprint = "1910.14617",
    archivePrefix = "arXiv",
    primaryClass = "hep-th",
    doi = "10.1007/JHEP10(2020)028",
    journal = "JHEP",
    volume = "10",
    pages = "028",
    year = "2020"
}

@article{Li:2019twz,
    author = "Li, Zhijin",
    title = "{Bootstrapping conformal four-point correlators with slightly broken higher spin symmetry and $3D$ bosonization}",
    eprint = "1906.05834",
    archivePrefix = "arXiv",
    primaryClass = "hep-th",
    doi = "10.1007/JHEP10(2020)007",
    journal = "JHEP",
    volume = "10",
    pages = "007",
    year = "2020"
}

@article{Silva:2021ece,
    author = "Silva, Joao A.",
    title = "{Four point functions in CFT\textquoteright{}s with slightly broken higher spin symmetry}",
    eprint = "2103.00275",
    archivePrefix = "arXiv",
    primaryClass = "hep-th",
    doi = "10.1007/JHEP05(2021)097",
    journal = "JHEP",
    volume = "05",
    pages = "097",
    year = "2021"
}

@article{Turiaci:2018nua,
    author = "Turiaci, Gustavo J. and Zhiboedov, Alexander",
    title = "{Veneziano Amplitude of Vasiliev Theory}",
    eprint = "1802.04390",
    archivePrefix = "arXiv",
    primaryClass = "hep-th",
    doi = "10.1007/JHEP10(2018)034",
    journal = "JHEP",
    volume = "10",
    pages = "034",
    year = "2018"
}

@article{Yacoby:2018yvy,
    author = "Yacoby, Ran",
    title = "{Scalar Correlators in Bosonic Chern-Simons Vector Models}",
    eprint = "1805.11627",
    archivePrefix = "arXiv",
    primaryClass = "hep-th",
    month = "5",
    year = "2018"
}

@article{Jain:2021gwa,
    author = "Jain, Sachin and John, Renjan Rajan",
    title = "{Relation between parity-even and parity-odd CFT correlation functions in three dimensions}",
    eprint = "2107.00695",
    archivePrefix = "arXiv",
    primaryClass = "hep-th",
    doi = "10.1007/JHEP12(2021)067",
    journal = "JHEP",
    volume = "12",
    pages = "067",
    year = "2021"
}

@article{Jain:2021whr,
    author = "Jain, Sachin and John, Renjan Rajan and Mehta, Abhishek and S, Dhruva K.",
    title = "{Constraining momentum space CFT correlators with consistent position space OPE limit and the collider bound}",
    eprint = "2111.08024",
    archivePrefix = "arXiv",
    primaryClass = "hep-th",
    doi = "10.1007/JHEP02(2022)084",
    journal = "JHEP",
    volume = "02",
    pages = "084",
    year = "2022"
}

@article{Skvortsov:2018uru,
    author = "Skvortsov, Evgeny",
    title = "{Light-Front Bootstrap for Chern-Simons Matter Theories}",
    eprint = "1811.12333",
    archivePrefix = "arXiv",
    primaryClass = "hep-th",
    doi = "10.1007/JHEP06(2019)058",
    journal = "JHEP",
    volume = "06",
    pages = "058",
    year = "2019"
}

@article{Skvortsov:2020wtf,
    author = "Skvortsov, Evgeny and Tran, Tung and Tsulaia, Mirian",
    title = "{More on Quantum Chiral Higher Spin Gravity}",
    eprint = "2002.08487",
    archivePrefix = "arXiv",
    primaryClass = "hep-th",
    doi = "10.1103/PhysRevD.101.106001",
    journal = "Phys. Rev. D",
    volume = "101",
    number = "10",
    pages = "106001",
    year = "2020"
}

@article{Baumann:2020dch,
    author = "Baumann, Daniel and Duaso Pueyo, Carlos and Joyce, Austin and Lee, Hayden and Pimentel, Guilherme L.",
    title = "{The Cosmological Bootstrap: Spinning Correlators from Symmetries and Factorization}",
    eprint = "2005.04234",
    archivePrefix = "arXiv",
    primaryClass = "hep-th",
    doi = "10.21468/SciPostPhys.11.3.071",
    journal = "SciPost Phys.",
    volume = "11",
    pages = "071",
    year = "2021"
}

@article{Jain:2021vrv,
    author = "Jain, Sachin and John, Renjan Rajan and Mehta, Abhishek and Nizami, Amin A. and Suresh, Adithya",
    title = "{Higher spin 3-point functions in 3d CFT using spinor-helicity variables}",
    eprint = "2106.00016",
    archivePrefix = "arXiv",
    primaryClass = "hep-th",
    doi = "10.1007/JHEP09(2021)041",
    journal = "JHEP",
    volume = "09",
    pages = "041",
    year = "2021"
}

@unpublished{Kukolj:2024,
  author       = {Trivko Kukolj},
  title        = {Higher-spin Ward Identities of Chern-Simons-matter theory},
  school       = {},
  year         = {},
  month        = {},
  note         = {M.Sc. Thesis submitted to the Weizmann Institute on 02.02.2024.}
}

@article{Sezgin:2003pt,
    author = "Sezgin, E. and Sundell, P.",
    title = "{Holography in 4D (super) higher spin theories and a test via cubic scalar couplings}",
    eprint = "hep-th/0305040",
    archivePrefix = "arXiv",
    reportNumber = "MIFP-03-09, UU-07-03",
    doi = "10.1088/1126-6708/2005/07/044",
    journal = "JHEP",
    volume = "07",
    pages = "044",
    year = "2005"
}

@article{Seiberg:2016gmd,
    author = "Seiberg, Nathan and Senthil, T. and Wang, Chong and Witten, Edward",
    title = "{A Duality Web in 2+1 Dimensions and Condensed Matter Physics}",
    eprint = "1606.01989",
    archivePrefix = "arXiv",
    primaryClass = "hep-th",
    doi = "10.1016/j.aop.2016.08.007",
    journal = "Annals Phys.",
    volume = "374",
    pages = "395--433",
    year = "2016"
}

@article{Metsaev:1991mt,
    author = "Metsaev, R. R.",
    title = "{Poincare invariant dynamics of massless higher spins: Fourth order analysis on mass shell}",
    doi = "10.1142/S0217732391000348",
    journal = "Mod. Phys. Lett. A",
    volume = "6",
    pages = "359--367",
    year = "1991"
}

@article{Didenko:2022qga,
    author = "Didenko, V. E.",
    title = "{On holomorphic sector of higher-spin theory}",
    eprint = "2209.01966",
    archivePrefix = "arXiv",
    primaryClass = "hep-th",
    doi = "10.1007/JHEP10(2022)191",
    journal = "JHEP",
    volume = "10",
    pages = "191",
    year = "2022"
}

@article{Sharapov:2022nps,
    author = "Sharapov, Alexey and Skvortsov, Evgeny and Sukhanov, Arseny and Van Dongen, Richard",
    title = "{More on Chiral Higher Spin Gravity and convex geometry}",
    eprint = "2209.15441",
    archivePrefix = "arXiv",
    primaryClass = "hep-th",
    doi = "10.1016/j.nuclphysb.2023.116152",
    journal = "Nucl. Phys. B",
    volume = "990",
    pages = "116152",
    year = "2023"
}

@article{Sharapov:2022wpz,
    author = "Sharapov, Alexey and Skvortsov, Evgeny and Van Dongen, Richard",
    title = "{Chiral higher spin gravity and convex geometry}",
    eprint = "2209.01796",
    archivePrefix = "arXiv",
    primaryClass = "hep-th",
    doi = "10.21468/SciPostPhys.14.6.162",
    journal = "SciPost Phys.",
    volume = "14",
    number = "6",
    pages = "162",
    year = "2023"
}

@article{Metsaev:1991nb,
    author = "Metsaev, R. R.",
    title = "{S matrix approach to massless higher spins theory. 2: The Case of internal symmetry}",
    doi = "10.1142/S0217732391002839",
    journal = "Mod. Phys. Lett. A",
    volume = "6",
    pages = "2411--2421",
    year = "1991"
}

@article{Ponomarev:2016lrm,
    author = "Ponomarev, Dmitry and Skvortsov, E. D.",
    title = "{Light-Front Higher-Spin Theories in Flat Space}",
    eprint = "1609.04655",
    archivePrefix = "arXiv",
    primaryClass = "hep-th",
    doi = "10.1088/1751-8121/aa56e7",
    journal = "J. Phys. A",
    volume = "50",
    number = "9",
    pages = "095401",
    year = "2017"
}

@article{Skvortsov:2018jea,
    author = "Skvortsov, Evgeny D. and Tran, Tung and Tsulaia, Mirian",
    title = "{Quantum Chiral Higher Spin Gravity}",
    eprint = "1805.00048",
    archivePrefix = "arXiv",
    primaryClass = "hep-th",
    doi = "10.1103/PhysRevLett.121.031601",
    journal = "Phys. Rev. Lett.",
    volume = "121",
    number = "3",
    pages = "031601",
    year = "2018"
}

@article{Skvortsov:2020gpn,
    author = "Skvortsov, Evgeny and Tran, Tung",
    title = "{One-loop Finiteness of Chiral Higher Spin Gravity}",
    eprint = "2004.10797",
    archivePrefix = "arXiv",
    primaryClass = "hep-th",
    doi = "10.1007/JHEP07(2020)021",
    journal = "JHEP",
    volume = "07",
    pages = "021",
    year = "2020"
}

@article{Skvortsov:2022syz,
    author = "Skvortsov, Evgeny and Van Dongen, Richard",
    title = "{Minimal models of field theories: Chiral higher spin gravity}",
    eprint = "2204.10285",
    archivePrefix = "arXiv",
    primaryClass = "hep-th",
    doi = "10.1103/PhysRevD.106.045006",
    journal = "Phys. Rev. D",
    volume = "106",
    number = "4",
    pages = "045006",
    year = "2022"
}

@article{Sharapov:2022faa,
    author = "Sharapov, Alexey and Sharapov, Alexey and Skvortsov, Evgeny and Skvortsov, Evgeny and Sukhanov, Arseny and Sukhanov, Arseny and Van Dongen, Richard and Van Dongen, Richard",
    title = "{Minimal model of Chiral Higher Spin Gravity}",
    eprint = "2205.07794",
    archivePrefix = "arXiv",
    primaryClass = "hep-th",
    doi = "10.1007/JHEP09(2022)134",
    journal = "JHEP",
    volume = "09",
    pages = "134",
    year = "2022",
    note = "[Erratum: JHEP 02, 183 (2023)]"
}

@article{Vasiliev:1990en,
    author = "Vasiliev, Mikhail A.",
    title = "{Consistent equation for interacting gauge fields of all spins in (3+1)-dimensions}",
    reportNumber = "LEBEDEV-90-29",
    doi = "10.1016/0370-2693(90)91400-6",
    journal = "Phys. Lett. B",
    volume = "243",
    pages = "378--382",
    year = "1990"
}

@article{Aharony:2024nqs,
    author = "Aharony, Ofer and Kalloor, Rohit R. and Kukolj, Trivko",
    title = "{A chiral limit for Chern-Simons-matter theories}",
    eprint = "2405.01647",
    archivePrefix = "arXiv",
    primaryClass = "hep-th",
    month = "5",
    year = "2024"
}

@article{Jain:2024bza,
    author = "Jain, Sachin and S, Dhruva K. and Skvortsov, Evgeny",
    title = "{Hidden sectors of Chern-Simons Matter theories and Exact Holography}",
    eprint = "2405.00773",
    archivePrefix = "arXiv",
    primaryClass = "hep-th",
    month = "5",
    year = "2024"
}

@article{Skvortsov:2022wzo,
    author = "Skvortsov, Evgeny and Yin, Yihao",
    title = "{On (spinor)-helicity and bosonization in AdS$_{4}$/CFT$_{3}$}",
    eprint = "2207.06976",
    archivePrefix = "arXiv",
    primaryClass = "hep-th",
    doi = "10.1007/JHEP03(2023)204",
    journal = "JHEP",
    volume = "03",
    pages = "204",
    year = "2023"
}
\end{document}